%%%%%%%%%%%%%%%%%%     PLAIN TEX FILE
%%%%%%%%%%%%%%%%%%
 %%%%%%%%%%%%%%%%%%  %%%%%%%%%%%%%%%%%%  %%%%%%%%%%%%%%%%%%  %%%%%%%%%%%%%%%%%%
 %%%%%%%%%%%%%%%%%%  %%%%%%%%%%%%%%%%%%  %%%%%%%%%%%%%%%%%%  %%%%%%%%%%%%%%%%%%
 %%%%%%%%%%%%%%%%%%  %%%%%%%%%%%%%%%%%%  %%%%%%%%%%%%%%%%%%  %%%%%%%%%%%%%%%%%%

 %%%%%%%%%%%%%%%%%%  hex macros for preprints, cm version %%%%%%%%%%%%%%
%                     (P. Ginsparg, last updated 9/91)
%                if confused, type `b' in response to query
%
%---------------------------------------------------------------------%
%% site dependent options:
%% \unredoffs and \redoffs define horizontal and vertical offsets
%% respectively for unreduced and reduced modes. \speclscape defines
%% the \special{} call that sets printer to landscape (sideways) mode.
%% from standard set below, leave uncommented as appropriate or redefine
%
%%% next 400dpi
%\def\unredoffs{} \def\redoffs{\voffset=-.31truein\hoffset=-.48truein}
%\def\speclscape{\special{landscape}}
%
%%% apple lw
\def\unredoffs{} \def\redoffs{\voffset=-.31truein\hoffset=-.59truein}
\def\speclscape{\special{ps: landscape}}
%
%%% qms lasergrafix:
%\def\unredoffs{} \def\redoffs{\voffset=-.4truein\hoffset=.125truein}
%\def\speclscape{\special{qms: landscape}}
%
%%% saclay A4 paper:
%\def\unredoffs{\hoffset-.14truein\voffset-.2truein}
%\def\redoffs{\voffset=-.55truein\hoffset=-.1truein} \def\speclscape{}
%
%---------------------------------------------------------------------%
%
\newbox\leftpage \newdimen\fullhsize \newdimen\hstitle \newdimen\hsbody
\tolerance=1000\hfuzz=2pt
\catcode`\@=11 % This allows us to modify PLAIN macros.
\def\bigans{b }
%\message{ big or little (b/l)? }\read-1 to\answ
\def\answ{b }
\ifx\answ\bigans\message{(This will come out unreduced.}
\magnification=1200\unredoffs\baselineskip=16pt plus 2pt minus 1pt
\hsbody=\hsize \hstitle=\hsize %take default values for unreduced format
\else\message{(This will be reduced.} \let\l@r=L
\magnification=1000\baselineskip=16pt plus 2pt minus 1pt \vsize=7truein
\redoffs \hstitle=8truein\hsbody=4.75truein\fullhsize=10truein\hsize=\hsbody
\output={\ifnum\pageno=0 %%% This is the HUTP version
  \shipout\vbox{\speclscape{\hsize\fullhsize\makeheadline}
    \hbox to \fullhsize{\hfill\pagebody\hfill}}\advancepageno
  \else
  \almostshipout{\leftline{\vbox{\pagebody\makefootline}}}\advancepageno
  \fi}
\def\almostshipout#1{\if L\l@r \count1=1 \message{[\the\count0.\the\count1]}
      \global\setbox\leftpage=#1 \global\let\l@r=R
 \else \count1=2
  \shipout\vbox{\speclscape{\hsize\fullhsize\makeheadline}
      \hbox to\fullhsize{\box\leftpage\hfil#1}}  \global\let\l@r=L\fi}
\fi
%---------------------------------------------------------------------
%
\newcount\yearltd\yearltd=\year\advance\yearltd by -1900

\def\Title#1#2{\nopagenumbers\abstractfont\hsize=\hstitle\rightline{#1}%
\vskip 1in\centerline{\titlefont #2}\abstractfont\vskip .5in\pageno=0}
\def\Date#1{\vfill\leftline{#1}\tenpoint\supereject\global\hsize=\hsbody%
\footline={\hss\tenrm\folio\hss}}%      restores pagenumbers
%
%       use following instead of \Date on the preliminary draft,
%       puts date/time on each page in big mode, writes labels in margins

\def\draftmode{\message{ DRAFTMODE }\def\draftdate{{\rm preliminary draft:
\number\month/\number\day/\number\yearltd\ \ \hourmin}}%
\headline={\hfil\draftdate}\writelabels\baselineskip=20pt plus 2pt minus 2pt
 {\count255=\time\divide\count255 by 60 \xdef\hourmin{\number\count255}
  \multiply\count255 by-60\advance\count255 by\time
  \xdef\hourmin{\hourmin:\ifnum\count255<10 0\fi\the\count255}}}
%       use \nolabels to get rid of eqn, ref, and fig labels in draft mode
\def\nolabels{\def\wrlabeL##1{}\def\eqlabeL##1{}\def\reflabeL##1{}}
\def\writelabels{\def\wrlabeL##1{\leavevmode\vadjust{\rlap{\smash%
{\line{{\escapechar=` \hfill\rlap{\sevenrm\hskip.03in\string##1}}}}}}}%
\def\eqlabeL##1{{\escapechar-1\rlap{\sevenrm\hskip.05in\string##1}}}%
\def\reflabeL##1{\noexpand\llap{\noexpand\sevenrm\string\string\string##1}}}
\nolabels
%
% tagged sec numbers
\global\newcount\secno \global\secno=0
\global\newcount\meqno \global\meqno=1
\def\newsec#1{\global\advance\secno by1\message{(\the\secno. #1)}
%\ifx\answ\bigans \vfill\eject \else \bigbreak\bigskip \fi  %if desired
\global\subsecno=0\eqnres@t\noindent{\bf\the\secno. #1}
\writetoca{{\secsym} {#1}}\par\nobreak\medskip\nobreak}
\def\eqnres@t{\xdef\secsym{\the\secno.}\global\meqno=1\bigbreak\bigskip}
\def\sequentialequations{\def\eqnres@t{\bigbreak}}\xdef\secsym{}
\global\newcount\subsecno \global\subsecno=0
\def\subsec#1{\global\advance\subsecno by1\message{(\secsym\the\subsecno.
#1)}
\ifnum\lastpenalty>9000\else\bigbreak\fi
\noindent{\it\secsym\the\subsecno. #1}\writetoca{\string\quad
{\secsym\the\subsecno.} {#1}}\par\nobreak\medskip\nobreak}
\def\appendix#1#2{\global\meqno=1\global\subsecno=0\xdef\secsym{\hbox{#1.}}
\bigbreak\bigskip\noindent{\bf Appendix #1. #2}\message{(#1. #2)}
\writetoca{Appendix {#1.} {#2}}\par\nobreak\medskip\nobreak}
%
%       \eqn\label{a+b=c}       gives displayed equation, numbered
%                               consecutively within sections.
%     \eqnn and \eqna define labels in advance (of eqalign?)
%
\def\eqnn#1{\xdef #1{(\secsym\the\meqno)}\writedef{#1\leftbracket#1}%
\global\advance\meqno by1\wrlabeL#1}
\def\eqna#1{\xdef #1##1{\hbox{$(\secsym\the\meqno##1)$}}
\writedef{#1\numbersign1\leftbracket#1{\numbersign1}}%
\global\advance\meqno by1\wrlabeL{#1$\{\}$}}
\def\eqn#1#2{\xdef #1{(\secsym\the\meqno)}\writedef{#1\leftbracket#1}%
\global\advance\meqno by1$$#2\eqno#1\eqlabeL#1$$}
%
%                            footnotes
\newskip\footskip\footskip14pt plus 1pt minus 1pt %sets footnote baselineskip
\def\footnotefont{\ninepoint}\def\f@t#1{\footnotefont #1\@foot}
\def\f@@t{\baselineskip\footskip\bgroup\footnotefont\aftergroup\@foot\let\next}
\setbox\strutbox=\hbox{\vrule height9.5pt depth4.5pt width0pt}
\global\newcount\ftno \global\ftno=0
\def\foot{\global\advance\ftno by1\footnote{$^{\the\ftno}$}}
%
%say \footend to put footnotes at end
%will cause problems if \ref used inside \foot, instead use \nref before
\newwrite\ftfile
\def\footend{\def\foot{\global\advance\ftno by1\chardef\wfile=\ftfile
$^{\the\ftno}$\ifnum\ftno=1\immediate\openout\ftfile=foots.tmp\fi%
\immediate\write\ftfile{\noexpand\smallskip%
\noexpand\item{f\the\ftno:\ }\pctsign}\findarg}%
\def\footatend{\vfill\eject\immediate\closeout\ftfile{\parindent=20pt
\centerline{\bf Footnotes}\nobreak\bigskip\input foots.tmp }}}
\def\footatend{}
%
%     \ref\label{text}
% generates a number, assigns it to \label, generates an entry.
% To list the refs on a separate page,  \listrefs
%
\global\newcount\refno \global\refno=1
\newwrite\rfile
\def\ref{[\the\refno]\nref}
\def\nref#1{\xdef#1{[\the\refno]}\writedef{#1\leftbracket#1}%
\ifnum\refno=1\immediate\openout\rfile=refs.tmp\fi
\global\advance\refno by1\chardef\wfile=\rfile\immediate
\write\rfile{\noexpand\item{#1\ }\reflabeL{#1\hskip.31in}\pctsign}\findarg}
%        horrible hack to sidestep tex \write limitation
\def\findarg#1#{\begingroup\obeylines\newlinechar=`\^^M\pass@rg}
{\obeylines\gdef\pass@rg#1{\writ@line\relax #1^^M\hbox{}^^M}%
\gdef\writ@line#1^^M{\expandafter\toks0\expandafter{\striprel@x #1}%
\edef\next{\the\toks0}\ifx\next\em@rk\let\next=\endgroup\else\ifx\next\empty%
\else\immediate\write\wfile{\the\toks0}\fi\let\next=\writ@line\fi\next\relax}}
\def\striprel@x#1{} \def\em@rk{\hbox{}}
\def\lref{\begingroup\obeylines\lr@f}
\def\lr@f#1#2{\gdef#1{\ref#1{#2}}\endgroup\unskip}
\def\semi{;\hfil\break}
\def\addref#1{\immediate\write\rfile{\noexpand\item{}#1}} %now unnecessary
\def\startrefs#1{\immediate\openout\rfile=refs.tmp\refno=#1}
\def\xref{\expandafter\xr@f}\def\xr@f[#1]{#1}
\def\refs#1{\count255=1[\r@fs #1{\hbox{}}]}
\def\r@fs#1{\ifx\und@fined#1\message{reflabel \string#1 is undefined.}%
\nref#1{need to supply reference \string#1.}\fi%
\vphantom{\hphantom{#1}}\edef\next{#1}\ifx\next\em@rk\def\next{}%
\else\ifx\next#1\ifodd\count255\relax\xref#1\count255=0\fi%
\else#1\count255=1\fi\let\next=\r@fs\fi\next}
%

%
% this is ugly, but moore insists
\newwrite\ffile\global\newcount\figno \global\figno=1
\def\fig{fig.~\the\figno\nfig}
\def\nfig#1{\xdef#1{fig.~\the\figno}%
\writedef{#1\leftbracket fig.\noexpand~\the\figno}%
\ifnum\figno=1\immediate\openout\ffile=figs.tmp\fi\chardef\wfile=\ffile%
\immediate\write\ffile{\noexpand\medskip\noexpand\item{Fig.\ \the\figno. }
\reflabeL{#1\hskip.55in}\pctsign}\global\advance\figno by1\findarg}
\def\xfig{\expandafter\xf@g}\def\xf@g fig.\penalty\@M\ {}
\def\figs#1{figs.~\f@gs #1{\hbox{}}}
\def\f@gs#1{\edef\next{#1}\ifx\next\em@rk\def\next{}\else
\ifx\next#1\xfig #1\else#1\fi\let\next=\f@gs\fi\next}
\newwrite\lfile
{\escapechar-1\xdef\pctsign{\string\%}\xdef\leftbracket{\string\{}
\xdef\rightbracket{\string\}}\xdef\numbersign{\string\#}}

\def\writestop{\def\writestoppt{\immediate\write\lfile{\string\pageno%
\the\pageno\string\startrefs\leftbracket\the\refno\rightbracket%
\string\def\string\secsym\leftbracket\secsym\rightbracket%
\string\secno\the\secno\string\meqno\the\meqno}\immediate\closeout\lfile}}
\def\writestoppt{}\def\writedef#1{}
\def\seclab#1{\xdef #1{\the\secno}\writedef{#1\leftbracket#1}\wrlabeL{#1=#1}}
\def\subseclab#1{\xdef #1{\secsym\the\subsecno}%
\writedef{#1\leftbracket#1}\wrlabeL{#1=#1}}
\newwrite\tfile \def\writetoca#1{}
\def\leaderfill{\leaders\hbox to 1em{\hss.\hss}\hfill}
%        use this to write file with table of contents
\def\writetoc{\immediate\openout\tfile=toc.tmp
   \def\writetoca##1{{\edef\next{\write\tfile{\noindent ##1
   \string\leaderfill {\noexpand\number\pageno} \par}}\next}}}
%       and this lists table of contents on second pass

%
\catcode`\@=12 % at signs are no longer letters
%
%        Unpleasantness in calling in abstract and title fonts
\edef\tfontsize{\ifx\answ\bigans scaled\magstep3\else scaled\magstep4\fi}
\font\titlerm=cmr10 \tfontsize \font\titlerms=cmr7 \tfontsize
\font\titlermss=cmr5 \tfontsize \font\titlei=cmmi10 \tfontsize
\font\titleis=cmmi7 \tfontsize \font\titleiss=cmmi5 \tfontsize
\font\titlesy=cmsy10 \tfontsize \font\titlesys=cmsy7 \tfontsize
\font\titlesyss=cmsy5 \tfontsize \font\titleit=cmti10 \tfontsize
\skewchar\titlei='177 \skewchar\titleis='177 \skewchar\titleiss='177
\skewchar\titlesy='60 \skewchar\titlesys='60 \skewchar\titlesyss='60
\def\titlefont{\def\rm{\fam0\titlerm}% switch to title font
\textfont0=\titlerm \scriptfont0=\titlerms \scriptscriptfont0=\titlermss
\textfont1=\titlei \scriptfont1=\titleis \scriptscriptfont1=\titleiss
\textfont2=\titlesy \scriptfont2=\titlesys \scriptscriptfont2=\titlesyss
\textfont\itfam=\titleit \def\it{\fam\itfam\titleit}\rm}
 \ifx\answ\bigans\else scaled\magstep1\fi
\ifx\answ\bigans\def\abstractfont{\tenpoint}\else
\font\abssl=cmsl10 scaled \magstep1
\font\absrm=cmr10 scaled\magstep1 \font\absrms=cmr7 scaled\magstep1
\font\absrmss=cmr5 scaled\magstep1 \font\absi=cmmi10 scaled\magstep1
\font\absis=cmmi7 scaled\magstep1 \font\absiss=cmmi5 scaled\magstep1
\font\abssy=cmsy10 scaled\magstep1 \font\abssys=cmsy7 scaled\magstep1
\font\abssyss=cmsy5 scaled\magstep1 \font\absbf=cmbx10 scaled\magstep1
\skewchar\absi='177 \skewchar\absis='177 \skewchar\absiss='177
\skewchar\abssy='60 \skewchar\abssys='60 \skewchar\abssyss='60
\def\abstractfont{\def\rm{\fam0\absrm}% switch to abstract font
\textfont0=\absrm \scriptfont0=\absrms \scriptscriptfont0=\absrmss
\textfont1=\absi \scriptfont1=\absis \scriptscriptfont1=\absiss
\textfont2=\abssy \scriptfont2=\abssys \scriptscriptfont2=\abssyss
\textfont\itfam=\bigit \def\it{\fam\itfam\bigit}\def\footnotefont{\tenpoint}%
\textfont\slfam=\abssl \def\sl{\fam\slfam\abssl}%
\textfont\bffam=\absbf \def\bf{\fam\bffam\absbf}\rm}\fi
\def\tenpoint{\def\rm{\fam0\tenrm}% switch back to 10-point type
\textfont0=\tenrm \scriptfont0=\sevenrm \scriptscriptfont0=\fiverm
\textfont1=\teni  \scriptfont1=\seveni  \scriptscriptfont1=\fivei
\textfont2=\tensy \scriptfont2=\sevensy \scriptscriptfont2=\fivesy
\textfont\itfam=\tenit
\def\it{\fam\itfam\tenit}\def\footnotefont{\ninepoint}%
\textfont\bffam=\tenbf \def\bf{\fam\bffam\tenbf}\def\sl{\fam\slfam\tensl}\rm}
\font\ninerm=cmr9 \font\sixrm=cmr6 \font\ninei=cmmi9 \font\sixi=cmmi6
\font\ninesy=cmsy9 \font\sixsy=cmsy6 \font\ninebf=cmbx9
\font\nineit=cmti9 \font\ninesl=cmsl9 \skewchar\ninei='177
\skewchar\sixi='177 \skewchar\ninesy='60 \skewchar\sixsy='60
\def\ninepoint{\def\rm{\fam0\ninerm}% switch to footnote font
\textfont0=\ninerm \scriptfont0=\sixrm \scriptscriptfont0=\fiverm
\textfont1=\ninei \scriptfont1=\sixi \scriptscriptfont1=\fivei
\textfont2=\ninesy \scriptfont2=\sixsy \scriptscriptfont2=\fivesy
\textfont\itfam=\ninei \def\it{\fam\itfam\nineit}\def\sl{\fam\slfam\ninesl}%
\textfont\bffam=\ninebf \def\bf{\fam\bffam\ninebf}\rm}
%
%---------------------------------------------------------------------
%

\hyphenation{anom-aly anom-alies coun-ter-term coun-ter-terms}
\def\inv{^{\raise.15ex\hbox{${\scriptscriptstyle -}$}\kern-.05em 1}}

\def\Dsl{\,\raise.15ex\hbox{/}\mkern-13.5mu D} %this one can be subscripted
\def\dsl{\raise.15ex\hbox{/}\kern-.57em\partial}

\def\tr{{\rm tr}} 
\font\bigit=cmti10 scaled \magstep1
 %pound sterling
\def\lspace{\ifx\answ\bigans{}\else\qquad\fi}
\def\lbspace{\ifx\answ\bigans{}\else\hskip-.2in\fi} % $$\lbspace...$$
\def\boxeqn#1{\vcenter{\vbox{\hrule\hbox{\vrule\kern3pt\vbox{\kern3pt
           \hbox{${\displaystyle #1}$}\kern3pt}\kern3pt\vrule}\hrule}}}
\def\mbox#1#2{\vcenter{\hrule \hbox{\vrule height#2in
               \kern#1in \vrule} \hrule}}  %e.g. \mbox{.1}{.1}
%       matters of taste
%\def\tilde{\widetilde} \def\bar{\overline} \def\hat{\widehat}
%
% some sample definitions
  %     curly letters

\def\e#1{{\rm e}^{^{\textstyle#1}}}

\def\darr#1{\raise1.5ex\hbox{$\leftrightarrow$}\mkern-16.5mu #1}
 %pound sterling

 %puts a small half in a displayed eqn
\def\roughly#1{\raise.3ex\hbox{$#1$\kern-.75em\lower1ex\hbox{$\sim$}}}

%\input harvmac.tex

%%temporary additional macros
% \input macros.tex
% April 16 -- NN

%%%%%%%%%%%%%%%%%%%%%  Rublenye bukvy   %%%%%%%%%%%%%%%%%%%%%%%%
\def\IB{\relax\hbox{$\inbar\kern-.3em{\rm B}$}}
\def\IC{\relax\hbox{$\inbar\kern-.3em{\rm C}$}}
\def\ID{\relax\hbox{$\inbar\kern-.3em{\rm D}$}}
\def\IE{\relax\hbox{$\inbar\kern-.3em{\rm E}$}}
\def\IF{\relax\hbox{$\inbar\kern-.3em{\rm F}$}}
\def\IG{\relax\hbox{$\inbar\kern-.3em{\rm G}$}}
\def\IGa{\relax\hbox{${\rm I}\kern-.18em\Gamma$}}
\def\IH{\relax{\rm I\kern-.18em H}}
\def\IK{\relax{\rm I\kern-.18em K}}
\def\II{\relax{\rm I\kern-.18em I}}
\def\IL{\relax{\rm I\kern-.18em L}}
\def\IP{\relax{\rm I\kern-.18em P}}
\def\IR{\relax{\rm I\kern-.18em R}}
\def\IZ{\relax\ifmmode\mathchoice {\hbox{\cmss Z\kern-.4em Z}}{\hbox{\cmss
Z\kern-.4em Z}} {\lower.9pt\hbox{\cmsss Z\kern-.4em Z}}
{\lower1.2pt\hbox{\cmsss Z\kern-.4em Z}}\else{\cmss Z\kern-.4em Z}\fi}

\def\IB{\relax{\rm I\kern-.18em B}}
\def\IC{{\relax\hbox{$\inbar\kern-.3em{\rm C}$}}}
\def\ID{\relax{\rm I\kern-.18em D}}
\def\IE{\relax{\rm I\kern-.18em E}}
\def\IF{\relax{\rm I\kern-.18em F}}

%%%%%%%%%%%%%%%%%%%% Calligraphic letters  %%%%%%%%%%%%%%%%%%%%%%%

\def\CW {{\cal W}}

%%%%%%%%%%%%%%%%%%%%%%%%%% Derivatives  %%%%%%%%%%%%%%%%%%%%%%%%
\def\p{\partial}

%%Beltrami

%%%%%%%%%%%%%%%%%%%% letters with bar %%%%%%%%%%%%%%%%%%%%%%%%%%

%%%%%%%%%%%%%%%%%%%%%%%%%%% Math symbols %%%%%%%%%%%%%%%%%%%%%%%

\def\s{\lies}

%%%%%%%%%%%%%%%%%%%%% Short Cuts %%%%%%%%%%%%%%%%%%%%%%%

%%%%%%%%%%%%%%%%%% Greek %%%%%%%%%%%%%%%%%%%%%%

\def\a{\alpha}
\def\b{\beta}
\def\g{\gamma}  \def\G{\Gamma}
\def\d{\delta}  
\def\m{\mu}
\def\n{\nu}

\def\l{\lambda} 
\def\k{\kappa}
\def\e{\epsilon}

%%%%%%%%%%%%%%%%%% Big ( )  %%%%%%%%%%%%%%%%%%%%%%
\def\|{\Big|}
\def\({\Big(}   \def\){\Big)}
\def\[{\Big[}   \def\]{\Big]}

%%%%%%%%%%%%%%%%%% Text %%%%%%%%%%%%%%%%%%%%%%

%%%%%%%%%%%%% References %%%%%%%%%%%%%%%%%%%%

\def\paper#1#2#3#4{#1, {\sl #2}, #3 {\tt #4}}
% refs with #1=authors, #2=title, #3=publ.ref, #4=hep no :
%\lref\NAME{\paper
%{Authors}{Title(in \it)}{\PLB{No.}{Year}{page},}
%{\hh 0206053 (in\tt)}.}

%\def\hh#1{hep-th/{\it #1}}
\def\hh{hep-th/}

% journal~{\bf no.} (year) page

\def\PLB#1#2#3{Phys. Lett.~{\bf B#1} (#2) #3}
\def\NPB#1#2#3{Nucl. Phys.~{\bf B#1} (#2) #3}
\def\PRL#1#2#3{Phys. Rev. Lett.~{\bf #1} (#2) #3}
\def\CMP#1#2#3{Comm. Math. Phys.~{\bf #1} (#2) #3}
\def\PRD#1#2#3{Phys. Rev.~{\bf D#1} (#2) #3}
\def\MPL#1#2#3{Mod. Phys. Lett.~{\bf #1} (#2) #3}
\def\IJMP#1#2#3{Int. Jour. Mod. Phys.~{\bf #1} (#2) #3}

%%%%%%%%%%%%%%%%%%% Something to deal with sub-sub-sections
%%%%%%%%%%%%%%%%%%%%%%%%%%%%%%%%%%%%%%%%%%%%%%%

\def\unlockat{\catcode`\@=11}
\def\lockat{\catcode`\@=12}

\unlockat

% Something to deal with sub-sub-sections

\def\newsec#1{\global\advance\secno by1\message{(\the\secno. #1)}
\global\subsecno=0\global\subsubsecno=0\eqnres@t\noindent {\bf\the\secno. #1}
\writetoca{{\secsym} {#1}}\par\nobreak\medskip\nobreak}
\global\newcount\subsecno \global\subsecno=0
\def\subsec#1{\global\advance\subsecno by1\message{(\secsym\the\subsecno.
#1)}
\ifnum\lastpenalty>9000\else\bigbreak\fi\global\subsubsecno=0
\noindent{\it\secsym\the\subsecno. #1}
\writetoca{\string\quad {\secsym\the\subsecno.} {#1}}
\par\nobreak\medskip\nobreak}
\global\newcount\subsubsecno \global\subsubsecno=0
\def\subsubsec#1{\global\advance\subsubsecno by1
\message{(\secsym\the\subsecno.\the\subsubsecno. #1)}
\ifnum\lastpenalty>9000\else\bigbreak\fi
\noindent\quad{\secsym\the\subsecno.\the\subsubsecno.}{#1}
\writetoca{\string\qquad{\secsym\the\subsecno.\the\subsubsecno.}{#1}}
\par\nobreak\medskip\nobreak}

\def\subsubseclab#1{\DefWarn#1\xdef #1{\noexpand\hyperref{}{subsubsection}%
{\secsym\the\subsecno.\the\subsubsecno}%
{\secsym\the\subsecno.\the\subsubsecno}}%
\writedef{#1\leftbracket#1}\wrlabeL{#1=#1}}% Macros for boxes
\lockat

%why???\font\manual=manfnt
\def\dbend{\lower3.5pt\hbox{\manual\char127}}

%%%%%%%%%%%%%%%%%%% Macros for boxes %%%%%%%%%%%%%%%%%%

\def\boxit#1{\vbox{\hrule\hbox{\vrule\kern8pt
\vbox{\hbox{\kern8pt}\hbox{\vbox{#1}}\hbox{\kern8pt}}
\kern8pt\vrule}\hrule}}

\def\mathboxit#1{\vbox{\hrule\hbox{\vrule\kern8pt\vbox{\kern8pt
\hbox{$\displaystyle #1$}\kern8pt}\kern8pt\vrule}\hrule}}

%%%%%%%%%%%%%%%%%%%% ANOTHER SET OF MACROS %%%%%%%%%%%%%%%%%%

\def\inbar{\,\vrule height1.5ex width.4pt depth0pt}

\font\cmss=cmss10 \font\cmsss=cmss10 at 7pt

%REFERENCES
%%%%%%%%%%%%%%%%%%%%%%%%%%%%%%%%%%%%%%%%%%%%%%%%%%%

\lref\simons{ J. Cheeger and J. Simons, {\it Differential Characters and
Geometric Invariants},  Stony Brook Preprint, (1973), unpublished.}

\lref\cargese{ L.~Baulieu, {\it Algebraic quantization of gauge theories},
Perspectives in fields and particles, Plenum Press, eds. Basdevant-Levy,
Cargese Lectures 1983}

\lref\antifields{ L. Baulieu, M. Bellon, S. Ouvry, C.Wallet, Phys.Letters
B252 (1990) 387; M.  Bocchichio, Phys. Lett. B187 (1987) 322;  Phys. Lett. B
192 (1987) 31; R.  Thorn    Nucl. Phys.   B257 (1987) 61. }

\lref\thompson{ George Thompson,  Annals Phys. 205 (1991) 130; J.M.F.
Labastida, M. Pernici, Phys. Lett. 212B  (1988) 56; D. Birmingham, M.Blau,
M. Rakowski and G.Thompson, Phys. Rept. 209 (1991) 129.}

\lref\tonin{ Tonin}

\lref\wittensix{ E.  Witten, {\it New  Gauge  Theories In Six Dimensions},
\hh{9710065}. }

\lref\orlando{ O. Alvarez, L. A. Ferreira and J. Sanchez Guillen, {\it  A New
Approach to Integrable Theories in any Dimension}, hep-th/9710147.}

\lref\wittentopo{ E.  Witten,  {\it  Topological Quantum Field Theory},
\hh9403195, Commun.  Math. Phys.  {117} (1988)353.  }

\lref\wittentwist{ E.  Witten, {\it Supersymmetric Yang--Mills theory on a
four-manifold}, J.  Math.  Phys.  {35} (1994) 5101.}

\lref\west{ L.~Baulieu, P.~West, {\it Six Dimensional TQFTs and  Self-dual
Two-Forms,} Phys.Lett. B {\bf 436 } (1998) 97, /hep-th/9805200}

\lref\bv{ I.A. Batalin and V.A. Vilkowisky,    Phys. Rev.   D28  (1983)
2567\semi M. Henneaux,  Phys. Rep.  126   (1985) 1\semi M. Henneaux and C.
Teitelboim, {\it Quantization of Gauge Systems}
  Princeton University Press,  Princeton (1992).}

\lref\kyoto{ L. Baulieu, E. Bergschoeff and E. Sezgin, Nucl. Phys.
B307(1988)348\semi L. Baulieu,   {\it Field Antifield Duality, p-Form Gauge
Fields
   and Topological Quantum Field Theories}, hep-th/9512026,
   Nucl. Phys. B478 (1996) 431.  }

\lref\sourlas{ G. Parisi and N. Sourlas, {\it Random Magnetic Fields,
Supersymmetry and Negative Dimensions}, Phys. Rev. Lett.  43 (1979) 744;
Nucl.  Phys.  B206 (1982) 321.  }

\lref\SalamSezgin{ A.  Salam  and  E.  Sezgin, {\it Supergravities in
diverse dimensions}, vol.  1, p. 119\semi P.  Howe, G.  Sierra and P.
Townsend, Nucl Phys B221 (1983) 331.}

\lref\nekrasov{ A. Losev, G. Moore, N. Nekrasov, S. Shatashvili, {\it
Four-Dimensional Avatars of Two-Dimensional RCFT},  hep-th/9509151, Nucl.
Phys.  Proc.  Suppl.   46 (1996) 130\semi L.  Baulieu, A.  Losev,
N.~Nekrasov  {\it Chern-Simons and Twisted Supersymmetry in Higher
Dimensions},  hep-th/9707174, to appear in Nucl.  Phys.  B.  }

\lref\WitDonagi{R.~ Donagi, E.~ Witten, ``Supersymmetric Yang--Mills Theory
and Integrable Systems'', hep-th/9510101, Nucl. Phys.{\bf B}460 (1996)
299-334}
\lref\Witfeb{E.~ Witten, ``Supersymmetric Yang--Mills Theory On A
Four-Manifold,''  hep-th/9403195; J. Math. Phys. {\bf 35} (1994) 5101.}
\lref\Witgrav{E.~ Witten, ``Topological Gravity'', Phys.Lett.206B:601, 1988}
\lref\witaffl{I. ~ Affleck, J.A.~ Harvey and E.~ Witten,
        ``Instantons and (Super)Symmetry Breaking
        in $2+1$ Dimensions'', Nucl. Phys. {\bf B}206 (1982) 413}
\lref\wittabl{E.~ Witten,  ``On $S$-Duality in Abelian Gauge Theory,''
hep-th/9505186; Selecta Mathematica {\bf 1} (1995) 383}
\lref\wittgr{E.~ Witten, ``The Verlinde Algebra And The Cohomology Of The
Grassmannian'',  hep-th/9312104}
\lref\wittenwzw{E. Witten, ``Non abelian bosonization in two dimensions,''
Commun. Math. Phys. {\bf 92} (1984)455 }
\lref\witgrsm{E. Witten, ``Quantum field theory, grassmannians and algebraic
curves,'' Commun.Math.Phys.113:529,1988}
\lref\wittjones{E. Witten, ``Quantum field theory and the Jones
polynomial,'' Commun.  Math. Phys., 121 (1989) 351. }
\lref\witttft{E.~ Witten, ``Topological Quantum Field Theory", Commun. Math.
Phys. {\bf 117} (1988) 353.}
\lref\wittmon{E.~ Witten, ``Monopoles and Four-Manifolds'', hep-th/9411102}
\lref\Witdgt{ E.~ Witten, ``On Quantum gauge theories in two dimensions,''
Commun. Math. Phys. {\bf  141}  (1991) 153}
\lref\witrevis{E.~ Witten,
 ``Two-dimensional gauge theories revisited'', hep-th/9204083; J. Geom.
Phys. 9 (1992) 303-368}
\lref\Witgenus{E.~ Witten, ``Elliptic Genera and Quantum Field Theory'',
Comm. Math. Phys. 109(1987) 525. }
\lref\OldZT{E. Witten, ``New Issues in Manifolds of SU(3) Holonomy,'' {\it
Nucl. Phys.} {\bf B268} (1986) 79 \semi J. Distler and B. Greene, ``Aspects
of (2,0) String Compactifications,'' {\it Nucl. Phys.} {\bf B304} (1988) 1
\semi B. Greene, ``Superconformal Compactifications in Weighted Projective
Space,'' {\it Comm. Math. Phys.} {\bf 130} (1990) 335.}
\lref\bagger{E.~ Witten and J. Bagger, Phys. Lett. {\bf 115B}(1982) 202}
\lref\witcurrent{E.~ Witten,``Global Aspects of Current Algebra'',
Nucl.Phys.B223 (1983) 422\semi ``Current Algebra, Baryons and Quark
Confinement'', Nucl.Phys. B223 (1993) 433}
\lref\Wittreiman{S.B. Treiman, E. Witten, R. Jackiw, B. Zumino, ``Current
Algebra and Anomalies'', Singapore, Singapore: World Scientific ( 1985) }
\lref\Witgravanom{L. Alvarez-Gaume, E.~ Witten, ``Gravitational Anomalies'',
Nucl.Phys.B234:269,1984. }

\lref\nicolai{\paper {H.~Nicolai}{New Linear Systems for 2D Poincar\'e
Supergravities}{\NPB{414}{1994}{299},}{\hh 9309052}.}

%%%%%%
%% References herein
%%%%%%%%%%%%%%%%%%%%%%%%%%%%%%%%%%%%%%%%%%%%%%%%

%%\lref\NAME{\paper
%%{Authors}{Title(in \sl)}{\PLB{No.}{Year}{page},}
%%{\hh 0206053 (in\tt)}.}

\lref\baex{\paper {L.~Baulieu, B.~Grossman}{Monopoles and Topological Field
Theory}{\PLB{214}{1988}{223}.}{}\paper {L.~Baulieu}{Chern-Simons
Three-Dimensional and
Yang--Mills-Higgs Two-Dimensional Systems as Four-Dimensional Topological
Quantum Field Theories}{\PLB{232}{1989}{473}.}{}}

\lref\bg{\paper {L.~Baulieu, B.~Grossman}{Monopoles and Topological Field
Theory}{\PLB{214}{1988}{223}.}{}}

\lref\seibergsix{\paper {N.~Seiberg}{Non-trivial Fixed Points of The
Renormalization Group in Six
 Dimensions}{\PLB{390}{1997}{169}}{\hh 9609161}\semi
\paper {O.J.~Ganor, D.R.~Morrison, N.~Seiberg}{
  Branes, Calabi-Yau Spaces, and Toroidal Compactification of the N=1
  Six-Dimensional $E_8$ Theory}{\NPB{487}{1997}{93}}{\hh 9610251}\semi
\paper {O.~Aharony, M.~Berkooz, N.~Seiberg}{Light-Cone
  Description of (2,0) Superconformal Theories in Six
  Dimensions}{Adv. Theor. Math. Phys. {\bf 2} (1998) 119}{\hh 9712117.}}

\lref\cs{\paper {L.~Baulieu}{Chern-Simons Three-Dimensional and
Yang--Mills-Higgs Two-Dimensional Systems as Four-Dimensional Topological
Quantum Field Theories}{\PLB{232}{1989}{473}.}{}}

\lref\beltrami{\paper {L.~Baulieu, M.~Bellon}{Beltrami Parametrization and
String Theory}{\PLB{196}{1987}{142}}{}\semi
\paper {L.~Baulieu, M.~Bellon, R.~Grimm}{Beltrami Parametrization For
Superstrings}{\PLB{198}{1987}{343}}{}\semi
\paper {R.~Grimm}{Left-Right Decomposition of Two-Dimensional Superspace
Geometry and Its BRS Structure}{Annals Phys. {\bf 200} (1990) 49.}{}}

\lref\bbg{\paper {L.~Baulieu, M.~Bellon, R.~Grimm}{Left-Right Asymmetric
Conformal Anomalies}{\PLB{228}{1989}{325}.}{}}

\lref\bonora{\paper {G.~Bonelli, L.~Bonora, F.~Nesti}{String Interactions
from Matrix String Theory}{\NPB{538}{1999}{100},}{\hh 9807232}\semi
\paper {G.~Bonelli, L.~Bonora, F.~Nesti, A.~Tomasiello}{Matrix String Theory
and its Moduli Space}{}{\hh 9901093.}}

\lref\corrigan{\paper {E.~Corrigan, C.~Devchand, D.B.~Fairlie,
J.~Nuyts}{First Order Equations for Gauge Fields in Spaces of Dimension
Greater Than Four}{\NPB{214}{452}{1983}.}{}}

\lref\acha{\paper {B.S.~Acharya, M.~O'Loughlin, B.~Spence}{Higher
Dimensional Analogues of Donaldson-Witten Theory}{\NPB{503}{1997}{657},}{\hh
9705138}\semi
\paper {B.S.~Acharya, J.M.~Figueroa-O'Farrill, M.~O'Loughlin,
B.~Spence}{Euclidean
  D-branes and Higher-Dimensional Gauge   Theory}{\NPB{514}{1998}{583},}{\hh
  9707118.}}

\lref\Witr{\paper{E.~Witten}{Introduction to Cohomological Field   Theories}
{Lectures at Workshop on Topological Methods in Physics (Trieste, Italy, Jun
11-25, 1990), \IJMP{A6}{1991}{2775}.}{}}

\lref\ohta{\paper {L.~Baulieu, N.~Ohta}{Worldsheets with Extended
Supersymmetry} {\PLB{391}{1997}{295},}{\hh 9609207}.}

\lref\gravity{\paper {L.~Baulieu}{Transmutation of Pure 2-D Supergravity
Into Topological 2-D Gravity and Other Conformal Theories}
{\PLB{288}{1992}{59},}{\hh 9206019.}}

\lref\wgravity{\paper {L.~Baulieu, M.~Bellon, R.~Grimm}{Some Remarks on  the
Gauging of the Virasoro and   $w_{1+\infty}$
Algebras}{\PLB{260}{1991}{63}.}{}}

\lref\fourd{\paper {E.~Witten}{Topological Quantum Field
Theory}{\CMP{117}{1988}{353}}{}\semi
\paper {L.~Baulieu, I.M.~Singer}{Topological Yang--Mills Symmetry}{Nucl.
Phys. Proc. Suppl. {\bf 15B} (1988) 12.}{}}

\lref\topo{\paper {L.~Baulieu}{On the Symmetries of Topological Quantum Field
Theories}{\IJMP{A10}{1995}{4483},}{\hh 9504015}\semi
\paper {R.~Dijkgraaf, G.~Moore}{Balanced Topological Field
Theories}{\CMP{185}{1997}{411},}{\hh 9608169.}}

\lref\wwgravity{\paper {I.~Bakas} {The Large $N$ Limit   of Extended
Conformal Symmetries}{\PLB{228}{1989}{57}.}{}}

\lref\wwwgravity{\paper {C.M.~Hull}{Lectures on $\CW$-Gravity,
$\CW$-Geometry and
$\CW$-Strings}{}{\hh 9302110}, and~references therein.}

\lref\wvgravity{\paper {A.~Bilal, V.~Fock, I.~Kogan}{On the origin of
$W$-algebras}{\NPB{359}{1991}{635}.}{}}

\lref\surprises{\paper {E.~Witten} {Surprises with Topological Field
Theories} {Lectures given at ``Strings 90'', Texas A\&M, 1990,}{Preprint
IASSNS-HEP-90/37.}}

\lref\stringsone{\paper {L.~Baulieu, M.B.~Green, E.~Rabinovici}{A Unifying
Topological Action for Heterotic and  Type II Superstring  Theories}
{\PLB{386}{1996}{91},}{\hh 9606080.}}

\lref\stringsN{\paper {L.~Baulieu, M.B.~Green, E.~Rabinovici}{Superstrings
from   Theories with $N>1$ World Sheet Supersymmetry}
{\NPB{498}{1997}{119},}{\hh 9611136.}}

\lref\bks{\paper {L.~Baulieu, H.~Kanno, I.~Singer}{Special Quantum Field
Theories in Eight and Other Dimensions}{\CMP{194}{1998}{149},}{\hh
9704167}\semi
\paper {L.~Baulieu, H.~Kanno, I.~Singer}{Cohomological Yang--Mills Theory
  in Eight Dimensions}{ Talk given at APCTP Winter School on Dualities in
String Theory (Sokcho, Korea, February 24-28, 1997),} {\hh 9705127.}}

\lref\witdyn{\paper {P.~Townsend}{The eleven dimensional supermembrane
revisited}{\PLB{350}{1995}{184},}{\hh9501068}\semi
\paper{E.~Witten}{String Theory Dynamics in Various Dimensions}
{\NPB{443}{1995}{85},}{\hh 9503124}.}

\lref\bfss{\paper {T.~Banks, W.Fischler, S.H.~Shenker,
L.~Susskind}{$M$-Theory as a Matrix Model~:
A~Conjecture}{\PRD{55}{1997}{5112},} {\hh9610043.}}

\lref\seiberg{\paper {N.~Seiberg}{Why is the Matrix Model
Correct?}{\PRL{79}{1997}{3577},} {\hh 9710009.}}

\lref\sen{\paper {A.~Sen}{$D0$ Branes on $T^n$ and Matrix Theory}{Adv.
Theor. Math. Phys.~{\bf 2} (1998) 51,} {\hh 9709220.}}

\lref\laroche{\paper {L.~Baulieu, C.~Laroche} {On Generalized Self-Duality
Equations Towards Supersymmetric   Quantum Field Theories Of
Forms}{\MPL{A13}{1998}{1115},}{\hh  9801014.}}

\lref\bsv{\paper {M.~Bershadsky, V.~Sadov, C.~Vafa} {$D$-Branes and
Topological Field Theories}{\NPB{463} {1996}{420},}{\hh9511222.}}

\lref\vafapuzz{\paper {C.~Vafa}{Puzzles at Large N}{}{\hh 9804172.}}

\lref\dvv{\paper {R.~Dijkgraaf, E.~Verlinde, H.~Verlinde} {Matrix String
Theory}{\NPB{500}{1997}{43},} {\hh9703030.}}

\lref\wynter{\paper {T.~Wynter}{Gauge Fields and Interactions in Matrix
String Theory}{\PLB{415}{1997}{349},}{\hh9709029.}}

\lref\kvh{\paper {I.~Kostov, P.~Vanhove}{Matrix String Partition
Functions}{}{\hh9809130.}}

\lref\ikkt{\paper {N.~Ishibashi, H.~Kawai, Y.~Kitazawa, A.~Tsuchiya} {A
Large $N$ Reduced Model as Superstring}{\NPB{498} {1997}{467},}{\hh
9612115.}}

\lref\ss{\paper {S.~Sethi, M.~Stern} {$D$-Brane Bound States
Redux}{\CMP{194}{1998} {675},}{\hh 9705046.}}

\lref\mns{\paper {G.~Moore, N.~Nekrasov, S.~Shatashvili} {$D$-particle Bound
States and Generalized Instantons}{} {\hh 9803265.}}

\lref\bsh{\paper {L.~Baulieu, S.~Shatashvili} {Duality from Topological
Symmetry}{} {\hh 9811198.}}

\lref\pawu{ {G.~Parisi, Y.S.~Wu,} {}{ Sci. Sinica  {\bf 24} {(1981)} {484}.}}

%%%%%%%%%%%%%%%%
\lref\lbpert{ {L.~Baulieu,}   {\it Pertrubative gauge theories}, {Physics
Reports {\bf 129 } (1985) 1.} {}}

\lref\kugoj{ {T.~Kugo and I.~Ojima,}   {\it Local covariant
operator formalism of non-Abelian gauge theories and quark confinement
problem} {Prof. Theor. Phys. Suppl.
{\bf 66} 1 (1979).} {}}

\lref\nakanoj{ {N.~Nakanishi and I.~Ojima,}   {\it Covariant operator
formalism of gauge theories and quantum gravity} {vol. 27 of Lecture Notes in
Physics (World Scientific 1990).} {}}

\lref\hirschfeld{ {P.~Hirschfeld,}   {\it } {Nucl. Phys.
{\bf 157} (1979) 37.} {}}

\lref\fleepr{ {R.~Friedberg, T.~D.~Lee, Y.~Pang, H.~C.~Ren,}   {\it A soluble
gauge model with Gribov-type copies}, {Ann. of Phys. {\bf 246 } (1996) 381.} {}}

\lref\coulomb{ {L.~Baulieu, D.~Zwanziger, }   {\it Renormalizable
Non-Covariant Gauges and Coulomb Gauge Limit}, {Nucl. Phys. B {\bf 548 }
(1999) 527-562.} {\hh 9807024}.}

\lref\coulham{ {D.~Zwanziger, }   {\it Lattice Coulomb hamiltonian and static
color-Coulomb field}, {Nucl. Phys. B {\bf 485 } (1997) 185-240.} {}}

\lref\rcoulomb{ {D.~Zwanziger, }   {\it Renormalization in the Coulomb
gauge and order parameter for confinement in QCD}, {Nucl.Phys. B {\bf 518
} (1998) 237-272.} {}}

\lref\kogsuss{ {J.~Kogut and L. Susskind, }   {\it } {Phys. Rev. D {\bf 11}
(1975) 395.} {}}

\lref\kugoojima{ {T.~Kugo and I. Ojima, }   {\it Local covariant
operator formalism of non-Abelian gauge theories and quark confinement
problem}, {Suppl. Prog. Theor. Phys. {\bf 66 } (1979) 1-130.} {}}

\lref\horne{ {J.H.~Horne, }   {\it
Superspace versions of Topological Theories}, {Nucl.Phys. B {\bf 318
} (1989) 22.} {}}

\lref\sto{ {S.~Ouvry, R.~Stora, P.~Van~Baal }   {\it
}, {Phys. Lett. B {\bf 220
} (1989) 159;} {}{ R.~Stora, {\it Exercises in   Equivariant Cohomology},
In Quabtum Fields and Quantum Space Time, Edited
by 't Hooft et al., Plenum Press, New York, 1997}            }

\lref\semenov{ {M. Semenov-Tyan-Shanskii and V. Franke},  {\it } {Zap. Nauch.
Sem. Leningrad. Otdeleniya Matematicheskogo Instituta im V. A. Steklov, AN
SSSR, vol 120, p 159, 1982 (English translation: New York: Plenum Press 1986.
{}}{}}

\lref\gfdadzinside{ {G. Dell'Antonio and D. Zwanziger},  {\it All gauge orbits
and some Gribov copies encompassed by the Gribov horizon,} Proceedings of the
NATO Advanced Workshop on Probabilistic Methods in Quantum Field Theory and
Quantum Gravity, Carg\`{e}se, August 21-27, 1989, Plenum (N.Y.), P. Damgaard,
H. H\"{u}ffel and A. Rosenblum, Eds. {\bf} {}}

\lref\dan{ {D.~Zwanziger},  {\it Covariant Quantization of Gauge
Fields without Gribov Ambiguity}, {Nucl. Phys. B {\bf   192}, (1981)
{259}.}{}}

\lref\danlau{ {L.~Baulieu, D.~Zwanziger, } {\it Equivalence of Stochastic
Quantization and the-Popov Ansatz,
  }{Nucl. Phys. B  {\bf 193 } (1981) {163}.}{}}

\lref\floratos{ {E.~Floratos, J~Iliopoulos, D.~Zwanziger, } {\it A
covariant ghost-free perturbation expansion for Yang-Mills theories,} {Nucl.
Phys. B  {\bf 241 } (1984) {221-227}.}{}}

\lref\dzgribreg{ {D.~Zwanziger},  {\it Non-perturbative modification of the
Faddeev-Popov formula and banishment of the naive vacuum}, {Nucl. Phys. B
{\bf   209}, (1982) {336}.}{}}

\lref\danzinn{  {J.~Zinn-Justin, D.~Zwanziger, } {}{Nucl. Phys. B  {\bf
295} (1988) {297}.}{}}

\lref\bodeker{  {D.~B\"{o}deker,} {}{hep-ph/9905239} {\bf} {}.}

\lref\dzvan{ {D.~Zwanziger, }   {\it Vanishing of zero-momentum lattice
gluon propagator and color confinement}, {Nucl.Phys. B {\bf 364 }
(1991) 127.} }

\lref\horizcon{ {D.~Zwanziger, }   {\it Renormalizability of the critical limit
of lattice gauge theory by BRS invariance,} {Nucl. Phys. B {\bf 399 } (1993)
477.} }

\lref\horizcona{ {D.~Zwanziger, }   {\it Fundamental modular region, Boltzmann
factor and area law in lattice theory }, {Nucl.Phys. B {\bf 412 }
(1994) 657.} }

\lref\horizpt{ {M.~Schaden and D.~Zwanziger, }   
{\it Horizon condition holds pointwise on finite lattice with free boundary
condition,} {hep-th/9410019.} }

\lref\czcoulf{ {A.~Cucchieri and D.~Zwanziger, }   {\it Static color-Coulomb
force}, {Phys. Rev. Lett. {\bf 78 } (1997) 3814} }

\lref\acthrdlan{ {A.~Cucchieri,}   {\it} 
{Phys. Rev. D {\bf 60 } 034508 (1999).} }

\lref\acfkppl{ {A.~Cucchieri, F.~Karsch, P~Petreczky}   {\it} 
{Phys. Lett. B {\bf 497 } 80 (2001).} }

\lref\acfkppa{ {A.~Cucchieri, F.~Karsch, P~Petreczky}   {\it} 
{Phys. Rev. D {\bf 64 } 036001 (2001).} }

\lref\cucchierigh{ {A.~Cucchieri,}   {\it} 
{Nucl. Phys. B {\bf 508 } 353 (1997).} }

\lref\actmdzgh{ {A.~Cucchieri, T. Mendes, D. Zwanziger}   {\it} 
{Nucl. Phys. B Proc. Suppl. {\bf 106 } 697 (2002).} }

\lref\acdland{ {A.~Cucchieri,}   {\it Infrared behavior of the gluon
propagator in lattice landau gauge: the three-dimensional case,} 
%{Nucl.Phys. B {\bf 412 } (1994) 657.} 
hep-lat/9902023CHECK IF SAME AS PHYS REV ARTICLE}

\lref\cznumstgl{ {A.~Cucchieri, and D.~Zwanziger}   {\it Numerical study of
gluon propagator and confinement scenario in minimal Coulomb gauge}, 
{Phys. Rev. D {\bf 65} 014001}}

\lref\czfitgrib{ {A.~Cucchieri, and D.~Zwanziger}   {\it Fit to gluon
propagator and Gribov formula}, hep-lat/0012024}

\lref\dznonpland{ {D.~Zwanziger}   {\it Non-perturbative Landau gauge and
infrared critical exponents in QCD},  Phys. Rev. D, {\bf 65} 094039 (2002) and
hep-th/0109224.}

\lref\leinweber{ {D.~B.~Leinweber, J.~I.~Skullerud, A.~G.~Williams,
and C. Parrinello}   {\it }  Phys. Rev. {\bf D58} (1998) 031501
{\it ibid} {\bf D60} (1999) 094507.}

\lref\bonneta{ {F.~Bonnet, P.~O.~Bowman, D.~B.~Leinweber, A.~G.~Williams,}  
{\it }  Phys. Rev. {\bf D62} (2000) 051501.}

\lref\bonnetb{ {F.~Bonnet, P.~O.~Bowman, D.~B.~Leinweber, A.~G.~Williams
and J. M. Zanotti} {\it }  Phys. Rev. {\bf D64} (2001) 034501.}

\lref\langfeld{ {K.~Langfeld, H. Reingardt, and J.~Gattnar}   {\it } 
Nucl. Phys. B {\bf 621} (2002) 131.}

\lref\suman{ {H.~Suman, and K.~Schilling}   {\it } 
Phys. Lett. {\bf B373} (1996) 314.}

\lref\alex{ {C. Alexandrou, P. de Forcrand, and E. Follana } {\it The gluon
propagator without lattice Gribov copies,}  Phys. Rev. {\bf D63:} 094504
(2001), and  {\it The gluon propagator with lattice Gribov copies on a finer
lattice,} Phys. Rev. {\bf D65:} 114508, 2002.}

\lref\bowman{ {Patrick O. Bowman, Urs M. Heller, Derek B.
Leinweber, Anthony G. Williams,} {\it Gluon propagator on coarse lattices in
laplacian gauges,} Phys. Rev. {\bf D66:} 074505, 2002.}

\lref\direnzo{ {F.~DiRenzo, L.~Scorzato,}  
{\it Lattice 99,}  Nucl. Phys. B (Proc. Suppl.) {\bf 83-84} (2000) 822.}

\lref\nakamuraa{ {M. Mizutani and A. Nakamura}   {\it }  Nucl. Phys. B (Proc.
Suppl.) {\bf 34} (1994) 253.}

\lref\nakamurab{ {F.~Shoji, T.~Suzuki, H.~Kodama, and A.~Nakamura,}  
{\it }  Phys. Lett. {\bf B476} (2000) 199.}

\lref\nakamurac{ {H. Aiso, M. Fukuda, T. Iwamiya, A. Nakamura, T. Nakamura,
and M. Yoshida }   {\it Gauge fixing and gluon propagators,}  Prog. Theor.
Physics. (Suppl.) {\bf 122} (1996) 123.}

\lref\nakamurad{ {H. Aiso, J. Fromm, M. Fukuda, T. Iwamiya, A. Nakamura, T.
Nakamura, M. Stingl and M. Yoshida }   {\it Towards understanding of
confinement of gluons,}  Nucl. Phys. B (Proc. Suppl.) {\bf 53} (1997) 570.}

\lref\nakamurae{ {F.~Shoji, T.~Suzuki, H.~Kodama, and A.~Nakamura,}  
{\it }  Phys. Lett. {\bf B476} (2000) 199.}

\lref\munoz{ { A.~Munoz Sudupe, R. F. Alvarez-Estrada, } {}
Phys. Lett. {\bf 164} (1985) 102; {} {\bf 166B} (1986) 186. }

\lref\okano{ { K.~Okano, } {}
Nucl. Phys. {\bf B289} (1987) 109; {} Prog. Theor. Phys.
suppl. {\bf 111} (1993) 203. }

\lref\baugros{ {L.~Baulieu, B.~Grossman, } {\it A topological Interpretation
of  Stochastic Quantization} {Phys. Lett. B {\bf  212} {(1988)} {351}.}}

\lref\bautop{ {L.~Baulieu}{ \it Stochastic and Topological Field Theories},
{Phys. Lett. B {\bf   232} (1989) {479}}{}; {}{ \it Topological Field Theories
And Gauge Invariance in Stochastic Quantization}, {Int. Jour. Mod.  Phys. A
{\bf  6} (1991) {2793}.}{}}

\lref\bautopr{  {L.~Baulieu, B.~Grossman, } {\it A topological Interpretation
of  Stochastic Quantization} {Phys. Lett. B {\bf  212} {(1988)} {351}};
 {L.~Baulieu}{ \it Stochastic and Topological Field Theories},
{Phys. Lett. B {\bf   232} (1989) {479}}{}; {}{ \it Topological Field Theories
And Gauge Invariance in Stochastic Quantization}, {Int. Jour. Mod.  Phys. A
{\bf  6} (1991) {2793}.}{}}

\lref\bautoprr{  {L.~Baulieu, B.~Grossman, } { } {Phys. Lett. B {\bf  212}
{(1988)} {351}};
 {L.~Baulieu}{ },
{Phys. Lett. B {\bf   232} (1989) {479}}{}; {}{  }, {Int. Jour. Mod.
Phys. A {\bf  6} (1991) {2793}.}{}}
\lref\samson{ {L.~Baulieu, S.L.~Shatashvili, { \it Duality from Topological
Symmetry}, {JHEP {\bf 9903} (1999) 011, hep-th/9811198.}}}{}

\lref\halperna{ {Z. Bern, M.B.~Halpern, L. Sadun, C. Taubes}{}, {Phys. Lett.
{\bf 165B,} 151, 1985.}}

\lref\halpernb{ {Z. Bern, M.B.~Halpern, L. Sadun, C. Taubes}{}, {Nucl. Phys.
{\bf B284,} 1, 1987.}}

\lref\halpernc{ {Z. Bern, M.B.~Halpern, L. Sadun, C. Taubes}{}, {Nucl. Phys.
{\bf B284,} 35, 1987.}}

\lref\halpernd{ {Z. Bern, M.B.~Halpern, L. Sadun}{}, {Nucl. Phys.
{\bf B284,} 92, 1987.}}

\lref\halperne{ {Z. Bern, M.B.~Halpern, L. Sadun}{}, {Z. Phys.
{\bf C35,} 255, 1987.}}

\lref\sadun{ {L. Sadun}{}, {Z. Phys. {\bf C36,} 467, 1987.}}

\lref\halpernr{ {M. B. Halpern}{}, {Prog. Theor. Phys. Suppl. {\bf 111,} 163,
1993.}}

\lref\halpern{ {H.S.~Chan, M.B.~Halpern}{}, {Phys. Rev. D {\bf   33} (1986)
{540}.}}

\lref\bern{ {Z. Bern, H.S.~Chan, M.B.~Halpern}{}, {Z. Phys. {\bf C35}
(1987) {255}.}}

\lref\yue{ {Yue-Yu}, {Phys. Rev. D {\bf   33} (1989) {540}.}}

\lref\neuberger{ {H.~Neuberger,} {\it Non-perturbative gauge Invariance},
{ Phys. Lett. B {\bf 175} (1986) {69}.}{}}

\lref\yangmills{  {C~N.~Yang and R.~L.~Mills,} {}{Phys. Rev. {\bf 96}
(1954) {191}.}{}}

\lref\gribov{  {V.N.~Gribov,} {}{Nucl. Phys. B {\bf 139} (1978) {1}.}{}}

\lref\huffel{ {P.H.~Daamgard, H. Huffel},  {}{Phys. Rep. {\bf 152} (1987)
{227}.}{}}

\lref\creutz{ {M.~Creutz},  {\it Quarks, Gluons and  Lattices, }  Cambridge
University Press 1983, pp 101-107.}

\lref\zinn{ {J. ~Zinn-Justin, }  {Nucl. Phys. B {\bf  275} (1986) {135}.}}

\lref\gozzi{ {E. ~Gozzi,} {\it Functional Integral approach to Parisi--Wu
Quantization: Scalar Theory,} { Phys. Rev. {\bf D28} (1983) {1922}.}}

\lref\singer{
 I.M. Singer, { Comm. of Math. Phys. {\bf 60} (1978) 7.}}

\lref\neu{ {H.~Neuberger,} {Phys. Lett. B {\bf 183}
(1987) {337}.}{}}

\lref\testa{ {M.~Testa,} {}{Phys. Lett. B {\bf 429}
(1998) {349}.}{}}

\lref\martin{ L.~Baulieu and M. Schaden, {\it Gauge Group TQFT and Improved
Perturbative Yang--Mills Theory}, {  Int. Jour. Mod.  Phys. A {\bf  13}
(1998) 985},   hep-th/9601039.}

\lref\ostseil { K.~Osterwalder and E.~Seiler, {\it Gauge field theories on
the lattice} {  Ann. Phys. {\bf 110} (1978) 440}.}

\lref\fradshen { E.~Fradken and S.~Shenker, {\it Phase diagrams of lattice
guage theories with Higgs fields} {  Phys. Rev. {\bf D19} (1979) 3682}.}

\lref\banksrab{ T.~Banks and E.~Rabinovici, {} {  Nucl. Phys. {\bf B160}
(1979) 349}.}

\lref\nielsen{N. K. Nielsen, {\it On The Gauge Dependence Of Spontaneous
Symmetry Breaking In Gauge Theories},
Nucl.\ Phys.\ {\bf B101}, 173 (1975)}

\lref\nadkarni{ S. Nadkarni, {\it The SU(2) adjoint Higgs model in three
dimensions } {  Nucl. Phys. {\bf B334} (1990) 559}.}

\lref\stackteper{ A.~Hart, O.~Philipsen, J.~D.~Stack, and M.~Teper,
{\it On the phase diagram of the SU(2) adjoint Higgs model in 2+1
dimensions } { hep-lat/9612021}.}

\lref\kajantie{K.~Kajantie, M.~Laine, K.~Rummujkainen, M.~Shaposhnikov
{\it 3D SU(N)+adjoint Higgs theory and finite temperature QCD }
{hep-ph/9704416}.}

\lref\batrouni{G. G. Batrouni, G. R. Katz, A. S. Kronfeld, G. P. Lepage,
B. Svetitsky and K. G. Wilson, {} {Phys. Rev. {\bf D32} (1985) 2736}}

\lref\davies{C. T. H. Davies, G. G. Batrouni, G. R. Katz, A. S. Kronfeld,
G. P. Lepage, K. G. Wilson, P. Rossi and B. Svetitsky, {} {Phys. Rev. {\bf
D41} (1990) 1953}}

\lref\fukugita{M. Fukugita, Y. Oyanagi and A. Ukawa, {}
{Phys. Rev. Lett. {\bf 57} (1986) 953;
Phys. Rev. {\bf D36} (1987) 824}}

\lref\kronfeld{A. S. Kronfeld, {\it Dynamics of Langevin Simulation} {Prog.
Theor. Phys. Suppl. {\bf 111} (1993) 293}}

\lref\polyakov{ A.~Polyakov, {} {Phys. Letts. {\bf B59} (1975) 82;
Nucl. Phys. {\bf B120} (1977) 429}; {\it Gauge fields and strings,}
ch. 4 (Harwood Academic Publishers, 1987).}

\lref\thooft{ G.~'t Hooft, {} {Nucl. Phys. {\bf B79} (1974) 276}; {}
Nucl. Phys. {\bf B190} (1981) 455 {}.}

\lref\elitzur{S.~Elitzur, {} {Phys. Rev. {\bf D12} (1975) 3978}}

%\polyakov \nadkarni \stackteper \kajantie

%%%%%%%%%%%%%%%%%%%%%%%%%%%%%%%%%%%%%%%%%%%%%%%%%%%%%%%%%%%%%%%%%
\lref\baugros{ {L.~Baulieu, B.~Grossman, } {\it A topological Interpretation
of  Stochastic Quantization} {Phys. Lett. B {\bf  212} {(1988)} {351}.}}

\lref\bautop{ {L.~Baulieu}{ \it Stochastic and Topological Field Theories},
{Phys. Lett. B {\bf   232} (1989) {479}}{}; {}{ \it Topological Field Theories
And Gauge Invariance in Stochastic Quantization}, {Int. Jour. Mod.  Phys. A
{\bf  6} (1991) {2793}.}{}}

\lref\bautopr{  {L.~Baulieu, B.~Grossman, } {\it A topological Interpretation
of  Stochastic Quantization} {Phys. Lett. B {\bf  212} {(1988)} {351}};
 {L.~Baulieu}{ \it Stochastic and Topological Field Theories},
{Phys. Lett. B {\bf   232} (1989) {479}}{}; {}{ \it Topological Field Theories
And Gauge Invariance in Stochastic Quantization}, {Int. Jour. Mod.  Phys. A
{\bf  6} (1991) {2793}.}{}}

\lref\samson{ {L.~Baulieu, S.L.~Shatashvili, { \it Duality from Topological
Symmetry}, {JHEP {\bf 9903} (1999) 011, hep-th/9811198.}}}{}

\lref\halpern{ {H.S.~Chan, M.B.~Halpern}{}, {Phys. Rev. D {\bf   33} (1986)
{540}.}}

\lref\yue{ {Yue-Yu}, {Phys. Rev. D {\bf   33} (1989) {540}.}}

\lref\neuberger{ {H.~Neuberger,} {\it Non-perturbative gauge Invariance},
{ Phys. Lett. B {\bf 175} (1986) {69}.}{}}

\lref\huffel{ {P.H.~Daamgard, H. Huffel},  {}{Phys. Rep. {\bf 152} (1987)
{227}.}{}}

\lref\creutz{ {M.~Creutz},  {\it Quarks, Gluons and  Lattices, }  Cambridge
University Press 1983, pp 101-107.}

\lref\zinn{ {J. ~Zinn-Justin, }  {Nucl. Phys. B {\bf  275} (1986) {135}.}}

\lref\shamir{  {Y.~Shamir,  } {\it Lattice Chiral Fermions
  }{ Nucl.  Phys.  Proc.  Suppl.  {\bf } 47 (1996) 212,  hep-lat/9509023;
V.~Furman, Y.~Shamir, Nucl.Phys. B {\bf 439 } (1995), hep-lat/9405004.}}

 \lref\kaplan{ {D.B.~Kaplan, }  {\it A Method for Simulating Chiral
Fermions on the Lattice,} Phys. Lett. B {\bf 288} (1992) 342; {\it Chiral
Fermions on the Lattice,}  {  Nucl. Phys. B, Proc. Suppl.  {\bf 30} (1993)
597.}}

\lref\neubergerr{ {H.~Neuberger, } {\it Chirality on the Lattice},
hep-lat/9808036.}

\lref\neubergers{ {Rajamani Narayanan, Herbert Neuberger,} {\it INFINITELY MANY
    REGULATOR FIELDS FOR CHIRAL FERMIONS.}
    Phys.Lett.B302:62-69,1993.
    [HEP-LAT 9212019]}

\lref\neubergert{ {Rajamani Narayanan, Herbert Neuberger,}{\it CHIRAL FERMIONS
    ON THE LATTICE.}
    Phys.Rev.Lett.71:3251-3254,1993.
    [HEP-LAT 9308011]}

\lref\neubergeru{ {Rajamani Narayanan, Herbert Neuberger,}{\it A CONSTRUCTION
OF LATTICE CHIRAL GAUGE THEORIES.}
    Nucl.Phys.B443:305-385,1995.
    [HEP-TH 9411108]}

\lref\neubergerv{ {Herbert Neuberger,}{\it EXACTLY MASSLESS QUARKS ON THE
    LATTICE.}
    Phys. Lett. B417 (1998) 141-144.
    [HEP-LAT 9707022]}

%The first 3 papers deal with chiral fermions in general, while the last
%with the
%particular case of vector like fermions. All these papers are quite well
%known.
%
%If you wish to quote reviews, the review by Shamir is seriously flawed.
%More recent
%reviews are available.  Surprisingly, I happen to like:

\lref\neubergerw{ {Herbert Neuberger,}{\it CHIRAL FERMIONS ON THE
LATTICE.}
    Nucl. Phys. B, Proc. Suppl. 83-84 (2000) 67-76.
    [HEP-LAT 9909042]}

\lref\zbgr {L.~Baulieu and D. Zwanziger, {\it QCD$_4$ From a
Five-Dimensional Point of View},    Nucl. Phys. {\bf B581} 2000, 604;
hep-th/9909006.}

\lref\bgz {P. A. Grassi, L.~Baulieu and D. Zwanziger, {\it Gauge and
Topological Symmetries in the Bulk Quantization of Gauge Theories},
Nucl. Phys. {\bf B597} 583, 2001 hep-th/0006036.}

\lref\bulkq {L.~Baulieu and D. Zwanziger, {\it From stochastic
quantization to bulk quantization; Schwinger-Dyson equations and
the S-matrix}, JHEP 08:016, 2001 hep-th/0012103.}

\lref\bulkqg {L.~Baulieu and D. Zwanziger, {\it Bulk quantization of gauge
theories: confined and Higgs phases}, JHEP 08:015, 2001 and
hep-th/0107074.}

\lref\equivstoch{L.~Baulieu and D. Zwanziger, {\it Equivalence of
stochastic quantization and the Faddeev-Popov Ansatz},
Nucl. Phys. B193 (1981) 163-172.}

 \lref\zbsd {L.~Baulieu and D. Zwanziger, {
\it From stochastic quantization to bulk quantization: Schwinger-Dyson
equations and S-matrix QCD$_4$}, hep-th/0012103.}

\lref\cuzwns {A.~Cucchieri and D.~Zwanziger, {\it Numerical study of
gluon propagator and confinement scenario in minimal Coulomb gauge},
hep-lat/0008026.}

\lref\fitgrib {A.~Cucchieri and D.~Zwanziger, {\it Fit to gluon
propagator and Gribov formula},    hep-th/0012024.}

\lref\vanish {D.~Zwanziger, {\it Vanishing of zero-momentum lattice
gluon propagator and color confinement},   Nucl. Phys. {\bf B364}
(1991) 127-161.}

\lref\gribov {V.~N.~Gribov, {\it Quantization of non-Abelian gauge
theories},   Nucl. Phys. {\bf B139} (1978)~1-19.}

\lref\singer {I.~Singer, {\it }   Comm. Math. Phys. {\bf 60}
(1978)~7.}

\lref\feynman {R.~P.~Feynman, {\it The qualitative behavior of
Yang-Mills theory in 2+1 dimensions},   Nucl. Phys. {\bf B188} (1981)
479-512.}

\lref\cutkosky {R.~E.~Cutkosky, {\it}   J. Math. Phys. {\bf 25} (1984)
939; R. E. Cutkosky and K. Wang, Phys. Rev. {\bf D37} (1988) 3024; R. E.
Cutkosky, Czech. J. Phys. {\bf 40} (1990) 252.}

\lref\vanbaal{J. Koller and P. van Baal, Nucl. Phys. {\bf B302} (1988)
1; P. van Baal, Acta Phys. Pol. {\bf B20} (1989) 295;
P. van Baal, Nucl. Phys. {\bf B369} (1992) 259;
P. van Baal and N. D. Hari Dass, Nucl. Phys. {\bf B385} (1992) 185.}

\lref\stingl {M. Stingl, {\it Propagation properties and condensate
formation of the confined Yang-Mills field},   Phys. Rev. D {\bf 34} (1986)
3863-3881.}

\lref\smekal{L.~von Smekal, A.~Hauck and R.~Alkofer,  {\it A Solution to
Coupled Dyson-Schwinger Equations in Gluons and Ghosts in Landau Gauge,}  
Ann. Phys. {\bf 267} (1998) 1; L. von Smekal, A. Hauck and R. Alkofer, {\it The
Infrared Behavior of Gluon and Ghost Propagators in Landau Gauge QCD,}  
Phys. Rev. Lett. {\bf 79} (1997) 3591; L. von Smekal {\it Perspectives for
hadronic physics from Dyson-Schwinger equations for the dynamics of quark and
glue,} Habilitationsschrift, Friedrich-Alexander Universit\"{a}t,
Erlangen-N\"{u}rnberg (1998).}

\lref\smekrev {R.~Alkofer and L.~von Smekal, {\it The infrared behavior of
QCD Green's functions},   Phys. Rept. {\bf 353}, 281 (2001).}

\lref\fischalk {C. S. Fischer and R.~Alkofer, {\it Infrared exponents and
running coupling of SU(N) Yang-Mills Theories},   Phys. Lett. B {\bf 536},
177 (2002).}

\lref\fischalkrein{C.~S.~Fischer, R.~Alkofer and H.~Reinhardt,
   {\it The elusiveness of infrared critical exponents in Landau gauge
   Yang-Mills theories,}
   Phys. Rev. D {\bf 65}, 094008 (2002)}

\lref\lerche {C. Lerche and L. von Smekal {\it On the infrared exponent for
gluon and ghost propagation in Landau gauge QCD}, hep-ph/0202194}

\lref\atkinsona {D.~Atkinson and J.~C.~R.~Bloch, {\it Running coupling in
non-perturbative QCD}   Phys. Rev. {\bf
D58} (1998) 094036.}

\lref\atkinsonb {D.~Atkinson and J.~C.~R.~Bloch, {\it QCD in the infrared with
exact angular integrations}   Mod. Phys. Lett. {\bf A13} (1998) 1055.}

\lref\brown {N.~Brown and M.~R. Pennington, {\it}   Phys. Rev. {\bf D38} (1988)
2266; Phys. Rev. {\bf D39} (1989) 2723.}

\lref\szczep {A.~P.~Szczepaniak and E.~S.~Swanson, {\it Coulomb Gauge QCD,
Confinement, and the Constituent Representation},   hep-ph/0107078.}

%%%%%%%%%%%%%%%%%%%%CAPTIONS%%%%%%%%%%%%%%%%%%%%%%%%%%%%%%%%%%%%%%%%%%%%%%%

\nfig\compar{Fig. 1.  Diagrammatic representation of the functional DS equation for the
quantum effective drift force, ${\cal Q}(A)$, in the presence of external
sources $A$, eq. \qudf.  The vertices are the tree-level vertices
of the drift force $K$.  The internal lines represent the exact gluon
propagator ${\cal D}(A)$ in the presence of the external source.  The
circle is the exact 3-gluon vertex of the quantum effective action $\Gamma(A)$
in the presence of the external source.}

\nfig\compar{Fig. 2.  Diagrammatic representation of the DS equation for the
gluon propagator, eq. \qudfone.  The vertices are the tree
level vertices of the drift force $K$.  The internal lines are the exact gluon
propagator $D$ with sources set to 0.  The circles represent the exact 3 and
4-gluon vertices of the quantum effective action $\Gamma$, with sources set
to~0.}

%%%%%%

%\draft

%%%%%%%%%

\Title{\vbox
{\baselineskip 10pt
\hbox{hep-th/0206053}
%\hbox{CERN-TH-00-??}
\hbox{LPTHE-00-50}
\hbox{NYU-TH-7.10.01}
 \hbox{   }
}}
{\vbox{\vskip -30 true pt
\centerline{
   }
\medskip
 \centerline{  }
\centerline{Time-independant stochastic quantization, DS}
\centerline{equations, and infrared critical exponents in QCD}
\medskip
\vskip4pt }}
\centerline{
%{\bf Laurent Baulieu}$^{  \dag     }$ and  
{\bf  Daniel Zwanziger}
%$^{ \ddag}$
}
\centerline{
%baulieu@lpthe.jussieu.fr, 
Daniel.Zwanziger@nyu.edu}
%pag5@nyu.edu
\vskip 0.5cm
%\centerline{\it $^{\dag}$LPTHE, Universit{\'e}s P. \& M. Curie (Paris~VI)
%et D. Diderot (Paris~VII), Paris,  France,}
%{\foot{UMR 7589 associ{\'e}e CNRS et
%Universit{\'e}s P. \&M. Curie (Paris~VI) et D. Diderot (Paris~VII)},
%Boite 126,
%4 place Jussieu, F-75252
%Paris Cedex 05, France.}
%\centerline{\it $^{\dag}$  Dept. of Physics, University
%of Rutgers, New Brunswick, NJ 60637, USA }
%\centerline{\it $^{\S}  $}
\centerline{\it 
%$^{\ddag}$   
Physics Department, New York University,
New-York,  NY 10003,  USA}

\medskip
\vskip  1cm
\noindent

	We derive the equations of time-independent stochastic quantization,
without reference to an unphysical 5th time, from the principle of gauge
equivalence.  It asserts that probability distributions
$P$ that give the same expectation values for gauge-invariant observables 
$\langle W \rangle = \int dA \ W \ P$ are physically indistiguishable.
This method escapes the Gribov critique.  We derive an exact system of
equations that closely resembles the Dyson-Schwinger equations of Faddeev-Popov
theory.  The system is truncated, and solved non-perturbatively,
by means of a power law Ansatz, for the critical exponents that
characterize the asymptotic form at $k \approx 0$ of the gluon propagator in
Landau gauge.  For the tranverse and longitudinal parts, we find respectively
$D^T \sim (k^2)^{-1-\a_T} \approx (k^2)^{0.043}$, suppressed and in fact
vanishing, though weakly, and $D^L \sim a \ (k^2)^{-1-\a_L} \approx a \
(k^2)^{-1.521}$, enhanced, with $\a_T = - 2\a_L$.  Although the longitudinal
part vanishes with the gauge parameter $a$ in the Landau-gauge limit, 
$a \to 0$, there are vertices of order $a^{-1}$ so, counter-intuitively, 
the longitudinal part of the gluon propagator does contribute in internal
lines in Landau gauge, replacing the ghost that occurs in Faddeev-Popov
theory.  We compare our results with the corresponding results in
Faddeev-Popov theory.

\Date{\ }

\def\e{\epsilon}

\def\a{\alpha}
\def\b{\beta}
\def\d{\delta}

\def\m{\mu}
\def\n{\nu}

\def\s{\sigma}
\def\l{\lambda}

\def\o{\omega}

\def\k{\kappa}

\newsec{Introduction}

\subsec{Some recent developments in non-perturbative QCD}
	The problem of the strong interaction presents an exciting challenge. 
One would like to understand how and why QCD describes a world of color-neutral
hadrons with a mass gap, even though it appears perturbatively to be a theory
of unconfined and massless gluons and quarks.  Clearly an understanding of
non-Abelian gauge theory at the non-perturbative level is required.  Happily,
there has recently developed a convergence of results
by different methods: (i) non-perturbative solutions of the truncated
Dyson-Schwinger (DS) equations in Faddeev-Popov theory, (ii) numerical
evaluation of gauge-fixed, lattice QCD propagators, and (iii) exact analytic
results.  The agreement between these very different methods 
almost 5 decades after the appearance of the original article of Yang
and Mills \yangmills, would indicate that by (ii) we are beginning to get
reliable values of the gluon propagator in the unbroken phase, and by (i) an
understanding of the mechanism that determines it.  This motivates the present
investigation in which we derive the DS equations of time-independent
stochastic quantization and solve them by truncation and a power-law Ansatz
for the gluon propagator in the asymptotic, low-momentum r\'{e}gime.  
In accordance with earlier results by methods (i), (ii), and (iii), 
we find that, compared to the free propagator~$1/k^2$,
 {\it the would-be physical, transverse component of the gluon
propagator is short range, while the unphysical, longitudinal component is
long range.}

	As concerns (i), solutions of the DS equations, the decisive step was taken
in~\smekal, where a solution of the truncated DS equations in Faddeev-Popov
quantization in Landau gauge was obtained for which {\it the transverse gluon
propagator is short range, while the ghost propagator is long range.}  These
properties were confirmed in subsequent DS calculations, using a variety of
approximations for the vertex \atkinsona, \atkinsonb, \lerche, and
\dznonpland.    More recent calculations extend the
asymptotic infrared and ultraviolet solutions to finite momentum $k$, without
angular approximation \fischalkrein, \fischalk.  All these
calculations\foot{Stingl \stingl\ had earlier obtained a solution of the DS
equation with the property that $D^T(k)$ vanishes at $k= 0$, without however
including the ghost loop, whereas the ghost loop gives the dominant
contribution in the infrared region in the recent solutions.} give a
transverse gluon propagator in Landau gauge $D^T(k)$ that is highly suppressed
in the infrared compared to the free massless  propagator $1/k^2$, and that in
fact {\it vanishes} $\lim_{k \to 0}D^T(k) = 0$, at $k = 0$, in some cases
weakly, like a small positive power of $k$.  Indeed, according to the present
calculation it vanishes like $(k^2)^{0.043}$.  A review of DS equations in QCD
may be found in~\smekrev.  In the present work we shall discover a close
conection between the ghost propagator in Faddeev-Popov theory and the
longitudinal part of the gluon propagator in time-independent stochastic
quantization.

	Concerning (ii), numerical studies, it is striking that an accumulation of
numerical evaluations of the gluon propagator in Landau gauge also show
qualitive suppression of the gluon propagator at low momentum, both in
3-dimensions on relatively large lattices, \acthrdlan, \acfkppl, \acfkppa, and
in 4 dimensions, 
\leinweber, \bonneta, \bonnetb, \langfeld.  Suppression of the gluon
propagator and enhancement of the ghost propagator at low momentum has been
reported by \suman, \cucchierigh, and \actmdzgh.  Similar numerical results were obtained in Coulomb gauge,
where an extrapolation to infinite lattice volume of the 3-dimensionally
transverse, would-be physical, equal-time gluon propagator 
$D_{ij}(\vec{k})$ was consistent with its vanishing at
$\vec{k} = 0$, \cznumstgl.  In QCD in the Coulomb gauge, the instantaneous
Coulomb propagator, $D_{44}$, is closely related to the ghost or
Faddeev-Popov propagator, and is a strong candidate for a confining
potential.  Significantly, $D_{44}$ was found to be long range \cznumstgl.

	A recent numerical calculation in the Landau gauge, \bonnetb, reports a finite
value of $D^T(k)$ at $k = 0$.  This is strongly suppressed compared to
$1/k^2$, and suffices to exclude a free massless gluon.  It might be thought
that the finite value of $D^T(0)$ reported in \bonnetb\ contradicts the zero
value, $D^T(0) =0$, found here.  However it is difficult to distinguish
numerically between a finite value at $k = 0$ and one that vanishes weakly,
like $(k^2)^\g$, with a small value for the infrared anomalous dimension such
as $\g = 0.043$ found here.  For this function is almost constant down to
very low $k$, and then veers toward zero with an infinite slope.  
Moreover a numerical determination of the continuum propagator at $k = 0$
requires an extrapolation to infinite lattice volume.  To establish a
discrepancy it would be necessary to take $\g$ as a fitting parameter, and
determine the numerical uncertainty in this quantity after extrapolation to
infinite lattice volume,  and this has not been done.  Present numerical and
analytic results are not inconsistent, within the considerable uncertainty of
the numerical extrapolation to infinite lattice volume, and both agree that
there is strong suppression compared to $1/k^2$.

	Infrared suppression of the gluon propagator $D(k)$ and
enhancement of the ghost propagator $G(k)$ in Landau gauge was first found by
Gribov, using avowedly rough approximations \gribov.  He obtained the
formulas, $D(k) = k^2/[(k^2)^2 + M^4]$, and, in the infrared,
$G(k) \sim 1/(k^2)^2$, by restricting the region of functional integration to the
interior of the Gribov horizon in order to avoid Gribov
copies.\foot{We remind the reader that
numerical gauge-fixing to the Landau gauge is achieved by minimizing, with
respect to local gauge transformations $g(x)$, a lattice analog of 
$F_A(g) = \int d^4x |{^g}A|^2$.  At any minimum, this functional is
(a)~stationary, and (b)~the matrix of second derivatives is non-negative. 
These conditions correspond to (a)~the Landau gauge condition $\p \cdot A =
0$, and (b)~the positivity of the Faddeev-Popov operator $-D(A) \cdot \p$
which, moreover is symmetric $-D(A) \cdot \p = -\p \cdot D(A)$, for $\p \cdot
A = 0$.  Condition~(b) defines the Gribov region, so numerical studies of the
Landau gauge automatically select configurations within the Gribov region.
Positivity of $-\p \cdot D(A)$ means that all its eigenvalues $\l_n$ are
positive, and the boundary of the Gribov region, known as the (first) Gribov
horizon, is where the first (non-trivial) eigenvalue vanishes.  Thus the
Faddeev-Popov determinent, $\det [-D(A) \cdot \p] = \prod_n \l_n$, 
which is the product of the eigenvalues, is positive
inside the Gribov horizon and vanishes on it.  These considerations do not
apply to numerical gauge fixing to the Laplacian gauge
\alex, \bowman.} 
In Coulomb gauge he also obtained a long-range Coulomb potential.  
Concerning (iii), exact analytic results, it was subsequently found, \horizcon\
and \vanish, that restriction to the interior of the Gribov horizon, enforced
by a  {\it horizon condition,} yields at $k = 0$, both the
vanishing of the gluon propagator $\lim_{k \to 0}D(k) = 0$ in
Landau and Coulomb gauge, and the enhancement of the ghost propagator
$\lim_{k \to 0}k^2G(k) = \infty$.\foot{It is noteworthy that the confinement
criterion of Kugo and Ojima \kugoj\ and \nakanoj\
also entails $\lim_{k \to 0}k^2G(k) = \infty$.}

	It was at first surprising that the solution of the DS equations obtained
in~\smekal, \atkinsona, \atkinsonb, and \lerche\ agreed with these exact
results that are a consequence of cutting off the functional
integral at the Gribov horizon, for this condition was not imposed in
solving the DS equations.  However it was subsequently pointed out~\dznonpland\
that the DS equations in Faddeev-Popov theory depend only on the
integrand, and the fact that the integral of a derivative vanishes {\it
provided only that the integrand vanishes on the boundary}.  The key point is
that {\it the integrand does vanish on the Gribov horizon} for the
Faddeev-Popov determinent, $\det[- D(A)\cdot \p]$, vanishes there (as explained
in footnote 2).  Thus
Gribov's prescription to cut off the functional integral at the (first) Gribov
horizon, is not a constraint that changes the DS equations, but rather it
resolves an ambiguity in the solution of these equations~\dznonpland.  The
cut-off at the first Gribov horizon assures that both the gluon and ghost
Euclidean propagators are positive, which is a property of the solutions
obtained for the truncated DS equations.   Moreover the solution of the DS
equations in Faddeev-Popov theory  with a cut-off at the Gribov horizon is the
only one for which a comparison with numerical gauge fixing to the lattice
Landau gauge is (approximately) justified.  For as explained in footnote 2,
numerical gauge fixing to the Landau gauge automatically produces a
configuration that lies inside the Gribov horizon. Thus a consistent picture
emerges of the gluon and ghost propagators in QCD using the different methods
(i), (ii) and~(iii).

\subsec{Diffulties of Faddeev-Popov method at non-perturbative level}
	
	The DS calculations \smekal\ -- \lerche\ rely on Faddeev-Popov theory 
which however is subject to the well-known critiques of Gribov \gribov\ and
Singer \singer.  At the perturbative level, Faddeev-Popov theory is
unexceptionable, and elegant BRST proofs are available of perturbative
renormalizability and perturbative unitarity \lbpert.  In lattice gauge theory
however the BRST method fails because the total number of Gribov
copies is even, but they contribute with opposite signs, leading to an exact
cancellation \neu, \testa.  In continuum gauge theory, the Faddeev-Popov-BRST
method may nevertheless be formally correct at the non-perturbative level
without a cut-off at the Gribov horizon, if ones sums over all signed Gribov
copies~\hirschfeld,~\fleepr.  However even if this is true, it would imply
large cancellations between copies, that may amplify the error of an
approximate non-perturbative calculation, and even the Euclidean gluon
propagator $D(k)$ is not necessarily positive.   Alternatively, one may choose
the solution  of the DS equations in Faddeev-Popov theory that corresponds to a
cut-off at the first Gribov horizon, which indeed is our interpretation of the
solutions of~\smekal\ ---~\fischalk.  Hopefully, this is an excellent
approximation.  But it remains an {\it ad hoc} prescription that is not
correct in principle because of the existence of Gribov copies inside the
Gribov horizon \semenov\ and~\gfdadzinside.

	Wilson's lattice gauge theory provides a quantization that is both
theoretically sound and well suited to numerical simulation.  It
also provides a simple analytic model of confinement in QCD by giving an area
law for Wilson loops in the strong-coupling limit.  A striking feature of
lattice gauge theory is that both the numerical simulations and the
strong-coupling expansion are manifestly gauge invariant.  This manifest gauge
invariance provides a paradigm for continuing efforts to understand
confinement in QCD.  Nevertheless it may be worthwile to pursue other
approaches.  The vexing problem of bound states in quantum
field theory is particularly urgent in QCD where confinement causes all
physical particles to be bound states of the fundamental quark and gluon
constituents.  In this regard it is noteworthy that even the simplest of all
bound-state problems, the Hydrogen atom, is not easily solved in a
gauge-invariant formulation.  

\subsec{Review of stochastic quantization}
	
	In order to avoid the difficulties just mentioned of the Faddeev-Popov method,
we turn to stochastic quantization of gauge fields for, as we
shall see, this method provides a correct continuum quantization at the
non-perturbative level.  Stochastic quantization has been developed by a
number of authors  \pawu, \dan, who have expressed the solution as a
functional integral \gozzi, and demonstrated the renormalizability of this
approach \zinn, \danzinn.  A systematic development is presented in \halperna,
\halpernb, \halpernc, \halpernd, \halperne, \sadun, reviewed in \halpernr, that
includes  the 4-and 5-dimensional Dyson-Schwinger equation for the quantum
effective action, an extension of the method to gravity, and gauge-invariant
regularization by smoothing in the 5th time.  Renormalizability has also been
established by an elaboration of BRST techniques~\bgz,~\bulkqg. Stochastic
quantization may be and has been exactly simulated numerically including on
rather large lattices, of volume $(48)^4$, \nakamuraa, \nakamurab, \nakamurac,
\nakamurad, \nakamurae.  This suggests the possibility of a promising
interplay of DS and numerical methods.

	In its original formulation \pawu, stochastic quantization relies on the
observation that the formal Euclidean proabability distribution
$P_0(A) = N \exp[-S_{\rm YM}(A)]$, with 4-dimensional Euclidean 
Yang-Mills action $S_{\rm YM}(A)$, is the equilibrium distribution of the
stochastic process defined by the equation, 
\eqn\fokplan{\eqalign{  
 { { \p P} \over { \p t } } = \int d^4x \ { { \d } \over { \d A_\m^a(x) } } 
     \Big( { { \d P } \over { \d A_\m^a(x) } } 
    + { { \d S_{\rm YM} } \over { \d A_\m^a(x) } } P \Big). 
}}
Indeed it is obvious that $P_0(A)$ is a time-independent solution of this
equation.  Here~$t$ is an artificial 5th time that is a continuum analog of
the number of sweeps in a Monte Carlo simulation of the Euclidean theory
defined by the action $S_{\rm YM}(A)$.  As explained in sec.~3, this equation
has the form of the diffusion equation with ``drift force" 
$ - { { \d S_{\rm YM} } \over { \d A_\m^a(x) } }$, and is known as the
Fokker-Planck equation.  The same stochastic process may equivalently be
represented by the Langevin equation
\eqn\langevin{\eqalign{
{ { \p A_\m^a } \over { \p t } } 
= - { { \d S_{\rm YM} } \over { \d A_\m^a } } + \eta_\m^a,
}}
where $A_\m^a = A_\m^a(x, t)$ depends on the artificial 5th time, and
corresponds in a Monte-Carlo simulation to the configuration on the lattice
with points $x_\m$, with $\m = 1,...4$ at sweep~$t$.  Here 
$\eta_\m^a = \eta_\m^a(x, t)$ is Gaussian white noise defined by 
$ \langle \eta_\m^a(x, t)\rangle = 0 $ and 
$ \langle \eta_\n^b(x, t) \eta_\m^a(y, t') \rangle 
  = 2 \d(x-y) \d_{\m\n} \d^{ab} \d(t - t') \rangle$. 
If $N \exp[-S_{\rm YM}(A)]$ were a normalizable probability distribution ---
which it is not --- every normalized solution to \fokplan\ would relax to it as
equilibrium distribution.   However the process defined by \fokplan\ or
\langevin\ does not provide a restoring force in gauge orbit directions, so
probability escapes to infinity along the gauge orbits, and as a result 
$P(A,t)$ does not relax to a well-defined limitiing distribution
$\lim_{t \to \infty} P(A,t) \neq  N \exp[-S_{\rm YM}(A)]$.  Nevertheless,
according to \pawu, expectation values 
$ \langle {\cal O}(A)\rangle_t $ of gauge-invariant quantities ${\cal O}(A)$ 
calculated at fixed but finite time $t$ according to either of the above
equations do relax to the desired Euclidean expectation value, 
$\langle {\cal O} \rangle = \lim_{t \to \infty}\langle {\cal O}\rangle_t $.

	Unfortunately the renormalization program cannot be carried out in this
scheme as stated, because that requires that gauge-non-invariant correlators
also be well defined.  A remedy is provided by the observation \dan\ that
the Langevin equation may be modified by the addition of an infinitesimal gauge
transformation, 
$D_\m^{ac} v^c = (\p_\m \d^{ac} + f^{abc}A_\m^b)v^c$,
\eqn\fixlangevin{\eqalign{
{ { \p A_\m^a } \over { \p t } } = - { { \d S} \over { \d A_\m^a } }
    + D_\m^{ac} v^c + \eta_\m^a.
}}
Clearly this cannot alter the expecation-value of gauge-invariant
quantities. Symmetry and power-counting arguments determine 
$v^a = a^{-1}\p_\l A_\l^a = a^{-1}\p \cdot A^a$, where $a$ is a free
parameter.  For $a > 0$, the new term, that is tangent to the gauge orbit,
provides a restoring force along gauge orbit directions, so
gauge-non-invariant correlators also exist.\foot{To establish 
that the new force is globally restoring, we note that the
hilbert norm of~$A$ is decreasing under the flow defined by the new force
alone, $\dot{A}_\m = a^{-1} D_\m \p \cdot A$.  We have
$\p ||A||^2 / \p t = 2(A_\m, \dot{A}_\m) = 2 a^{-1}(A_\m, D_\m \p \cdot A)
= 2 a^{-1}(A_\m, \p_\m \p \cdot A) = - 2 a^{-1}||\p \cdot A||^2 \leq 0$.  
This also shows that the region of equilibrium under this force is the set of
transverse configurations, $\p \cdot A = 0$.  Similarly, from
$\p ||\p \cdot A||^2 / \p t = 2 (\p \cdot A, \p \cdot \dot{A}) 
= 2 a^{-1}(\p \cdot A, \p \cdot D \p \cdot A)$ it follows that this equilibrium
is {\it stable} inside the Gribov horizon, where $- \p \cdot D$ is a positive
operator, and {\it unstable} outside it. }  The new scheme is renormalizable.
Only a harmless gauge-transformation has been introduced, so the Gribov problem
of globally correct gauge-fixing is by-passed, and a continuum quantization of
gauge fields that is correct at the non-perturbative level has been achieved.

	The modified  Langevin  equation is equivalent to the modified Fokker-Planck
equation 
\eqn\fixfokplan{\eqalign{  
 { { \p P} \over { \p t } } = \int d^4x{ { \d } \over { \d A_\m^a(x) } } 
     \Big( { { \d P } \over { \d A_\m^a(x) } } 
    - K_\m^a(x) P \Big),  
}}
where the ``drift force" now includes the infinitesimal gauge transformation
\dan,
\eqn\drift{\eqalign{       
  K_\m^a(x)  \equiv - { { \d S_{\rm YM} } \over { \d A_\m^a(x) } } 
       + a^{-1}D_\m^{ac} \p \cdot A^c(x).  
}} 
The additional ``force" is not conservative, and cannot be written, like the
first term, as the gradient of some 4-dimensional gauge-fixing action,
$a^{-1}D_\m^{ac} \p \cdot A^c(x) 
\neq - { { \d S_{\rm gf} } \over { \d A_\m^a(x) } }$.
With this term, the normalized solutions $P(A,t)$ to \fixfokplan\ do relax to
an equilibrium distribution
$\lim_{t \to \infty} P(A,t) = P(A)$, and Euclidean expectation values are given
by the 4-dimensional functional integral, 
$\langle {\cal O} \rangle = \int dA \ {\cal O}(A) P(A)$.
Although we cannot
write $P(A)$ explicitly because the force is not conservative, we do
know that it is the normalized solution of the time-independent Fokker-Planck
equation
\eqn\tifokplan{\eqalign{        
  H_{\rm FP} P  \equiv \int d^4x{ { \d } \over { \d A_\m^a(x) } } 
     \Big( - { { \d P  } \over { \d A_\m^a(x) } } + K_\m^a P \Big)  = 0 .  
}}
This equation defines what we call ``time-independent stochastic
quantization", and $H_{\rm FP}$ is called the
``Fokker-Planck hamiltonian".  The solution $P(A)$ of this equation provides a
satisfactory non-perturbative quantization of gauge fields.

	[To avoid possible confusion of terminology, we note that stochastic
quantization, whether in the time-dependent or time-independent
formulation,  --- where ``time" is the artificial 5th time ---
increases the number of dimensions by one as compared to the
corresponding standard Faddeev-Popov formulation of gauge field theory.
Thus the solution of the time-dependent Fokker-Planck equation \fixfokplan\
can be usefully represented \gozzi\ as a functional integral with a {\it local
5-dimensional} action $I = \int dt d^4x {\cal L}_5$, whereas 
in Faddeev-Popov theory, expectation values may be calculated by a functional
integral with a    {\it local 4-dimensional} action 
$S = \int d^4x [(1/4)F_{\m\n}^2 + ...]$.   Likewise the Fokker-Planck
``hamiltonian" $H_{\rm FP}$ determines, by the time-independent Fokker-Planck
equation $H_{\rm FP}P = 0$, a Euclidean probability distribution $P(A)$ whose
argument is a field $A(x)$ that is a function in {\it 4-dimensional} 
space-time with points $x_\m$, $\mu = 1,...4$.  By comparison the quantum
mechanical hamiltonian $H_{\rm QM}$ in ordinary quantum field theory
determines, by the time-independent Schr\"{o}dinger equation 
$H_{\rm QM}\Psi = E\Psi$,  a wave-functional $\Psi(A)$, whose argument is a
field $A(\vec{x})$ that is a function in ordinary {\it 3-space} 
$\vec{x} = (x_1,x_2,x_3)$.  Thus $H_{\rm FP}$ is not a quantum mechanical
hamiltonian at all, but rather, it claims the name ``hamiltonian" as the
generator of time translations in the time-dependent Fokker-Planck equation
\fixfokplan, where the ``time" is the artificial 5th time.  Unlike the
quantum-mechanical hamiltonian formulation, time-independent
stochastic quantization is 4-dimensionally Lorentz (Euclidean) covariant.]

	Despite the development of stochastic quantization in \pawu\ --- \bulkqg\ 
it has apparently not so far been used for non-perturbative calculations in
QCD, apart from \dznonpland.\foot{The equations of stochastic quantization have
however been applied to dissipative problems in QCD, where $t$ is the physical
time, and $x$ physical 3-space, \bodeker.}   This may possibly be due to the
complication caused by the extra ``time" variable.  Although the
time-dependent formulation allows an elegant representation, with a local
5-dimensional action, it has the complication in practice that the gluon
propagator depends on two invariants
$D(k^2, \o)$ instead of only one $D(k^2)$.  This prevents a simple power
Ansatz for the infrared behavior $1/(k^2)^{1+\a}$ that allows one to determine
the infrared behavior of the 4-dimensional theory self-consistently.  
For this reason we turn to time-independent stochastic quantization, 
where the correlators have the same number of invariants as in Faddeev-Popov
theory.    

\subsec{Outline of the present article}

	We shall not use the 5-dimensional formulation here, but
only the 4-dimensional, time-independent Fokker-Planck equation \tifokplan. 
The solution $P(A)$ to this equation cannot be represented as a
functional integral over a local 4-dimensional action.  Nor shall we
attempt to construct an explicit solution to \tifokplan.  Our strategy
instead will be to convert it into a system of tractable DS equations for the
correlators.  

	As a first step, we convert \tifokplan\ into the DS
equation,~(6.3) below, for the quantum effective action~$\G$.  The DS equation
for $\G$ appears relatively complicated, with a {\it second-order} structure
inherited from the second-order operator in~\tifokplan.  The main
methodological innovation of the present approach is that the second-order
equation for $\Gamma$ is replaced, in secs.~6 and~7,  by the much simpler DS
equation~(6.6) for a quantity, 
${\cal Q}_\m^a(x)$, that we call ``the quantum effective drift force".  Indeed
the new equation
${\cal Q}_\m^a(x) = K_\m^a(x) \ + $ (loop integrals), where 
$K_\m^a(x) = - { { \d S_{\rm YM} } \over { \d A_\m^a(x) } } 
+ a^{-1} (D_\m \p \cdot A)^c$, has the same structure
as the {\it first-order} DS equation for $\G$ in Faddeev-Popov theory, 
${ {\d \G} \over {\d A_\m^a(x)} } = { {\d S} \over {\d A_\m^a(x)} }
\ + $ (loop integrals).  In both of these equations, the leading term may be
interpreted as a drift force and, most helpfully for the renormalization
program, it is local in~$A(x)$.

	In the present work we give an improved treatment, as compared to \dznonpland,
of the longitudinal degrees of freedom in the Landau-gauge limit $a \to 0$. 
In that work we integrated out the longitudinal degrees of freedom in the
Landau-gauge limit
$a \to 0$.  This gave a time-independent Fokker-Planck equation for the
transverse degrees of freedom only, with an effective drift force that was
however non-polynomial and non-local.\foot{This was in turn decomposed into a
conservative force that reproduced the Faddeev-Popov determinent, plus a
second term that was neglected in the solution found in
\dznonpland.}  By contrast, in the present work, the difficulty of a
non-polynomial drift force is avoided by retaining the longitudinal
degrees of freedom.  Of course the longitudinal part of the propagator vanishes
with the gauge-parameter $a$ in the Landau-gauge limit $\lim \to 0$.  However
the drift force \drift\ gives a vertex that diverges like~$1/a$ and so,
counter-intuitively, the longitudinal part of the propagator in the Landau
gauge limit gives a finite contribution in internal loops, somewhat  like the
ghost in Faddeev-Popov theory.  

	We shall be satisfied here to calculate only the infrared asymptotic form of
the propagator, because that is where the challenging, non-perturbative
confining phenomena manifest themselves.  At high momentum, QCD is
perturbative, and it has been verified to one-loop order by various methods
\munoz and \halperne, including the background field method \okano, that
stochastic quantization yields the standard $\beta$-function. We leave for
another occasion a numerical calculation which would be necessary to connect
the high- and low-momentum limits.

	Since we use only the time-independent formulation here, we present, in
secs.~2 and~3, a new derivation of eq.~\tifokplan\ that does not refer to the
unphysical 5th time.  At the end of sec.~3 the Minkowskian form of
time-independent stochastic quantization is presented.  [Some readers may
prefer to go directly to sec.~4, which begins with \tifokplan.]  The new
derivation is more powerful, and yields new results, in particular, the Ward
identity of Appendix~C, and the proof in Appendix~A that the kernel of the
Fokker-Planck hamiltonian for quarks depends on gauge parameters only.  We
shall derive it from the obvious principle of {\it gauge equivalence} which
asserts that probability distributions $P(A)$ that give the same expectation
values for gauge-invariant observables $\langle W \rangle = \int dA \ W(A) \
P(A)$ are physically indistiguishable. We show that time-independent
stochastic quantization provides a class of positive, normalized proabability
distributions $P(A, a)$, parametrized by a gauge parameter $a$ that are gauge
equivalent $P(A, a_1)
\sim P(A, a_2)$, and that includes includes the Yang-Mills distribution $N
\exp(-S_{\rm YM})$ as a limiting case.  This method of quantization of gauge
fields, in which the unphysical degrees of freedom are retained but
controlled, is closely related to the physics of our solution of the DS
equations.  Indeed we find that the physical degrees of freedom are short
range, whereas the unphysical degrees of freedom are not only present but of
long range.  In Appendix~A, we extend the method to include quarks, and in
Appendix~B, to lattice gauge theory.  In Appendix~C, we derive a Ward identity
that controls the divergences of the theory.

	In sec.~8 we derive the explicit form of the DS equation for the gluon
propagator.  In secs.~9 -- 11  we adopt a simple truncation scheme, and
by means of a power-law Ansatz we solve for the infrared critical exponents
that characterize the gluon propagator in Landau gauge asymptotically, at low
momentum.  The transverse part of the gluon propagator is short
range, and the longitudinal part long range.  In the concluding section we
compare our results with calculations in Faddeev-Popov theory, and we
interpret their qualitative features in a confinement scenario. We also
suggest some challenging open problems, and possibilities for comparison with
numerical simulation in lattice gauge theory.

\newsec{Gauge equivalence}

	We first consider Euclidean gauge theory and later the Minkowskian case.		
Non-abelian gauge theories are described by the Yang-Mills action
$S_{\rm YM}(A) = (1/4) \int d^4x (F_{\m\n}^a)^2$, where
$F_{\m\n}^a = \p_\m A_\n^a - \p_\n A_\m^a + gf^{abc}A_\m^b A_\n^c$. 
The Euclidean quantum field theory
is formally defined by the probability distribution $P_{\rm YM}(A) = N
\exp[-S_{\rm YM}(A)]$, and by the expectation-values
$\langle W \rangle = \int dA \ W(A) \ P_{\rm YM}(A)$, normalized so 
$\langle 1 \rangle = 1$.    The challenge of quantizing a non-abelian gauge
theory is that $P_{\rm YM}(A)$ is not really normalizable because of the
infinite volume of the local gauge group.  

	The challenge would be hopeless, but for the fact that we are interested only
in observables that are invariant under local gauge
transformations, $W({^g A}) = W(A)$, for all  $g(x)$, where
${^g A}_\m = g^{-1}A_\m g + g^{-1} \p_\m g$. 
This suggests the notion of {\it gauge-equivalent} probability
distributions.  Two probability distributions are  gauge equivalent, 
$P_1(A) \sim P_2(A)$, if and only if $\langle W \rangle_1 = \langle W
\rangle_2$, for all gauge-invariant observables $W$, where 
$\langle W \rangle_i = \int dA \ W(A) \  P_i(A)$.  Gauge equivalence of
probability distributions is dual to gauge invariance of observables.
Distributions that are gauge-equivalent are physically indistinguishable.
Our solution to the
quantization problem will be to replace the formal probability distribution 
$N \exp[-S_{\rm YM}(A)]$, by a normalizable distribution that is
gauge-equivalent to it.  More precisely we shall exhibit a class of
gauge-equivalent normalized distributions that includes 
$N \exp[-S_{\rm YM}(A)]$ as a limiting case.

\newsec{A machine that makes gauge-equivalent probability distributions}

	The construction of gauge-equivalent probability distributions relies on
an equation that has the same form as the time-independent Fokker-Planck
equation that is used to describe diffusion in the presence of a drift force. 
In this section, for simplicity, we deal with continuum gauge fields, or
gluodynamics, only.  The extension to quarks is given in Appendix A, and to
lattice gauge theory in Appendix B.

In order to simplify the appearance of various equations, we shall, as
convenient, use the index notation $A_x$, instead of $A_\m^a(x)$, where the
subscript $x$ represents the triplet $x, \m, a$.  We use  discrete notation and
the summation convention on the new index so, for example, 
${{\p J_x} \over {\p A_x}}$ replaces
$ \int d^4x  \ {{\d J_\m^a(x)} \over {\d A_\m^a(x)}}$.

	 Let $P(A)$ be a positive, $P(A) > 0$, normalized, $\int dA \ P(A) = 1$,
probability distribution or concentration. In simple diffusion theory there is
associated with this distribution a current,
\eqn\current{\eqalign{   
   J_x = - \ \hbar {{\p P} \over {\p A_x}} + K_x P,
}}
that is composed of a diffusive term, 
$ - \ \hbar {{\p P} \over {\p A_x}}$, proportional to the gradient of the
concentration, with diffusion constant $\hbar$, and a drift term, $K_x P$. 
Here $K_x$ is the drift force, as in Ohm's law with unit conductivity.  
We have introduced $\hbar$ for future convenience for a loop expansion
which is an expansion in powers of $\hbar$.   
Conservation of probability is expressed by the equation of continuity
${{\p P} \over {\p t}} = - {{\p J_x} \over {\p A_x}}$.  The analogy of interest
to us here is associated with the time-independent situation
only.\foot{Stochastic quantization \pawu, including a drift force tangent to
the gauge orbit \dan, has traditionally been based on the time-dependent
Fokker-Planck equation 
${{\p P} \over {\p t}} = - H_{\rm FP}P$,  and relied on relaxation of the
stochastic process to an equilibrium distribution that satisfies 
$H_{\rm FP}P = 0$.  Here
$t$ is an additional, unphysical time variable that corresponds to computer
time in a Monte Carlo simulation.  By contrast, in the present article, the
quantization of the non-Abelian gauge field follows from the geometrical
principle of gauge equivalence, from which we derive
the time-independent equation $H_{\rm FP}P = 0$ directly, without reference to
the additional time variable.}  
In this case
the current is divergenceless
\eqn\conserve{\eqalign{   
   {{\p J_x} \over {\p A_x}} = 0,
}}
which reads
\eqn\diffeqc{\eqalign{
 H_{\rm FP}P \equiv {{\p } \over {\p A_x}} \
   ( - \ \hbar {{\p } \over {\p A_x}} + K_x ) \ P= 0.
}}
This is the time-independent diffusion equation with drift force $K_x$.  We
call the linear operator defined here the Fokker-Planck ``hamiltonian",
although $H_{\rm FP}$ is not hermitian, as it would be in quantum mechanics,
and it certainly is not the quantum mechanical hamiltonian of the gauge field.

	We must be sure to choose a drift force that is restoring, so this equation
determines a positive normalized distribution $P(A)$.
	If the drift force were conservative
$K_x = - \ {{\p S_{\rm YM}} \over {\p A_x}}$,
then the normalized solution would be $P(A) = N \exp(-S_{\rm YM})$. 
Gauge invariance of the Yang-Mills action,
$D_\m^a{{\d S_{\rm YM}} \over {\d A_\m^a(x)}} = 0,$ means however that the
conservative drift force 
$- {{\d S_{\rm YM}} \over {\d A_\m(x)}}$ provides no restoring force in
gauge-orbit directions.  This is remedied by introducing an additional force
$K_{{\rm gt},\m}^a(x) = D_\m^{ac} v^c$ 
that is an infinitesimal gauge transformation, so the drift force is 
made of a conservative piece and a piece that is tangent to the gauge orbit, 
\eqn\driftforce{\eqalign{
K_\m^a(x) & = - {{\d S_{\rm YM}} \over {\d A_\m^a(x)}} 
    + K_{{\rm gt},\m}^a(x)        \cr
  & =  - {{\d S_{\rm YM}} \over {\d A_\m^a(x)}}
 + D_\m^{ac} v^c. 
}}
Geometrically, the drift force is a vector field or flow, and it is
intuitively clear that a flow that is tangent to the gauge orbit has
no effect on gauge-invariant observables.  We will not fail to choose 
$v^a(x; A)$ so that $D_\m v$ is a restoring force, to insure that \diffeqc\
possesses a positive, normalized solution.  Apart from this restoring
property, $v^a(x; A)$ may in principle be an arbitrary functional of~$A$.  The
time-independent Fokker-Planck equation reads explicitly  
\eqn\fpeq{\eqalign{
H_{\rm FP}P \equiv \int d^4x \ {{\d} \over {\d A_\m^a(x)}}
  \Big[ - \ \hbar {{\d P} \over {\d A_\m^a(x)}}  
    + \Big( - {{\d S_{\rm YM}} \over {\d A_\m^a(x)}}
 + D_\m^{ac} v^c\Big)P  \Big]  = 0.
}}
This equation is a machine that produces normalized probability distributions
$P_v(A)$ that are gauge equivalent to $N\exp(-S_{\rm YM})$.

	We now prove the basic result.  {\it Positive, normalized solutions of the
diffusion equation \fpeq\ for different $v$ are gauge equivalent 
$P_v \sim P_{v'}$, and include $N \exp[-S_{\rm YM}(A)]$ as a limiting case.}
Our solution to the problem of quantizing a gauge field is to use any one of
the $P_v(A)$ to calculate expectation-values of gauge-invariant observables.
We consider observables that are invariant under infinitesimal local gauge
transformations, namely that satisfy
$G^a(x)W = 0$. Here 
$G^a(x) \equiv - \ D_\m^{ac} { {\d } \over {\d A_\m^c(x)} }$, is the generator
of an infinitesimal gauge transformation, with local Lie algebra,
$[G^a(x), G^b(y)] = \d(x-y) f^{abc} G^c(x)$, and
$(D_\m X)^a \equiv \p_\m X^a + gf^{abc}A_\m^b X^c$ is the gauge-covariant
derivative in the adjoint representation.

	The proof relies upon the decomposition of $H_{\rm FP}$,
\eqn\dech{\eqalign{
H_{\rm FP} & = H_{\rm inv} - (v, G)^{\dag}    \cr
H_{\rm inv} & \equiv \int d^4x \ {{\d} \over {\d A_\m^a(x)}}
\Big[ - \ \hbar {{\d } \over {\d A_\m^a(x)}} -  
   {{\d S_{\rm YM} } \over {\d A_\m^a(x)}} \Big]    \cr
   (v, G) & \equiv 
   - \ \int d^4x  \ v^a D_\m^{ac}{{\d} \over {\d A_\m^c(x)}} 
   = \int d^4x  \ (D_\m v)^a {{\d} \over {\d A_\m^a(x)}}    \cr
 - \ (v, G)^{\dag} & = (G, v) = \int d^4x 
    \ {{\d} \over {\d A_\m^a(x)}} (D_\m v)^a   
}}    
where $\dag$ is the adjoint with respect to the inner product defined by 
$\int dA$, and $(v,G)$ is the generator of the local gauge transformation
$v^a(x)$. Note that $H_{\rm inv}$ is a gauge-invariant
operator, $[G^a(x), H_{\rm inv}] = 0$, that has
$\exp(- S_{\rm YM})$ as a null vector,
$H_{\rm inv}\exp(- S_{\rm YM}) = 0$.  Let $P(A)$ be the normalized solution of 
$H_{\rm FP}P = 0$ for given $v$.
It is sufficient to show that
$\langle W \rangle = \int dA \ W(A) P(A)$ is
independent of $v^a(x)$ for gauge-invariant observables $W$.  
Let $\d v^a(x)$ be an arbitrary infinitesimal variation of $v^a(x)$.  The
corresponding change in $P(A)$ satisfies 
$\d H_{\rm FP} P +  H_{\rm FP} \d P = 0$, 
where $\d H_{\rm FP} = (G, \d v)$, so
\eqn\delp{\eqalign{    
\d P = - \ H_{\rm FP}^{-1} \  \d H_{\rm FP} \ P.
}}
Note that $\d H_{\rm FP} \ P$ has the form of a divergence, so it is
orthogonal to the null space of~$H_{\rm FP}$.  This change in $P$ induces the
change in expectation value 
\eqn\delw{\eqalign{   
\d \langle W \rangle & = \int dA \ \d P \ W     \cr
   & = - \ \int dA \  (H_{\rm FP}^{-1} \  \d H_{\rm FP} \ P) \ W  \cr
    & = - \  \int dA \ P \ 
  [\d H_{\rm FP}^{\dag} \ (H_{\rm FP}^{\dag})^{-1} \ W]    \cr
   & = \  \int dA \ P \ 
    [(\d v, G) \ (H_{\rm FP}^{\dag})^{-1} \ W], 
}}
where $H_{\rm FP}^{\dag} = H_{\rm inv}^{\dag} - (v,G)$.  It is sufficient to
show that
$\d \langle W \rangle = 0$.  The proof is almost immediate, but we must verify
that the dependence of $v^a(x; A)$ and $\d v^a(x; A)$ on~$A$ does not cause any
problem. Recall that $W$ is gauge invariant,    
$G^a(x) W = 0$, so we have $H_{\rm FP}^{\dag}W = H_{\rm inv}^{\dag}W$,
which implies that $H_{\rm FP}^{\dag}W$ is gauge invariant,
$$G^a(x) \ H_{\rm FP}^{\dag}W = G^a(x) \ H_{\rm inv}^{\dag}W = 0.$$
It follows by induction that 
$(H_{\rm FP}^{\dag})^n W = (H_{\rm inv}^{\dag})^n W$ 
is gauge invariant for any integer $n$,
$G^a(x) \ (H_{\rm FP}^{\dag})^n W = 0,$ 
which implies that for any analytic function, 
$f(H_{\rm FP}^{\dag})W = f(H_{\rm inv}^{\dag})W$  is gauge invariant 
$G^a(x) \ f(H_{\rm FP}^{\dag}) \ W = 0$.  This holds in particular for 
$f(z) = 1/z$, and we have 
$G^a(x) \ (H_{\rm FP}^{\dag})^{-1} \ W = 0$.  This implies that
$\d \langle W \rangle = 0$ as asserted.  Note also that if $v = 0$, then the
formal solution is $P = N\exp( - S_{\rm YM})/\hbar$.  

	This proof does not rely on Faddeev-Popov gauge-fixing  which would require
a gauge choice that selects a single representative on each gauge orbit.  The
Gribov critique is by-passed, and Singer's theorem \singer\ does not apply. 
Gauge equivalence is a weaker condition than gauge fixing, but sufficient for
physics.  In the present approach we do not attempt to eliminate 
``unphysical" variables and keep only ``physical" degrees of freedom. 
Rather we work in the full $A$-space, keeping all variables, but taming the
gauge degrees of freedom by exploiting the freedom of gauge equivalence.  It
is the unphysical degrees of freedom that provide a long-range correlator,
and a strong candidate for a confining potential. 

	Another way to obtain a gauge-equivalent probability distribution is by
gauge transformation.  If our class of gauge-equivalent probability
distributions $P_v(A)$ is large enough, then it is possible to absorb an
infinitesimal gauge transformation
$\d A_\m =  D_\m \e$ by an appropriate change $\d v$ of $v$, 
$P_v(A + D_\m \e) = P_{v + \d v}(A)$.  This is true and leads to a useful Ward
identity that is derived in Appendix~C.

	There remains to choose $v$ so it has a globally restoring property.  An
optimal way to do this is to require that the force $D_\m v$,
that is tangent to the gauge orbit, points along the direction of steepest
descent, restricted to gauge-orbit directions, of a conveniently chosen
functional. For the minimizing functional, we take the Hilbert norm-square,
$F(A) = || A ||^2 = \int d^4x \ |A|^2$, and we consider a variation
$\d A_\m = \eta D_\m v$ that is tangent to the gauge orbit in the
$v$-direction, where $\eta$ is an infinitesimal parameter.
We have
\eqn\optimal{\eqalign{ 
\d F & = 2(A, \d A) = 2\eta(A, Dv) 
  = 2\eta \int d^4x \ A_\m^a \ (\p_\m v^a + f^{abc}A_\m^b v^c)  \cr  
  &  = - \ 2\eta \int d^4x \ \p_\m A_\m^a \ v^a.    
}}
Thus the direction of steepest descent, restricted to gauge orbit directions,
is given by 
\eqn\specv{\eqalign{ 
v^a = a^{-1} \p_\m A_\m^a,
}} 
where $a > 0$ is a positive constant.  We shall take this optimal choice for
$v$, so the total drift force that appears in the diffusion equation is given
by\foot{An alternative choice suitable for the Higgs phase was proposed in
\bulkqg.}
\eqn\driftfa{\eqalign{
K_\l^a(x;A) = D_\m^{ac} F_{\m\l}^c
 + a^{-1}D_\l^{ac} \p_\m A_\m^c. 
}}
Here $a$ is a dimensionless gauge parameter.  This completes the specification
of the time-independent stochastic quantization.

	The drift force $a^{-1} D_\l \p_\m A_\m$ tends to concentrate the probability
distribution $P(A)$ close to its region of stable equilibrium, especially if
$a$ is small.  Let us find the region of stable equilibrium.  From
\optimal\ we see that $\d F < 0$ unless $A$ satisfies $\p_\m A_\m = 0$.  This
defines the region of equilibrium, which may be stable or unstable.  The
region of (local) stable equilibrium is determined by the additional condition
that the second variation be non-negative
$\d^2 F > 0$, for all variations $\d A$ tangent to the gauge orbit, namely 
$\d A = D_\m \e$, for arbitrary $\e^a(x)$.  We have just found that the first
variation is given by $\d F = - 2 (\e, \p_\m A_\m)$.  So we have, for the
second variation,
$\d^2 F = - 2 (\e, \p_\m \d A_\m) = - 2 (\e, \p_\m D_\m \e)$.  Thus
the region of {\it stable} equilibrium is determined by the two conditions
$\p_\m A_\m = 0$ and $ - \ \p_\m D_\m(A) > 0$, namely transverse configurations
$A$, for which the Faddeev-Popov operator $ - \ \p_\m D_\m(A)$ is positive. 
These two conditions define the Gribov region.  We expect that in the limit 
$a \to 0$, both conditions will be satisfied.  This is the Landau gauge,
with probability restricted to the interior of the first Gribov horizon.

	So far we have discussed Euclidean quantum field theory, which is
characterized by elliptic differential operators.  However the above
considerations also apply to the Minkowski case.  Here the formal
weight is $Q(A) = N \exp[iS_{\rm YM}]$, where 
$S_{\rm YM} = (-1/4) \int d^4x F^{\m\n}F_{\m\n}$
is the Minkowskian Yang-Mills action, where indices are raised and lowered by
the metric 
$g_{\l\m} = g^{\l\m} = {\rm diag}(1,1,1,-1)$.
Expectation-values of 
gauge-invariant time-ordered observables, are given by the Feynman path
integral
$\langle W \rangle = \int dA \ W(A) \ Q(A)$, with 
$\langle 1 \rangle = 1$.  
Instead of eq.~\fpeq, we take gauge-equivalent
configurations that are solution of the equation
\eqn\minkfpeq{\eqalign{
H_{\rm M} Q = 0,
}}
where $H_{\rm M}$ is the corresponding Minkowskian ``hamiltonian"
\eqn\minkham{\eqalign{
H_{\rm M} \equiv  \int d^4x \ (i) {{\d} \over {\d A_\k(x)}} g_{\k\l}
\Big[ (i\hbar) {{\d} \over {\d A_\l(x)}} + K^\l(x;A) \Big] \,
}}
and the ``drift force" is given by
\eqn\driftforce{\eqalign{
K^\l(x;A) & \equiv  {{\d S_{\rm YM}} \over {\d A_\l(x)}}
 + a^{-1} D^\l \p \cdot A   \cr
 & = D_\m  F^{\m\l}
 + a^{-1} D^\l \p^\m A_\m. 
}}
The linear part of this force is 
$$ \p_\k \ g^{\k\m} (\p_\m A_\n - \p_\n A_\m) 
+ a^{-1} \p_\n \ g^{\k\m} \p_k A_\m$$
which, for $a > 0$ defines a regular hyperbolic operator that is invertible
with Feynman boundary conditions.  As above, one may
show that the solutions to this equation for different values of the
gauge-parameter $a$ are gauge equivalent to each other, and for $a \to
\infty$, one regains the formal weight $N \exp(iS_{\rm YM})/\hbar$.

	The drift force $D_\l \p \cdot A$ is not conservative, so one cannot write
down an exact solution to the time-independent Fokker-Planck
equation $H_{\rm FP}P = 0$. Nor can one express the solution as a functional
integral over a local 4-dimensional action.  However we shall, by successive
changes of variable, transform this equation into an equation of
Dyson-Schwinger type that may be used for perturbative expansion and
non-perturbative solution.

\newsec{Quantum effective action in stochastic quantization}

 The partition function $Z(J)$, which is the generating
functional of correlation functions with source $J$, is defined by 
\eqn\partfunct{\eqalign{
Z(J) \equiv \int dA \ \exp(J_x A_x/ \hbar) \ P(A).
}}
It is the fourier transform (with respect to $iJ_x$) of
the probability distribution $P(A)$, and satisfies the fourier-transformed
time-independent Fokker-Planck equation,
\eqn\ftdiffeq{\eqalign{
 J_x
\Big[ J_x - K_x(\hbar {{\p} \over {\p J}}) \Big] \
Z(J) = 0. }}
Here $K_x(\hbar {{\p} \over {\p J}})$ is the local cubic polynomial in its 
argument $\hbar {{\p} \over {\p J}}$ that is defined in \driftfa.  We set
$Z(J) = \exp[W(J)/\hbar]$, where the ``free energy" $W(J)$ is the
generating functional of connected correlation functions, in terms of which
the time-independent Fokker-Planck equation reads
\eqn\diffeqw{\eqalign{
 J_x \Big[ J_x 
- K_x \Big({{\p W} \over {\p J}} + \hbar {{\p} \over {\p J}}\Big)1 
 \Big] = 0. 
}}

	The quantum effective action 
\eqn\legendre{\eqalign{ 
\Gamma(A_{\rm cl}) = J_x A_{{\rm cl}, x} - W(J)
}}
is obtained by Legendre transformation from $W(J)$, by inverting
\eqn\aclassical{\eqalign{ 
A_{{\rm cl},x}(J) \equiv 
 { { \p W } \over {\p J_x} } 
 = { { \hbar } \over {Z} } { { \p Z } \over {\p J_x} }
 = \langle A_x \rangle_J,
}}
to obtain $J_x = J_x(A_{\rm cl})$.
In the following we shall write $\Gamma(A)$ instead of 
$\Gamma(A_{\rm cl})$ when there is no ambiguity caused by using the same
symbol for the quantum Euclidean field $A$ and the classical source 
$A = A_{\rm cl}$. The gluon propagator
in the presence of the source $J$ is given by
\eqn\twopt{\eqalign{
{\cal D}_{xy}(J) & \equiv 
\hbar^{-1}  \langle \ (A_x - \langle A_x\rangle_J) 
 \ (A_y - \langle A_y\rangle_J) \ \rangle_J  \cr
& = {{ \p A_y } \over {\p J_x}} =
{{\p^2 W  } \over {\p J_x \p J_y}}.
}} 
We note in passing that the gluon propagator ${\cal D}_{xy}(J)$ 
in the presence of the source $J$ is a
positive matrix, since one has, for any $f_x$,
$\sum_{xy}f_x{\cal D}_{xy}(J)f_y = 
\hbar^{-1}  \langle \ X^2 \ \rangle_J \geq 0$,
where $X \equiv \sum_x f_x(A_x - \langle A_x\rangle_J)$,
which is positive since it is the expectation-value of a square.
It is expressed in terms of the Legendre-transformed
variables $A$ and $\Gamma(A)$ by
\eqn\propgamma{\eqalign{
{{\cal D}^{-1}}_{xy}(A) 
= {{\p^2 \Gamma(A)  } \over {\p A_x \p A_y}}. 
}}
Expectation-values of functionals $O = O(A)$ are expressed 
in terms of $Z(J)$, $W(J)$ or $\Gamma(A)$ by
\eqn\expval{\eqalign{
\langle O \rangle_J & = Z^{-1} \ O\Big(\hbar {{\p }\over {\p J }}\Big) \ Z \cr
& = O\Big({{ \p W } \over {\p J }} + \hbar {{ \p } \over {\p J }} \Big)1 \cr
\langle O \rangle_A & = O\Big(A + \hbar{\cal D}(A){{ \p } \over {\p A}} \Big)1,
}}
where the subscript indicates that the expectation-value is calculated in the
presence of the source $J$ or $A$.  In the last line, the argument of $O$ is
written in matrix notation, and reads explicitly 
$A_x +  \ \hbar{\cal D}_{xy}(A){{ \p } \over {\p A_y}}$.

	The gluon propagator ${\cal D}_{xy}(J)$ is a positive matrix, as is its
inverse ${{\cal D}^{-1}}_{xy}(A)$, so both $W(J)$ and $\Gamma(A)$ are convex
functionals.  Physics is regained when the source $J$ is set to 0, namely 
$J_x = {{ \p \Gamma } \over {\p A_x }} = 0$.  
Since $\Gamma(A)$ is a convex functional, the point 
${{ \p \Gamma } \over {\p A_x }} = 0$
is an absolute minimum of $\Gamma$.  In the absence of spontaneous symmetry
breaking, this minimum is unique and defines the quantum vacuum.  Thus physics
is regained at the absolute minimum of~$\Gamma(A)$, which justifies the name
`quantum effective action'.  

	In terms	of the Legendre-transformed variables, the
time-independent Fokker-Planck equation \diffeqw\ reads
\eqn\diffeqga{\eqalign{
 {{ \p \Gamma } \over {\p A_x }}
\Big[ {{ \p \Gamma } \over {\p A_x}}
 +  K_x \Big(A + \hbar{\cal D}(A){{ \p } \over {\p A}}\Big) 1 \Big]  = 0.
 }}
Here ${\cal D}(A)$ is expressed in terms of $\Gamma(A)$ by \propgamma, and
$K_x\Big(A + \hbar{\cal D}(A){{ \p } \over {\p A}}\Big) 1$
is evaluated next.

\newsec{Quantum effective drift force}

	We call
\eqn\qudfa{\eqalign{
{\cal Q}_x(A) \equiv 
  K_x(A + \hbar{\cal D}(A){{ \p } \over {\p A}}\Big) 1.
}}
the `quantum effective drift force'.  It is the expectation-value 
\eqn\defqedf{\eqalign{
{\cal Q}_x(A_{\rm cl}) = \langle K_x \rangle_{A_{\rm cl}}
}}
of the drift force  \driftfa\ in the presence of the source $A_{\rm cl}$, as
one sees from \expval. To evaluate it, we expand $K_x(A)$ in terms of its
coefficient functions
\eqn\expkc{\eqalign{
K_x(A)  & = K_{xy}^{(1)}A_y 
    + (2!)^{-1} K^{(2)}_{x;yz} A_y A_z    
   + (3!)^{-1}  \ K^{(3)}_{xyzw} A_y A_z A_w.
}}
The coefficient functions are found from \driftfa, and are given in the
explicit notation by 
\eqn\kone{\eqalign{  
{K^{(1)}}_{\k\l}^{ab}(x,y) & = - \ {S_{\rm YM}^{(2)}}_{\k\l}^{ab}(x,y)   \cr
   & = \d^{ab} 
\ [ \ (\p^2 \d_{\k\l} - \p_\k \p_\l) + a^{-1} \p_\k \p_\l \ ] \d(x-y)  
}}
\eqn\ktwo{\eqalign{  
{K^{(2)}}_{\k\l\m}^{abc}(x;y,z) = 
   - \ {S_{\rm YM}^{(3)}}_{\k\l\m}^{abc}(x,y,z) 
   + a^{-1}{K_{\rm gt}^{(2)}}_{\k\l\m}^{abc}(x;y,z)
}}
\eqn\ktwoa{\eqalign{  
 - \ {S_{\rm YM}^{(3)}}_{\k\l\m}^{abc}(x,y,z)  = g f^{abc} \ 
\Big( & \p_{[\l} \d(x-y) \d_{\m]\k} \ \d(y-z)  
  + \p_{[\m} \d(y-z) \d_{\k]\l} \ \d(z-x)  \cr
 &  \ \ \ \ \ \ \ \ \ + \ \p_{[\k} \d(z-x) \d_{\l]\m} \ \d(x-y)   \Big)  
}}
\eqn\ktwob{\eqalign{  
{K_{\rm gt}^{(2)}}_{\k\l\m}^{abc}(x;y,z) =  g f^{abc} \ 
   \p_{[\m} \d(x-z) \d_{\l]\k} \ \d(x-y)   
}}
\eqn\kthree{\eqalign{ 
{K^{(3)}}_{\k\l\m\n}^{abcd}(x,y,z,w) 
    & = - \ {S_{\rm YM}^{(4)}}_{\k\l\m\n}^{abcd}(x,y,z,w)   \cr
& = - \ g^2\Big( \ f^{abe} f^{cde}\d_{\k[\m}\d_{\n]\l}
    + f^{ace} f^{bde}\d_{\k[\l}\d_{\n]\m}   \cr
 & \ \ \ \ \ \ \ \ \ \ + f^{ade} f^{cbe}\d_{\k[\m}\d_{\l]\n}  \ \Big)
      \ \d(x-y)\d(x-z)\d(x-w),
}}
where $\d_{\k[\l}\d_{\n]\m} \equiv \d_{\k\l}\d_{\n\m} - \d_{\k\n}\d_{\l\m}$
etc. The contribution to each
coefficient $K^{(n)}$ from $S_{\rm YM}$ is symmetric in all its arguments,
including the first.  Thus 
${K^{(3)}}_{\k\l\m\n}^{abcd}(x,y,z,w)$ is symmetric under permutations of
its 4 arguments. Moreover ${K^{(1)}}_{\k\l}^{ab}(x,y)$ is manifestly symmetric
in its arguments.  On the other hand the first argument of
${K_{\rm gt}^{(2)}}_{\k\l\m}^{abc}(x;y,z)$ is distinguished.

 The evaluation of the quantum effective drift force,
${\cal Q}_x(A) = K_x(A + \hbar {\cal D}{ {\p} \over {\p A}})1$, is
straightforward.  By substitution into \expkc\ we have
\eqn\qcom{\eqalign{
{\cal Q}_x(A) & = K_{xy}^{(1)}A_y 
    + (2!)^{-1} K^{(2)}_{x;yz} 
  \Big(A_y + \hbar {\cal D}_{yu}{ {\p} \over {\p A_u} } \Big) A_z    \cr
   & + (3!)^{-1}  \ K^{(3)}_{xyzw} 
  \Big(A_y + \hbar {\cal D}_{yu}{ {\p} \over {\p A_u} } \Big) 
  \Big(A_z + \hbar {\cal D}_{zv}{ {\p} \over {\p A_v} } \Big) A_w   \cr
& = K_{xy}^{(1)}A_y 
    + (2!)^{-1} K^{(2)}_{x;yz}(A_y A_z + \hbar {\cal D}_{yz})    \cr
   & + (3!)^{-1}  \ K^{(3)}_{xyzw} 
  \Big(A_y + \hbar {\cal D}_{yu}{ {\p} \over {\p A_u} } \Big) 
  (A_z A_w + \hbar {\cal D}_{zw}).
}}
Use of the identity,
\eqn\dervertex{\eqalign{
{{ \p {\cal D}_{zw}(A) } \over {\p A_r}}     
 = - {\cal D}_{zs}(A) {\cal D}_{wt}(A)
 \ { {\p^3 \Gamma(A)} \over { \p A_r \p A_s \p A_t } },
}}
that follows from 
$({\cal D}^{-1})_{z,w}(A) 
= { {\p^2 \Gamma(A)} \over { \p A_z \p A_w } }$, gives the formula for 
${\cal Q}_x(A)$ that is the first equation of next section.

\newsec{Basic equations for ${\cal Q}$ and $\Gamma$}

	The first basic equation of the present method is the formula, just derived,
for the quantum effective drift force,
\eqn\qudf{\eqalign{
{\cal Q}_x(A)  = K_x(A)
    & + \hbar \ (2!)^{-1} K^{(2)}_{x;yz}{\cal D}_{yz}   
    + \hbar \ (2!)^{-1}  \ K^{(3)}_{xyzw}{\cal D}_{yz}A_w   \cr
   & - \ \hbar^2 \  (3!)^{-1}  \ K^{(3)}_{xyzw} 
   {\cal D}_{yr}{\cal D}_{zs} {\cal D}_{wt}
       \ { {\p^3 \Gamma(A)} \over { \p A_r \p A_s \p A_t } },
}}
where ${\cal D} = {\cal D}(A)$ is the gluon propagator 
in the presence of the source $A$, and is expressed in
terms of $\Gamma(A)$ by 
$({\cal D}^{-1})_{z,w} = { {\p^2 \Gamma(A)} \over { \p A_z \p A_w }}$.  
This equation is represented graphically in fig.~1.  The terms of order
$\hbar$ and $\hbar^2$ correspond to one and two loops in the figure, 
and we write
\eqn\qudfloopa{\eqalign{
{\cal Q}_x = K_x + \hbar {\cal Q}_{{\rm 1loop},x}(\Gamma)
   + \hbar^2 {\cal Q}_{{\rm 2loop},x}(\Gamma).
}}
The second basic equation of the present approach is obtained by writing the
time-independent Fokker-Planck equation \diffeqga, satisfied by the quantum
effective action, $\Gamma$, in terms of the quantum effective drift force,
${\cal Q}_x(A)$,
\eqn\diffeqg{\eqalign{
  {{ \p \Gamma } \over {\p A_x }}
\Big[ {{ \p \Gamma } \over {\p A_x}}
 +  {\cal Q}_x(A) \Big]  = 0.
 }}
This equation is of classic Hamilton-Jacobi type, with energy $E = 0$,
and hamiltonian 
$H(p, A) = p_x[p_x + {\cal Q}_x(A)]$.

	The pair of equations \qudf\ and \diffeqg\ forms the basis of the present
approach and allows a systematic calculation of the correlation functions.
Equation \qudf\ resembles the DS equation for the gluon field
in Faddeev-Popov theory
namely ${{ \p \Gamma } \over {\p A_x}} 
=  {{ \p S} \over {\p A_x }}(A + \hbar{\cal D}{{ \p } \over {\p A}})1$,
where 
$S = S_{\rm YM} + S_{\rm gf} + S_{\rm gh}$, and $S_{\rm gh}$ is the ghost
action.  Indeed the same expressions appear in both equations, as is seen most
easily from fig.~1, except that the contribution from the ghost action,
${{ \d S_{\rm gh}} \over {\d A_\m }}(A + \hbar{\cal D}{{ \d } \over {\d A}})1$,
is replaced by the term proportional to $a^{-1}$ in the gluon vertex~$K^{(2)}$.

		In the functional equations \qudf\ and \diffeqg, satisfied by ${\cal Q}_x(A)$
and $\Gamma(A)$, $A$ is a dummy variable, and each of these functional
equations represents a set of equations satisfied by the
coefficient functions that appear in the expansions in powers of $A$,
\eqn\expq{\eqalign{
{\cal Q}_x(A) =  Q^{(1)}_{x;y} A_y 
+ (2!)^{-1} \ Q^{(2)}_{x;y_1,y_2} A_{y_1}A_{y_2} + ...
}}
\eqn\expg{\eqalign{
\Gamma(A) = & \ (2!)^{-1}  \Gamma^{(2)}_{y_1,y_2} A_{y_1}A_{y_2}   
 + (3!)^{-1} \Gamma^{(3)}_{y_1,y_2,y_3}  A_{y_1}A_{y_2} A_{y_3} + ... \ \ ,
}}
where $\Gamma^{(n)}$ is the proper $n$-vertex.  The individual equations for
the coefficient functions are conveniently obtained by differentiating \qudf\
and \diffeqg\ $n$ times with respect to $A_z$, and then setting 
$A = 0$. 

	We now come to an important point.  The time-independent Fokker-Planck
equation \diffeqga\ satisfied by $\Gamma(A)$ is equivalent to the pair of
coupled equations
\qudf\ and \diffeqg\ that is satisfied by the pair $\Gamma(A)$ and 
${\cal Q}_x(A)$.  Indeed {\it every solution of \qudf\ and \diffeqg\ yields a
solution of \diffeqga\ and conversely.}  This remark is the key to transforming
the time-independent Fokker-Planck equation into an equation of DS type.  For
it turns out that the Hamilton-Jacobi equation \diffeqg\ may be
solved exactly and explicitly for the coefficient functions $\Gamma^{(n)}$ 
of $\Gamma(A)$ in terms of the coefficient functions
$Q_x^{(m)}$ of 
${\cal Q}_x(A)$, where $m < n$.  In fact we shall obtain a simple
algebraic -- indeed, rational -- formula for 
$\Gamma^{(n)} = \Gamma^{(n)}(Q)$ for every $n$. 
This allows us to change variable from the quantum effective action,
$\Gamma = \Gamma({\cal Q})$, to the quantum effective drift force,
${\cal Q}_x$.  It will be the last in our series of changes of variable, $P(A)
\to Z(J) \to W(J) \to \Gamma(A) \to {\cal Q}_x(A)$.  

	Neither the Hamilton-Jacobi equation \diffeqg\ nor its solution 
$\Gamma = \Gamma({\cal Q})$ contains $\hbar$.  When the solution 
of \diffeqg, $\Gamma = \Gamma({\cal Q})$, is substituted into \qudfloopa,
one obtains an equation of the form 
\eqn\qudfloop{\eqalign{
{\cal Q}_x  = K_x + \hbar {\cal Q}_{{\rm 1loop},x}({\cal Q})
    + \hbar^2 {\cal Q}_{{\rm 2loop},x}({\cal Q}).
}}
This is an equation of DS type for the quantum effective drift force
${\cal Q}_x$.  By iteration, it provides the $\hbar$-expansion of 
${\cal Q}_x$.  The zeroth-order term $K_x$, given in \driftfa, 
is a local function of $A$.  We shall find an approximate, non-perturbative
solution of this equation.  But first we must find
$\Gamma({\cal Q})$. 

\newsec{Solution for quantum effective action $\Gamma({\cal Q})$} 

	In this section we solve \diffeqg\ for the coefficient functions
$\Gamma^{(2)} = \Gamma^{(2)}(Q)$ and
$\Gamma^{(3)} = \Gamma^{(3)}(Q)$.  
The solution for $\Gamma^{(4)}$ and $\Gamma^{(n)}$
for arbitrary~$n$ is found in Appendix~D.
	
	The solution for $\Gamma^{(2)}$, reads simply
\eqn\simpres{\eqalign{
\Gamma^{(2)}_{x_1x_2} = - \ Q^{(1)}_{x_1;x_2}.
}}
Note:  By definition, 
$\Gamma^{(n)}_{x_1 x_2...x_n}$ is symmetric in its $n$ arguments, whereas 
$Q^{(n-1)}_{x_1;x_2,...x_n}$ has a
distinguished first argument and is symmetric only in the remaining $n-1$
arguments, so in general the equation 
$\Gamma^{(n)}_{x_1 x_2...x_n} = - \ Q^{(n-1)}_{x_1;x_2,...x_n}$ 
would not be consistent.  However symmetries in fact constrain 
$Q^{(1)}_{x_1;x_2}$ to be symmetric, 
$Q^{(1)}_{x_1;x_2} = Q^{(1)}_{x_2;x_1}$, as we will see, so \simpres\ is in
fact consistent.
	
	To prove \simpres, we differentiate \diffeqg\ with respect to
$A_{x_1}$ and $A_{x_2}$, and obtain, after setting $A = 0$,
\eqn\hjtwo{\eqalign{
    \Gamma^{(2)}_{x_1 x} \ ( \Gamma^{(2)}_{x x_2}  +  Q^{(1)}_{x;x_2} )  
       + (x_1 \leftrightarrow x_2) = 0.
 }}
To solve this equation for $\Gamma^{(2)}$, we diagonalize all the matrices by
taking fourier transforms.  In the extended notation this equation reads,
\eqn\hjtwo{\eqalign{
 \int d^4x \   \Big( \  {\Gamma^{(2)}}_{\m_1 \m}^{a_1 a}(x_1,x) 
  [ {\Gamma^{(2)}}_{\m \m_2}^{a a_2}(x,x_2) \
 & +  {Q^{(1)}}_{\m \m_2}^{a a_2}(x;x_2) ]   \cr
& + [(x_1, \m_1, a_1) \leftrightarrow (x_2, \m_2, a_2)] \ \Big) = 0,
 }}
and we take fourier transforms, 
\eqn\fourier{\eqalign{
{Q^{(1)}}_{\l\m}^{ab}(x;y) & = \d^{ab} \ (2\pi)^{-4}
\int d^4k \exp[i k \cdot (x-y)]  \ \tilde{Q}_{\l\m}^{(1)}(k)  \cr
{\Gamma^{(2)}}_{\l\m}^{ab}(x,y) & = \d^{ab} \ (2\pi)^{-4}
\int d^4k \exp[i k \cdot (x-y)] \ \tilde{\Gamma}_{\l\m}^{(2)}(k).
}}
Color, translation, and Lorentz invariance, and use of the transverse and
longitudinal projectors, 
$P_{\l\m}^{\rm T}(k) = (\d_{\l\m} - k_\l k_\m/k^2)$
and
$P_{\l\m}^{\rm L}(k) = k_\l k_\m/k^2$
give the decomposition of these quantities into their transverse and
longitudinal invariant functions,
\eqn\lorentz{\eqalign{
\tilde{Q}_{\l\m}^{(1)}(k) 
    & = Q^{(1){\rm T}}(k^2) \ P_{\l\m}^{\rm T}(k) 
   +  Q^{(1){\rm L}}(k^2) \ P_{\l\m}^{\rm L}(k) \cr
\tilde{\Gamma}_{\l\m}^{(2)}(k) 
     & = T(k^2) \ P_{\l\m}^{\rm T}(k) +  a^{-1}L(k^2) \ P_{\l\m}^{\rm L}(k).
}}
The coefficient $a^{-1}$ is introduced here for later
convenience.  In terms of the fourier transforms, \hjtwo\ reads
\eqn\hjtwof{\eqalign{
   \tilde{\Gamma}_{\m_1\m}^{(2)}(k) \ 
[ \tilde{\Gamma}_{\m\m_2}^{(2)}(k) +  \tilde{Q}_{\m\m_2}^{(1)}(k) ] \ 
+ (k, \m_1, \m_2 \leftrightarrow - \ k, \m_2, \m_1) = 0.
 }}
Color and Lorentz symmetries, as expressed in \lorentz, constrain
$\tilde{Q}_{\l\n}^{(1)}(k)$ to be a symmetric tensor that is even in~$k$, 
$\tilde{Q}_{\l\n}^{(1)}(k) = \tilde{Q}_{\n\l}^{(1)}(k) =
\tilde{Q}_{\l\n}^{(1)}(-k)$, as is $\tilde{\Gamma}_{\l\n}^{(2)}(k)$.  Products
of such tensors have the same property, and as a result, the two terms in
\hjtwof\ are equal, and we have
\eqn\hjtwofa{\eqalign{
   \tilde{\Gamma}_{\m_1\m}^{(2)}(k) \ 
[ \tilde{\Gamma}_{\m\m_2}^{(2)}(k) +  \tilde{Q}_{\m\m_2}^{(1)}(k) ] \ 
     = 0,
 }} 
which proves the assertion~\simpres.  For future reference, we note
\eqn\simpresi{\eqalign{
   (\tilde{D}^{-1})_{\l\m}(k)
  & = \tilde{\Gamma}_{\l\m}^{(2)}(k) = - \ \tilde{Q}_{\l\m}^{(1)}(k)    \cr
  & = T(k^2) \ P_{\l\m}^{\rm T}(k) +  a^{-1}L(k^2) \ P_{\l\m}^{\rm L}(k).
}}
Here we have introduced the usual gluon propagator, with sources set to 0, 
$D_{xy} = {\cal D}_{xy}(A)|_{A=0}$.  It is given in terms of $\Gamma$ by
$(D^{-1})_{xy} = { {\p^2 \Gamma} \over {\p A_x \p A_y}}|_{A=0} 
   = \Gamma_{xy}^{(2)}$.

	We next find  $\Gamma^{(3)}$.  For this purpose we differentiate \diffeqg\
with respect to $A_{x_1}$, $A_{x_2}$ and $A_{x_3}$, and obtain, after setting
$A = 0$,
\eqn\hjthree{\eqalign{
    \Gamma^{(2)}_{x_1 x} \ 
( \Gamma^{(3)}_{x x_2 x_3} + Q^{(2)}_{x; x_2 x_3} ) \ 
+ \ (\rm cyclic)   = 0,
 }}
where we have used $\Gamma^{(2)} = - \ Q^{(1)}$, and (cyclic) represents the
cyclic permutations of~(1,2,3). 
A novelty of the stochastic method is now apparent.
For $Q^{(2)}_{x;x_2,x_3}$, unlike $Q^{(1)}_{x;x_2}$,
is not completely symmetric in all its arguments as it would be if the drift
force were conservative.  As a result, the equation
$\Gamma^{(3)}_{x,x_2,x_3} + Q^{(2)}_{x;x_2,x_3} = 0$
has no solution.  This is already apparent to zero order in $\hbar$, where
$Q^{(2)}_{x;x_2,x_3} = K^{(2)}_{x;x_2,x_3}$, 
but $K^{(2)}_{x;x_2,x_3}$ is not symmetric in its 3 arguments, as noted
above.  

	To solve \hjthree\ for $\Gamma^{(3)}_{x x_2 x_3}$, we again diagonalize the
matrix $\Gamma^{(2)}_{x_1 x}$ by Fourier transformation.  To do so, we
write the last equation in the extended notation,
\eqn\hjthreea{\eqalign{
 \int d^4x \  & \Big(
   \ {\Gamma^{(2)}}_{\m_1\m}^{a_1 a}(x_1,x) \ 
[ {\Gamma^{(3)}}_{\m\m_2\m_3}^{a a_2 a_3}(x,x_2,x_3) 
+  {Q^{(2)}}_{\m\m_2\m_3}^{a a_2 a_3}(x;x_2,x_3) ] \ 
+ \ (\rm cyclic)  \  \Big) = 0,
 }}
and take fourier transforms,
\eqn\fouriera{\eqalign{
{Q^{(2)}}_{\m_1\m_2\m_3}^{a_1 a_2 a_3}(x_1;x_2,x_3)  = (2\pi)^{-8}
& \int d^4k_1 d^4k_3 d^4k_3   
 \exp(i k_1 \cdot x_1 + i k_2 \cdot x_2 + i k_3 \cdot x_3)   \cr
& \times \d(k_1 + k_2 + k_3) \ 
 {\tilde{Q}^{(2)}}{_{\m_1 \m_2 \m_3}^{a_1 a_2 a_3}}(k_1;k_2,k_3)  
}}
\eqn\fourierb{\eqalign{
{\Gamma^{(3)}}_{\m_1\m_2\m_3}^{a_1 a_2 a_3}(x_1,x_2,x_3)  = (2\pi)^{-8}
& \int d^4k_1 d^4k_3 d^4k_3   
 \exp(i k_1 \cdot x_1 + i k_2 \cdot x_2 + i k_3 \cdot x_3)   \cr
& \times \d(k_1 + k_2 + k_3) \ 
{\tilde{\Gamma}^{(3)}}{_{\m_1 \m_2 \m_3}^{a_1 a_2 a_3}}(k_1,k_2,k_3),  
}}
where 
${\tilde{Q}^{(2)}}{_{\m_1 \m_2 \m_3}^{a_1 a_2 a_3}}(k_1;k_2,k_3)$ and
${\tilde{\Gamma}^{(3)}}{_{\m_1 \m_2 \m_3}^{a_1 a_2 a_3}}(k_1,k_2,k_3)$ 
are defined only for $k_1 + k_2 + k_3 = 0$.
This gives
\eqn\hjthreea{\eqalign{
    \tilde{\Gamma}_{\m_1\m}^{(2)}(k_1) \ 
[ {\tilde{\Gamma}^{(3)}}{_{\m\m_2\m_3}^{a_1 a_2 a_3}}(k_1,k_2,k_3) 
+  {\tilde{Q}^{(2)}}{_{\m\m_2\m_3}^{a_1 a_2 a_3}}(k_1;k_2,k_3) ]    
 \ +   (\rm cyclic)     = 0.
 }}
We use the symmetry of 
${\tilde{\Gamma}^{(3)}}{_{\m_1\m_2\m_3}^{a_1 a_2 a_3}}(k_1,k_2,k_3)$
in its three arguments to write this as
\eqn\hjthreeb{\eqalign{
    [\tilde{\Gamma}^{(2)}_{\m_1\n_1}(k_1) \d_{\m_2 \n_2}  \d_{\m_3 \n_3} 
   + ({\rm cyclic}) ]   \ 
  {\tilde{\Gamma}^{(3)}}{_{\n_1\n_2\n_3}^{a_1 a_2 a_3}}(k_1,k_2,k_3)  
   = - \ {H^{(3)}}{_{\m_1\m_2\m_3}^{a_1 a_2 a_3}}(k_1,k_2,k_3),
 }}
where
\eqn\defsofq{\eqalign{
{H^{(3)}}{_{\m_1\m_2\m_3}^{a_1 a_2 a_3}}(k_1,k_2,k_3) \equiv
 \tilde{\Gamma}^{(2)}_{\m_1\m}(k_1) \ 
{\tilde{Q}^{(2)}}{_{\m\m_2\m_3}^{a_1 a_2 a_3}}(k_1;k_2,k_3)  
   +  (\rm cyclic ) .
}}

	To complete the diagonalization of $\tilde{\Gamma}^{(2)}_{\l\m}(k)$, and solve
\hjthreeb\ for 
${\tilde{\Gamma}^{(3)}}{_{\m_1\m_2\m_3}^{a_1 a_2 a_3}}(k_1,k_2,k_3)$,
we apply a transverse or longitudinal projector to each of
its three arguments, and use the transverse and longitudinal decomposition
of $\tilde{\Gamma}^{(2)}_{\l\m}(k)$ given in \simpresi. One obtains
${\tilde{\Gamma}^{(3)}}{_{\m_1\m_2\m_3}^{a_1 a_2 a_3}}(k_1,k_2,k_3)$ in terms
of its transverse and longitudinal projections, defined by
$X_\m^T(k) \equiv P_{\m\n}^T(k)X_\n(k)$
and $X_\m^L(k) \equiv P_{\m\n}^L(k)X_\n(k)$,
\eqn\solvegthr{\eqalign{
{\tilde{\Gamma}^{(3){\rm T T T}}}
  {_{\m_1\m_2\m_3}^{a_1 a_2 a_3}}(k_1,k_2,k_3) = - \
  [T(k_1^2) + T(k_2^2) + T(k_3^2)]^{-1}  
  \ {H^{(3){\rm T T T}}}{_{\m_1\m_2\m_3}^{a_1 a_2 a_3}}(k_1,k_2,k_3)
}}
\eqn\solvegthra{\eqalign{
{\tilde{\Gamma}^{(3){\rm L T T}}}
  {_{\m_1\m_2\m_3}^{a_1 a_2 a_3}}(k_1,k_2,k_3) = - \
  [a^{-1}L(k_1^2) + T(k_2^2) + T(k_3^2)]^{-1}  
  \ {H^{(3){\rm L T T}}}{_{\m_1\m_2\m_3}^{a_1 a_2 a_3}}(k_1,k_2,k_3)
}}
etc.  The corresponding formulas for $\tilde{\Gamma}^{(4)}$ and
$\tilde{\Gamma}^{(n)}$ are found in Appendix~D.

\newsec{Dyson-Schwinger equation for the gluon propagator}

	We have solved the second basic equation \diffeqg\ for the coefficient
functions $\tilde{\Gamma}^{(n)}$, and expressed them in terms of the
$\tilde{Q}^{(m)}$, for $m < n$.  We now turn to the first basic
equation~\qudf, and derive the equations for the coefficient functions
${Q}^{(m)}$ by the same method of taking derivatives and setting $A$ = 0.  To
derive the equation for $Q^{(1)}$, we differentiate \qudf\ with respect to
$A_y$, and obtain, after setting $A = 0$,
\eqn\qudfone{\eqalign{ 
Q^{(1)}_{x; y} = & \ K^{(1)}_{x; y}  
    - \ \hbar \ (2!)^{-1}  \  K^{(2)}_{x;x_1,x_2}   
      D_{x_1 y_1} \ D_{x_2 y_2}  \ \Gamma^{(3)}_{y_1 y_2 y}  
  + \ \hbar \ (2!)^{-1}  \ K^{(3)}_{x x_1 x_2 y} \ D_{x_1 x_2} \cr 
	&  - \hbar^2  \ (3!)^{-1} K^{(3)}_{x x_1 x_2 x_3} \ D_{x_1 y_1} 
     \ D_{x_2 y_2} \ \ D_{x_3 y_3} \ \Gamma^{(4)}_{y_1 y_2 y_3 y} \cr
& + \hbar^2  \ (2!)^{-1} \ K^{(3)}_{x x_1 x_2 x_3} 
    D_{x_1 z_1}  D_{x_2 z_2}
 \  \Gamma^{(3)}_{z_1 z_2 z_3}  
 D_{z_3 y_1} \ D_{x_3 y_2} \ \Gamma^{(3)}_{y_1,y_2,y},    
}}
where we have again used \dervertex. This equation is represented
diagrammatically in fig.~2.

	In momentum space the coefficients \kone\ -- \kthree\ of the drift force read
\eqn\fourierko{\eqalign{
{K^{(1)}}_{\l\m}^{ab}(x;y) = \d^{ab} \ (2\pi)^{-4}
\int d^4k \exp[i k \cdot (x-y)] \ \tilde{K}_{\l\m}^{(1)}(k),
}}
\eqn\fourierktw{\eqalign{
{K^{(2)}}_{\m_1\m_2\m_3}^{a_1 a_2 a_3}(x_1;x_2,x_3)  
= f^{a_1 a_2 a_3} \ 
(2\pi)^{-8} & \int d^4k_1 d^4k_3 d^4k_3   
 \exp(i k_1 \cdot x_1 + i k_2 \cdot x_2 + i k_3 \cdot x_3)   \cr
& \times \d(k_1 + k_2 + k_3) \ 
 {\tilde{K}^{(2)}}{_{\m_1 \m_2 \m_3}}(k_1;k_2,k_3)  
}}
\eqn\fourierkth{\eqalign{
{K^{(3)}}_{\m_1\m_2\m_3\m_4}^{a_1 a_2 a_3 a_4}(x_1, x_2,x_3, x_4)  
  = (2\pi)^{-12}
& \int d^4k_1 d^4k_3 d^4k_3 d^4k_4 
 \exp(i \sum_{i=1}^4 k_i \cdot x_i)  \cr 
& \times \d(k_1 + k_2 + k_3 + k_4) \ 
{\tilde{K}^{(3)}}{_{\m_1 \m_2 \m_3 \m_4}^{a_1 a_2 a_3 a_4}},  
}}
where
\eqn\tkone{\eqalign{  
- \ \tilde{K}^{(1)}_{\l\m}(k) 
   = [ \ (k^2 \d_{\l\m} - k_\l k_\m) + a^{-1} k_\l k_\m \ ] 
}}
\eqn\tktwo{\eqalign{  
{\tilde{K}^{(2)}}{_{\m_1 \m_2 \m_3}}(k_1;k_2,k_3)    
  & =    - \ {\tilde{S}_{\rm YM}^{(3)}}{_{\m_1 \m_2 \m_3}}(k_1,k_2,k_3)
 + a^{-1}{\tilde{K}_{\rm gt}^{(2)}}{_{\m_1 \m_2 \m_3}}(k_1;k_2,k_3),  \cr
- \ {\tilde{S}_{\rm YM}^{(3)}}{_{\m_1 \m_2 \m_3}}(k_1,k_2,k_3) 
  & \equiv    ig \ [ \ 
  (k_1)_{[\m_2} \d_{\m_3]\m_1}  + ({\rm cyclic}) \ ]  \cr
{\tilde{K}_{\rm gt}^{(2)}}{_{\m_1 \m_2 \m_3}}(k_1;k_2,k_3) 
 & \equiv     i \ g \ [\ (k_3)_{\m_3} \d_{\m_1\m_2}
              - (2 \leftrightarrow 3) \ ].
}}
\eqn\tkthree{\eqalign{ 
- \ {\tilde{K}^{(3)}}{_{\m_1 \m_2 \m_3 \m_4}^{a_1 a_2 a_3 a_4}}
     = g^2 \ ( \ f^{a_1a_2e} f^{a_3a_4e} \ \d_{\m_1[\m_3}\d_{\m_4]\m_2}   
   & + f^{a_1a_3e} f^{a_2a_4e} \ \d_{\m_1[\m_2}\d_{\m_4]\m_3}   \cr
  & + f^{a_1a_4e} f^{a_3a_2e} \ \d_{\m_1[\m_3}\d_{\m_2]\m_4} \ ). 
}}
With  $\tilde{Q}^{(1)} = - \tilde{D}^{-1}$, we obtain finally the DS equation
for the gluon propagator
\eqn\qudfonem{\eqalign{
- \ \d^{ab} \ (\tilde{D}^{-1})_{\l\m}(k) 
= & \ - \d^{ab} \ [(k^2 \d_{\k\l} - k_\k k_\l) + a^{-1} k_\k k_\l ]   \cr
    & - \hbar \ f^{a a_1 a_2} \ (2!)^{-1} (2\pi)^{-4}
  \int dk_1 \ 
\tilde{K}^{(2)}_{\l\l_1\l_2}(k;-k_1,k_1-k)   \cr
    & \ \ \ \ \ \ \ \ \ \ \ \ \ \ \ \ \ \ \ \times
  \tilde{D}{_{\l_1\m_1}}(k_1) 
  \ \tilde{D}{_{\l_2\m_2}}(k-k_1)  
  \ {\tilde{\Gamma}^{(3)}}{_{\m_1\m_2\m}^{a_1 a_2 b}}(k_1,k-k_1,-k)  \cr 
  & + \hbar \ (2!)^{-1} (2\pi)^{-4} \int dk_1
  \ {\tilde{K}^{(3)}}{_{\l \l_1 \l_2 \m}^{a \ c \ c  \ b}}
   \ \tilde{D}{_{\l_1 \l_2}}(k_1)  \cr
	& + \d^{ab} \ \tilde{Q}_{{\rm 2l},\l\m}^{(1)}(k),
}}
where the two-loop term is given by
\eqn\qonetwolm{\eqalign{
\d^{ab} \ \tilde{Q}_{{\rm 2l},\l\m}^{(1)}(k) 
    & \equiv -  \ \hbar^2  \ (3!)^{-1} 
  (2\pi)^{-8}\int dk_1 dk_2 
    \ {\tilde{K}^{(3)}}{_{\l\l_1\l_2\l_3}^{a a_1 a_2 a_3}}
   \ \tilde{D}_{\l_1 \m_1}(k_1)  \ \tilde{D}_{\l_2 \m_2}(k_2)   \cr
& \ \ \ \ \ \ \ \ \ \ \ \ \ \ \ \ \ \ \ \  
   \times   \tilde{D}_{\l_3 \m_3}(k - k_1 - k_2) 
 \ {\tilde{\Gamma}^{(4)}}
        {_{\m_1 \m_2 \m_3 \m}^{a_1 a_2 a_3 b}}(k_1,k_2,k-k_1-k_2,-k)  \cr 
& + \hbar^2  \ (2!)^{-1}  
				(2\pi)^{-8} \int dk_1 dk_2  \  
   \ {\tilde{K}^{(3)}}{_{\l\l_1\l_2\l_3}^{a a_1 a_2 a_3}}  
   \ \tilde{D}_{\l_1 \n_1}(k_1)      \cr
& \ \ \ \ \ \ \ \ \ \ \ \ \ \ \ \ \ \ \ \ \ \ 
\times  
    \ \tilde{D}_{\l_2 \n_2}(k_2)
 \   {\tilde{\Gamma}^{(3)}}{_{\n_1 \n_2 \n_3}^{a_1 a_2 b_1}}(k_1,k_2,-k_1-k_2)
\   \tilde{D}_{\n_3 \m_1}(k_1+k_2)    \cr
& \ \ \ \ \ \ \ \ \ \ \ \ \ \ \ \ \ \ \ \ \ \     \times 
   \   \tilde{D}_{\l_3 \m_2}(k-k_1-k_2)
\ {\tilde{\Gamma}^{(3)}}{_{\m_1 \m_2 \m}^{b_1 a_3 b}}(k_1+k_2,k-k_1-k_2,-k).   
}}

\newsec{Truncation scheme}

	To obtain a non-perturbative solution of the DS equations, it is
necessary to truncate them in some way.  Needless to say,
truncation remains an uncontrolled approximation until it is tested by varying
the scheme, or by comparison with numerical simulation, as discussed in the
Introduction and Conclusion.  Moreover the truncation scheme is gauge
dependent.  This situation is familiar in atomic physics where bound state
calculations are done in the Coulomb gauge.  We shall
ultimately solve the truncated system in the Landau-gauge limit.  

	As a first step we neglect the two-loop contribution in eq.~\qudfonem.  We
shall also not retain the tadpole term, which in any case gets absorbed in the
renormalization.  The 3-vertex that we will
obtain 
\eqn\colorg{\eqalign{
{\tilde{\Gamma}^{(3)}}{_{\m_1\m_2\m_3}^{a_1 a_2 a_3}}(k_1,k_2,k_3)
 = f^{a_1 a_2 a_3} \ \tilde{\Gamma}_{\m_1\m_2\m_3}^{(3)}(k_1,k_2,k_3),
}}
defined for $k_1+k_2+k_3 = 0$, has the color dependence
that allows us to use the identity
$f^{a a_1 a_2} f^{a_1 a_2 b} = N \d^{ab}$ for $SU(N)$ color group.  As a
result, the DS equation \qudfonem\ simplifies to
\eqn\truncqudf{\eqalign{
 \ (\tilde{D}^{-1})_{\l\m}(k) 
= & \ \ (k^2 \d_{\k\l} - k_\k k_\l) + a^{-1} k_\k k_\l    \cr
    & + \hbar N \  (2!)^{-1} (2\pi)^{-4}
  \int dk_1 \ 
\tilde{K}^{(2)}_{\l\l_1\l_2}(k;-k_1,k_1-k),   \cr
    & \ \ \ \ \ \ \ \ \ \ \ \ \ \ \ \ \ \ \ \times
  \tilde{D}{_{\l_1\m_1}}(k_1) 
  \ \tilde{D}{_{\l_2\m_2}}(k-k_1)  
  \ \tilde{\Gamma}_{\m_1\m_2\m}^{(3)}(k_1,k-k_1,-k).  
}} 

	We convert this into a DS equation for the invariant propagator
functions $T(k^2)$ and $L(k^2)$.  The gluon propagator is given by
	\eqn\asprop{\eqalign{
\tilde{D}_{\l\m}(k) = { {P_{\l\m}^T(k)} \over {T(k^2)} } 
         + a \ { {P_{\l\m}^L(k)} \over {L(k^2)} }.
}}
To get the DS equation for $T(k^2)$, we apply projectors 
$P_{\k,\n}^T(k)$ to both
free indices of \truncqudf, and obtain
$[P^T(k) \tilde{D}^{-1}(k)P^T(k)]_{\l\m} = T(k^2)P^T_{\l\m}(k)$ 
on the left hand side.  We take the trace on Lorentz indices in $d$
space-time dimensions, and use $P^T_{\l\l}(k^2) = d-1$, to obtain the DS
equation for $T(k^2)$, 
\eqn\dst{\eqalign{
 T(k^2) 
= \  k^2 + 
    {  {  \hbar N } \over  {  2 (d-1) (2\pi)^d  }  }
  \int d^dk_1 \ [I^{T,TT}(k_1,k) + 2 I^{T,TL}(k_1,k) + I^{T,LL}(k_1,k)],  
}} 
where 
\eqn\ittt{\eqalign{
I^{T,TT}(k_1,k) =  
{   {  \ {\tilde{K}^{(2)}}
         {_{\l \l_1 \l_2}^{T T \ T}}(k,-k_1,-k_2)     
  \ {\tilde{\Gamma}^{(3)}}
       {_{\l_1 \l_2 \l}^{T \ T \ T}}(k_1,k_2,-k)  }
       \over
      { T(k_1^2) \ T(k_2^2) } }    
}}
\eqn\ittl{\eqalign{
I^{T,TL}(k_1,k)  =  a \     
{  {  {\tilde{K}^{(2)}}{_{\l\l_1\l_2}^{\rm T T \ L}}(k;-k_1,-k_2)  
   \ {\tilde{\Gamma}^{(3)}}{_{\l_1 \l_2 \l}^{\rm T \ L \ T}}(k_1,k_2,-k) }
					\over
     {  T(k_1^2) \ L(k_2^2)  }  }
}}
\eqn\itll{\eqalign{
I^{T,LL}(k_1,k)  = a^2 \     
{  {  {\tilde{K}^{(2)}}{_{\l\l_1\l_2}^{\rm T L \ L}}(k;-k_1,-k_2)
   \ {\tilde{\Gamma}^{(3)}}{_{\l_1 \l_2 \l}^{\rm L \ L \ T}}(k_1,k_2,-k) }
														\over
  {  L(k_1^2) \ L(k_2^2)  }  },
}}
$k_2 = k - k_1$, and the transverse and longitudinal projections are defined in
sec.~7.

	Similarly, to get the DS equation for $L(k^2)$, we apply projectors 
$P_{\k,\n}^L(k)$ to both free indices of \truncqudf, and obtain
$[P^L(k) \tilde{D}^{-1}(k)P^L(k)]_{\l\m} = a^{-1}L(k^2)P^L_{\l\m}(k)$ 
on the left hand side.  We take the trace on Lorentz indices in $d$
space-time dimensions, and use $P^L_{\l\l}(k^2) = 1$, to obtain the DS
equation for $L(k^2)$, 
\eqn\dsl{\eqalign{
 a^{-1}L(k^2) 
= \  a^{-1}k^2 + 
    {  {  \hbar N } \over  {  2 (2\pi)^d  }  }
  \int d^dk_1 \ [I^{L,TT}(k_1,k) + 2 I^{L,TL}(k_1,k) + I^{L,LL}(k_1,k)],  
}} 
where 
\eqn\iltt{\eqalign{
I^{L,TT}(k_1,k) =  
{   {  \ {\tilde{K}^{(2)}}
         {_{\l \l_1 \l_2}^{L T \ T}}(k,-k_1,-k_2)     
  \ {\tilde{\Gamma}^{(3)}}
       {_{\l_1 \l_2 \l}^{T \ T \ L}}(k_1,k_2,-k)  }
       \over
      { T(k_1^2) \ T(k_2^2) } }    
}}
\eqn\iltl{\eqalign{
I^{L,TL}(k_1,k)  =  a \     
{  {  {\tilde{K}^{(2)}}{_{\l\l_1\l_2}^{\rm L T \ L}}(k;-k_1,-k_2)  
    \ {\tilde{\Gamma}^{(3)}}{_{\l_1 \l_2 \l}^{\rm T \ L \ L}}(k_1,k_2,-k) }
					\over
     {  T(k_1^2) \ L(k_2^2)  }  }
}}
\eqn\illl{\eqalign{
I^{L,LL}(k_1,k)  = a^2 \     
{  {  {\tilde{K}^{(2)}}{_{\l\l_1\l_2}^{\rm L L \ L}}(k;-k_1,-k_2)
   \ {\tilde{\Gamma}^{(3)}}{_{\l_1 \l_2 \l}^{\rm L \ L \ L}}(k_1,k_2,-k) }
														\over
  {  L(k_1^2) \ L(k_2^2)  }  }.
}}

	The vertex ${\tilde{K}^{(2)}}$ is given in \tktwo. To complete the truncation
scheme and obtain closed equations for $T(k^2)$ and $L(k^2)$, we need an
approximation for the vertex $\tilde{\Gamma}^{(3)}$.  We will approximate 
$\tilde{\Gamma}^{(3)}$ by its value to zero-order in $\hbar$.  This vertex is
expressed linearly in terms of $\tilde{Q}^{(2)}$ by the exact formulas of
sec.~7, which may be written
$\tilde{\Gamma}^{(3)} = M\tilde{Q}^{(2)}$, where $M = M(\tilde{D})$.
At tree level, $\tilde{Q}^{(2)}$ is given by
\eqn\exptktwo{\eqalign{  
\tilde{Q}^{(2)} & = \tilde{K}^{(2)}  \cr
		& = - \ \tilde{S}_{\rm YM}^{(3)} + a^{-1}\tilde{K}_{\rm gt}^{(2)}
}}
where we have used \qudfloopa\ and \tktwo.  Each of these terms contributes
additively to $\tilde{\Gamma}^{(3)} = M \tilde{Q}^{(2)}$.  
Moreover 
$\tilde{S}_{\rm YM}^{(3)}$, is symmetric in all its arguments.  As a
result, it contributes unchanged to $\tilde{\Gamma}^{(3)}$, as one sees from
\hjthree, and we have
\eqn\truncgthr{\eqalign{  
\tilde{\Gamma}^{(3)}_{\m_1 \m_2 \m_3}(k_1,k_2,k_3) 
   =  & \ {\tilde{S}_{\rm YM}^{(3)}}{_{\m_1 \m_2 \m_3}}(k_1,k_2,k_3)  
    + {\tilde{\Gamma}_{\rm gt}^{(3)}}{_{\m_1 \m_2 \m_3}}(k_1,k_2,k_3),   
}} 
where 
$\tilde{\Gamma}_{\rm gt}^{(3)} = M\tilde{K}_{\rm gt}^{(2)}$ 
is obtained from \defsofq\ -- \solvegthra\ by the substitutions 
\eqn\subst{\eqalign{
\tilde{Q}_{\m_1\m_2\m_3}^{(2)}(k_1;k_2,k_3) & \to
   a^{-1}\tilde{K}_{{\rm gt} \ \m_1\m_2\m_3}^{(2)}(k_1;k_2,k_3)  \cr 
\tilde{\Gamma}_{\m_1\m_2\m_3}^{(3)}(k_1,k_2,k_3) & \to 
      \tilde{\Gamma}_{{\rm gt} \ \m_1\m_2\m_3}^{(3)}(k_1,k_2,k_3).
}}
Finally, to obtain
$\tilde{\Gamma}_{\rm gt}^{(3)}$ to zero-order in $\hbar$, we substitute the
tree-level propagators 
\eqn\tree{\eqalign{
  T(k^2) \to k^2, \ \ \ \ \ \ \ \ \ \ L(k^2) \to k^2,
}}
into the formulas of sec.~7.  This is done in Appendix E, and gives for the
vertex
${\tilde{\Gamma}_{\rm gt}^{(3)}}$, 
\eqn\gthrgtta{\eqalign{
{\tilde{\Gamma}_{\rm gt}^{(3)}}
  {_{\m_1\m_2\m_3}^{\rm T \ T \ T}}(k_1,k_2,k_3) & = 0    \cr
{\tilde{\Gamma}_{\rm gt}^{(3)}}
  {_{\m_1\m_2\m_3}^{\rm T \ T \ L}}(k_1,k_2,k_3) 
    & = - \ i g \ 
   { {k^2_1 - k^2_2} \over {ak^2_1 + ak^2_2 + k^2_3} }
   \ (k_3)_{\m_3}   \ [P^{\rm T}(k_1)P^{\rm T}(k_2)]_{\m_1\m_2}          \cr
{\tilde{\Gamma}_{\rm gt}^{(3)}}
    {_{\m_1\m_2\m_3}^{\rm T \ L \ L}}(k_1,k_2,k_3) 
   & = - \ ia^{-1}g  \ \Big( \ 
    { {k^2_3 - ak^2_1} \over {ak^2_1 + k^2_2 + k^2_3} }
   \ (k_2)_{\m_2} \ [P^{\rm T}(k_1)P^{\rm L}(k_3)]_{\m_1\m_3} 
         - (2 \leftrightarrow 3)   \ \Big)    \cr
{\tilde{\Gamma}_{\rm gt}^{(3)}}
     {_{\m_1\m_2\m_3}^{\rm L \ L \ L}}(k_1,k_2,k_3) & = 
    - \ i a^{-1}g \ \Big( \  
 { {k^2_2 - k^2_3} \over {k^2_1 + k^2_2 + k^2_3} }
    \ (k_1)_{\m_1} \ [P^{\rm L}(k_2)P^{\rm L}(k_3)]_{\m_2\m_3} 
  + ({\rm cyclic}) \Big),
}}
valid to zero-order in $\hbar$.  Because of the denominators, the vertex
${\tilde{\Gamma}_{\rm gt}^{(3)}}$ is non-local even to this order.  Equations 
\truncgthr\ and \gthrgtta\ complete the specification of 
${\tilde{\Gamma}^{(3)}}$ that appears in the truncated DS
equations \dst\ and \dsl\ for the 2 invariant propagator functions~$T(k^2)$
and~$L(k^2)$.  

	In Faddeev-Popov theory there are, by contrast, 3 invariant propagator
functions, namely, these 2 plus the ghost propagator.  However in
Faddeev-Popov theory, the Slavnov-Taylor identity in its BRST version implies
that the gluon self-energy is transverse, so there are finally only 2
independent invariant propagator functions in Faddeev-Popov theory also,
namely, the transverse part of the inverse gluon propagator and the ghost
propagator.\foot{In practice the truncated DS equations in Faddeev-Popov
theory violate the Slavnov-Taylor identities to some extent.}   In the present
theory, the longitudinal part of the gluon propagator replaces the ghost
propagator as the second invariant propagator function.  There is no
BRST symmetry in the present theory, but it possesses a Ward identity, 
derived in Appendix~C, that expresses the effect of a gauge transformation and
constrains the form of divergences.

\newsec{Landau gauge limit} 

	We now specialize to the Landau gauge limit $a \to 0$.  We cannot directly set
$a = 0$ in the DS equations \dst\ and \dsl\ because both vertices contain terms
of order $a^{-1}$.  With the gluon propagator given by \asprop, we take as an
Ansatz that the invariant propagator  functions $T(k^2)$ and
$L(k^2)$ remain finite in the limit $a \to 0$.  This accords with the behavior
obtained in \dznonpland\ by a Born-Oppenheimer type calculation.  At 
$a = 0$, the propagator is indeed transverse, which is the defining condition
for the Landau gauge, and $L(k^2)$ does drop out of the propagator. However the
vertices contain terms of order $a^{-1}$, and, remarkably, the
longitudinal propagator function~$L(k^2)$ does not decouple at 
$a = 0$, but remains an essential component of the dynamics!

	We next determine the $a$-dependence of the vertices asymptotically, at
small $a$.  By \tktwo, we have
$K^{(2)} = - S^{(3)}_{\rm YM} + a^{-1} K_{\rm gt}^{(2)}$, so this vertex
contains a term of order $a^0$ and a term of order $a^{-1}$.  We take the
asymptotic limit of \gthrgtta\ at small $a$, and obtain the interesting
$a$-dependence
\eqn\gthrgtas{\eqalign{
{\tilde{\Gamma}_{\rm gt}^{(3)}}
  {_{\m_1\m_2\m_3}^{\rm T \ T \ T}}(k_1,k_2,k_3) & = 0    \cr
{\tilde{\Gamma}_{\rm gt}^{(3)}}
  {_{\m_1\m_2\m_3}^{\rm T \ T \ L}}(k_1,k_2,k_3)  
    & =  \tilde{\g}_{\m_1\m_2\m_3}^{\rm T \ T \ L}(k_1,k_2,k_3)   \cr
{\tilde{\Gamma}_{\rm gt}^{(3)}}
    {_{\m_1\m_2\m_3}^{\rm T \ L \ L}}(k_1,k_2,k_3) 
   & = a^{-1} \ \tilde{\g}_{\m_1\m_2\m_3}^{\rm T \ L \ L}(k_1,k_2,k_3)    \cr
{\tilde{\Gamma}_{\rm gt}^{(3)}}
     {_{\m_1\m_2\m_3}^{\rm L \ L \ L}}(k_1,k_2,k_3) & = 
    a^{-1} \ \tilde{\g}_{\m_1\m_2\m_3}^{\rm L \ L \ L}(k_1,k_2,k_3),
}}
where
\eqn\gas{\eqalign{
\tilde{\g}_{\m_1\m_2\m_3}^{\rm T \ T \ L}(k_1,k_2,k_3)
    & = - \ i g \ 
   { {k^2_1 - k^2_2} \over {k^2_3} }
   \ (k_3)_{\m_3}   \ [P^{\rm T}(k_1)P^{\rm T}(k_2)]_{\m_1\m_2}          \cr
\tilde{\g}_{\m_1\m_2\m_3}^{\rm T \ L \ L}(k_1,k_2,k_3)
   & = - \ ig \ 
    { {k^2_3} \over {k^2_2 + k^2_3} }
   \ (k_2)_{\m_2} \ [P^{\rm T}(k_1)P^{\rm L}(k_3)]_{\m_1\m_3} 
      - \ (2 \leftrightarrow 3)     \cr
\tilde{\g}_{\m_1\m_2\m_3}^{\rm L \ L \ L}(k_1,k_2,k_3) & = 
    - \ i g  \  
 { {k^2_2 - k^2_3} \over {k^2_1 + k^2_2 + k^2_3} }
    \ (k_1)_{\m_1} \ [P^{\rm L}(k_2)P^{\rm L}(k_3)]_{\m_2\m_3} 
    + ({\rm cyclic}) 
}}
are independent of $a$.  These quantities are anti-symmetric in their three
arguments so, for example,
$\tilde{\g}_{\m_1\m_2\m_3}^{\rm T \ T \ L}(k_1,k_2,k_3)      
  =  -\tilde{\g}_{\m_1\m_3\m_2}^{\rm T \ L \ T}(k_1,k_3,k_2)$, etc.
We see that $\tilde{\Gamma}^{(3)}$ also contains a term of order
$a^0$ and a term of order $a^{-1}$.

	The	DS equation for $L(k^2)$, eq.~\dsl, is consistent with our Ansatz 
in the Landau gauge limit only if the leading term on the right is also of
order $a^{-1}$.  This is non-trivial, because both vertices contain terms of
order $a^{-1}$, so in principle terms of order $a^{-2}$ could appear on the
right hand side which would invalidate our Ansatz.  

	We now derive the DS equation for $L(k^2)$ in the Landau-gauge limit by
evaluating in succession the terms (i) $I^{\rm L, TT}$, (ii) $I^{\rm L, LL}$,
and (iii) $I^{\rm L, TL}$ that appear on the right hand side of \dsl, in the
limit $a \to 0$.

	(i) Consider eq.~\iltt\ for $I^{\rm L, TT}$.  It contains no explicit
powers of $a$. Moreover the vertex 
${\tilde{\Gamma}^{(3)}}{_{\m_1\m_2\m}^{\rm T \ T \ L}}(k_1,k_2,-k)$, 
given in \gthrgtas, is of order~$a^0$.  Thus the inconsistency of a term of
order $a^{-2}$ is avoided, and this intermediate state will give a
contribution of required order $a^{-1}$ only if the vertex,
${\tilde{K}^{(3)}}{_{\l\l_1\l_2}^{\rm L T \ T}}(k;-k_1,-k_2)$,
gives a contribution of order $a^{-1}$.  The term
$a^{-1}{\tilde{K}_{\rm gt}^{(3)}}
{_{\l\l_1\l_2}}(k;-k_1,-k_2)$
in \tktwo\ is in fact of this order.  The projected components of
$\tilde{K}_{\rm gt}^{(2)}$ are easily read off \tktwo\
by writing
$\d_{\l\m} = [P^T(k) + P^L(k)]_{\l\m}$, which gives
\eqn\projtwogt{\eqalign{  
{\tilde{K}_{\rm gt}^{(2)}}{_{\m_1 \m_2 \m_3}}(k_1;k_2,k_3) 
   & = i g \ (k_3)_{\m_3} \ \Big( \ [ \ P^T(k_1) + P^L(k_1) \ ] 
   [ \ P^T(k_2) + P^L(k_2) \ ] \ \Big)_{\m_1,\m_2}     \cr
	    & \ \ \  -  \ (2 \leftrightarrow 3) .
}}
The polarization vector $(k_3)_{\m_3}$ is purely longitudinal, as is
$(k_2)_{\m_2}$, and this implies
\eqn\vanktt{\eqalign{
{\tilde{K}_{\rm gt}^{(2)}}
{_{\l\l_1\l_2}^{\rm L \ T \ T}}(k;-k_1,-k_2)
= {\tilde{K}_{\rm gt}^{(2)}}
{_{\l\l_1\l_2}^{\rm T \ T \ T}}(k;-k_1,-k_2) = 0.
}}
Thus there is no contribution of the required order $a^{-1}$ from 
$I^{\rm L,TT}$.

	(ii) Consider eq.~\illl\ for $I^{\rm L, LL}$.  It has the coefficient $a^2$,
so there is no contribution of the required order $a^{-1}$ from 
$I^{\rm L,LL}$ either.  

	(iii) Now consider eq.~\iltl\ for $I^{\rm L, TL}$.  It has the
coefficient~$a$.  So  when each vertex is of order $a^{-1}$ there is an overall
contribution to $a^{-1} L(k^2)$ of the required order $a^{-1}$.  As a result,
the DS equation \dsl\ for $L(k^2)$ simplifies in Landau gauge to
\eqn\dslland{\eqalign{
 L(k^2)  = & \  k^2 + \hbar N \   (2\pi)^{-4} \int d^4k_1 \ \ 
    {  { {\tilde{K}_{\rm gt}^{(2)}}
{_{\l\l_1\l_2}^{\rm L \ T \ L}}(k;-k_1,-k_2)
           \ \tilde{\g}_{\l_1\l_2\l}^{\rm T \ L \ L}(k_1,k_2,-k)  }  
	          \over
     { T(k_1^2) \ L(k_2^2)  }  }.  
}} 
By \projtwogt\ and \gas, we have
\eqn\lvertices{\eqalign{
{\tilde{K}_{\rm gt}^{(2)}}{_{\l\l_1\l_2}^{\rm L \ T \ L}}(k;-k_1,-k_2)
        \ \tilde{\g}_{\l_1\l_2\l}^{\rm T \ L \ L}(k_1,k_2,-k)  
  & = - \ { {g^2 \ k_2^2} \over {k_2^2 + k^2} }
   \ [ k \cdot P^T(k_1) \cdot k + k \cdot P^T(k_1) \cdot k_2]   \cr
  & = - \ 2 \ { {g^2 \ k_2^2} \over {k_2^2 + k^2} }
   \  k \cdot P^T(k_1) \cdot k ,
}}
where we have used $k_2 = k - k_1$.  Note that a factor of the external
momentum $k$ appears at each vertex.  This corresponds to the factorization of
external ghost momentum in the Landau gauge in Faddeev-Popov theory.  This
gives the truncated DS equation for~$L(k^2)$ in Landau gauge,
\eqn\dsllanda{\eqalign{
 L(k^2) =  \  k^2 - { {  2 \hbar g^2N  } \over  { (2\pi)^4  } }
   \int d^4k_1 \ 
 {  { k_2^2 \ [ k^2 k_1^2 - (k \cdot k_1)^2 ]  }
		    \over
     {  k_1^2 \ (k_2^2 + k^2) \ T(k_1^2) \  L(k_2^2)   }  }.
}}

	The	DS equation for $T(k^2)$, eq.~\dst, is consistent with our Ansatz 
in the Landau gauge limit only if the leading term on the right is also of
order $a^0$.  This is non-trivial, because both vertices contain terms of
order $a^{-1}$, so in principle terms of order $a^{-1}$ and
$a^{-2}$ could appear on the right hand side which would invalidate our
Ansatz.  

	We now derive the DS equation for $T(k^2)$ in the Landau-gauge limit by
evaluating in succession the terms (i) $I^{\rm T, TT}$, (ii) $I^{\rm T, TL}$,
and (iii) $I^{\rm T, LL}$ that appear on the right hand side of \dst, in the
limit $a \to 0$.

	(i) Consider eq.~\ittt\ for $I^{\rm T, TT}$.  It contains no explicit powers
of $a$.  By \gthrgtas\ and \vanktt, the vertices from 
${\tilde{K}_{\rm gt}^{(2)}}{_{\l\l_1\l_2}^{\rm T T \ T}}(k;-k_1,-k_2)$
and
${\tilde{\Gamma}_{\rm gt}^{(3)}}{_{\m_1\m_2\m}^{\rm T \ T \ T}}(k_1,k_2,-k)$
vanish, and we obtain from \tktwo\ and \truncgthr,
\eqn\exptkttt{\eqalign{  
{\tilde{K}^{(2)}}{_{\l \l_1 \l_2}^{T \ T \ T}}(k;-k_1,-k_2) 
  =    - \ {\tilde{S}_{\rm YM}^{(3)}}
         {_{\l \l_1 \l_2}^{T T \ T}}(k,-k_1,-k_2)
}}
and
\eqn\truncgthra{\eqalign{  
{\tilde{\Gamma}^{(3)}}{_{\m_1 \m_2 \m}^{T \ T \ T}}(k_1,k_2,-k) 
   =  & \ {\tilde{S}_{\rm YM}^{(3)}}
       {_{\m_1 \m_2 \m}^{T \ T \ T}}(k_1,k_2,-k).   
}}
This gives a contribution of the required order.  

	(ii) Next consider eq.~\ittl\ for $I^{\rm T, TL}$.  It has coefficient~$a$.  
By \truncgthr\ and \gthrgtas, we have
\eqn\truncgthrb{\eqalign{  
{\tilde{\Gamma}^{(3)}}{_{\m_1\m_2\m}^{\rm T \ L \ T}}(k_1,k_2,-k) 
= {\tilde{S}^{(3)}}{_{\m_1\m_2\m}^{\rm T \ L \ T}}(k_1,k_2,-k)
+ {\tilde{\g}^{(3)}}{_{\m_1\m_2\m}^{\rm T \ L \ T}}(k_1,k_2,-k),
}}
which is of order $a^0$.  Thus only that part of the vertex
${\tilde{K}^{(2)}}{_{\l\l_1\l_2}^{\rm T T \ L}}(k;-k_1,-k_2)$ that is of 
order $a^{-1}$ will contribute to the desired order $a^0$. However from 
\gas\ for 
${\tilde{\g}^{(3)}}{_{\m_1\m_2\m_3}^{\rm T \ T \ L}}(k_1,k_2,k_3)$,
and by evaluation of 
${\tilde{S}^{(3)}}{_{\m_1\m_2\m}^{\rm T \ L \ T}}(k_1,k_2,-k)$,
one obtains
\eqn\truncgthrc{\eqalign{  
{\tilde{\Gamma}^{(3)}}{_{\m_1\m_2\m}^{\rm T \ L \ T}}(k_1,k_2,-k) 
= 0.
}}

	(iii) Finally consider eq.~\itll\ for $I^{\rm T, LL}$.  It has
coefficient~$a^2$.  To get a net contribution of order $a^0$, we make the
substitutions of the relevant projected vertices,
\eqn\substb{\eqalign{ 
{\tilde{K}^{(2)}}{_{\l\l_1\l_2}^{\rm T L \ L}}(k;-k_1,-k_2) & \to
  a^{-1}{\tilde{K}_{\rm gt}^{(2)}}{_{\l\l_1\l_2}^{\rm T L \ L}}(k;-k_1,-k_2)  
                  \cr 
{\tilde{\Gamma}^{(3)}}{_{\m_1\m_2\m}^{\rm L \ L \ T}}(k_1,k_2,-k) & \to 
 a^{-1}{\tilde{\g}^{(3)}}{_{\m_1\m_2\m}^{\rm L \ L \ T}}(k_1,k_2,-k),
}}
by \tktwo, \truncgthr, and \gthrgtas.  Again the conclusion is
consistent with our Ansatz.  

	We have now found all the terms on the right hand
side of \dst\ that contribute to $T(k^2)$ in the Landau gauge limit, namely,
\eqn\ittta{\eqalign{
I^{T,TT}(k_1,k) =  
{   { - \ {\tilde{S}_{\rm YM}^{(3)}}
         {_{\l \l_1 \l_2}^{T T \ T}}(k,-k_1,-k_2)     
  \ {\tilde{S}_{\rm YM}^{(3)}}
       {_{\l_1 \l_2 \l}^{T \ T \ T}}(k_1,k_2,-k)  }
       \over
      { T(k_1^2) \ T(k_2^2) } }    
}}
\eqn\ittla{\eqalign{
I^{T,TL}(k_1,k)  = 0
}}
\eqn\itlla{\eqalign{
I^{T,LL}(k_1,k)  =        
{  {  {\tilde{K}_{\rm gt}^{(2)}}{_{\l\l_1\l_2}^{\rm T L \ L}}(k;-k_1,-k_2)
   \ {\tilde{\g}^{(3)}}{_{\l_1 \l_2 \l}^{\rm L \ L \ T}}(k_1,k_2,-k) }
														\over
  {  L(k_1^2) \ L(k_2^2)  }  }.
}}
and $k_2 = k - k_1$.  The last term is given explicitly by
\eqn\evitll{\eqalign{
I^{T,LL}(k_1,k) 
    = - \ g^2 \ {  {     
       (k_1^2 + k_2^2) \ k_1 \cdot P^T(k) \cdot k_2
     - k_2^2 \ k_1 \cdot P^T(k) \cdot k_1
     - k_1^2 \ k_2 \cdot P^T(k) \cdot k_2   }  
                    \over
       {   (k_1^2 + k_2^2) \ L(k_1^2) \ L(k_2^2)   }  }
}}
\eqn\evitlla{\eqalign{
I^{T,LL}(k_1,k) 
     = \ 2 \ g^2 
     {  { \ k_1^2 \ k^2 - (k_1 \cdot k)^2  }
      \over
      {  k^2 \ L(k_1^2) \ L(k_2^2)  }  },
}}
where we have used
$k_1 \cdot P^T(k) \cdot k_1 = k_2 \cdot P^T(k) \cdot k_2
= - k_1 \cdot P^T(k) \cdot k_2$.
The non-local denominator $(k_1^2 + k_2^2)^{-1}$ has cancelled out of this
expression.

	We have obtained a consistent Landau gauge limit of the truncated DS
equations for the invariant propagator functions $T(k^2)$ and $L(k^2)$.
As asserted, the invariant longitudinal propagator function
$L(k^2)$ does not decouple in this limit.  The reader will have noticed a
striking similarity to the corresponding equations in Faddeev-Popov theory,
with the longitudinal propagator replacing the ghost propagator.

\newsec{Infrared critical exponents}

	We shall solve the the DS equations \dst\ and \dsllanda\ for the
asymptotic forms of $T(k^2)$ and $L(k^2)$ in Landau gauge at low momentum.  We
suppose that at asymptotically small
$k$, they obey simple power laws, 
\eqn\critexp{\eqalign{
T(k^2) \sim C_T \ (k^2)^{1 + \a_T}   \cr
L(k^2) \sim C_L \ (k^2)^{1 + \a_L},
}}
where $\a_T$ and $\a_L$ are infrared critical exponents whose value we wish to
determine.  Canonical dimension corresponds to $\a_T = \a_L = 0$.  As
explained in sec.~3, we know that in the Landau gauge limit, $a \to 0$, the
gauge field $A$ is constrained to be transverse $\p \cdot A = 0$, and to
lie inside the Gribov horizon, that is to say, where the Faddeev-Popov operator
is positive, $ - \p \cdot D(A) > 0$.  The transversality condition is satisfied
by our Ansatz.  As has been shown many times \gribov, \vanish, the positivity
condition strongly suppresses the low-momentum components of $\tilde{A}(k)$.
Recalling that the transverse part of the gluon propgator is given by
$D^T(k^2) = \langle |\tilde{A}(k)|^2 \rangle$, we look for a solution for
which $D^T(k^2) = 1/T(k^2)$ is suppressed at low $k$, so $T(k^2)$ is enhanced
at small $k$ compared to the canonical power 
$T(k^2) = k^2$.  This means $\a_T < 0$.  

	We now estimate the power of $k$ of the various terms in the DS equation~\dst\
for~$T(k^2)$.  The analysis is similar to the Faddeev-Popov case \smekal,
\atkinsona, \atkinsonb, \lerche, \dznonpland.  The left hand side has the
power $(k^2)^{1 + \a_T}$.  The tree-level term is $k^2$, so with $\a_T < 0$,
the tree level term is subdominant in the infrared and may be neglected.  
To evaluate the loop integral $\int d^4 k_1$, asymptotically at low external
momentum $k$ we take $k$ to be small compared to a QCD mass scale, 
$|k| << \Lambda_{\rm QCD}$, and we rescale the variable of integration
according to $k_1^\m = |k|x^\m$.  We now have a dimensionless integral in which
the QCD mass scale appears only in the very small ratio 
$|k|/\Lambda_{\rm QCD}$.  In the asymptotic infrared limit, this ratio goes
to~0, and everywhere in the integrand we use the asymptotic forms
\critexp.  This is equivalent to using the asymptotic forms \critexp\
everywhere in the original integral.  We shall see that the resulting integral
is convergent, which means that the integral is effectively cut off at
momentum $k_1 \sim k$.     

	We now estimate the contributions of the terms
$I^{T,TT}$ and $I^{T,LL}$, eqs.~\ittta\ and \itlla,
to the right hand side of the DS equation \dst,
by simply counting powers of $k$ and $k_1$.  One finds that,
after integration $\int d^4k_1$, these terms are of order 
$(k^2)^{1 - 2\a_T}$ and $(k^2)^{1 - 2\a_L}$
respectively, while the left-hand side is of order 
$(k^2)^{1 + \a_T}$, with $\a_T < 0$.  The powers match on both sides only if 
$\a_L > 0$.  In this case, $I^{T,LL}$ is the dominant term on the right,
and by equating powers of $k$, one obtains 
\eqn\powers{\eqalign{
\a_T = - \ 2\a_L, 
}}
and $\a_L > 0$. 
We retain only the dominant term $I^{T,LL}$ on the right in \dst, which
simplifies, for arbitrary space-time dimension~$d$, to
\eqn\dstas{\eqalign{
 C_T \ (k^2)^{1 + \a_T} 
 =  {  { \hbar g^2 N  } \over { (d-1) \ C_L^{2} \ (2\pi)^d \ k^2}  }
  \int d^dk_1    
 \  { { \  k_1^2 \ k^2 - (k_1 \cdot k)^2 \  } \over 
     \ { \  (k_1^2)^{1 + \a_L} \ (k_2^2)^{1 + \a_L} \ } }  ,
}}
where $k_2 = k - k_1$, and we have used \evitlla.  This agrees with
eq.~(6.14) of \dznonpland\ in Faddeev-Popov theory.  We write it as
\eqn\dstasit{\eqalign{
{ {C_T C_L^2} \over \hbar g^2 N}  = I_T,
}}
where $I_T$ is evaluated in Appendix~F.  We have generalized to
arbitrary space-time dimenion $d$, and we take $d$ in the range $2 < d \leq
4$.  By equating powers of $k$ for arbitrary $d$, we find that the critical
exponents are related by 
\eqn\relcritexp{\eqalign{
\a_T + 2\a_L = - \ (4-d)/2.  
}}
The last integral is ultraviolet convergent provided that $\a_L > (d-2)/4$,
which corresponds to $\a_T < -1$.  For $d = 4$, we obtain $\a_L > 1/2$ as the
condition for convergence of the integral.

	Now consider the DS equation \dsllanda\ for $L(k^2)$ in the infrared
asymptotic limit, 
\eqn\dsllandas{\eqalign{
 C_L(k^2)^{1 + \a_L} =  k^2 \ 
  -  \ {  { 2 \hbar g^2N  } \over { C_T C_L \ (2\pi)^4  }  } 
   \int d^4k_1 \ { {   k^2 k_1^2 - (k \cdot k_1)^2 } \over
     { \ (k_1^2)^{2 + \a_T}  \ (k_2^2)^{\a_L} 
      \ (k_2^2 + k^2) } } ,   
}} 
for $d = 4$.  By power counting, the integral on the right has
the power $(k^2)^{1 - \a_T - \a_L}$.  This agrees with the power on the left,
provided $\a_T = 2\a_L$, which is identical to the previous equation.
However we have also previously found $\a_L > 0$.  In this case, the tree level
term $k^2$ is dominant in the infrared, and the equation appears
inconsistent.  However the degree of divergence of the integral is $2\a_L$, so
the integral diverges for $\a_L > 0$, and a subtraction is required.  The
integral contains an explicit factor of
$k^2$, and the divergence is of the form $B k^2$, where $B$ is an infinite
constant.  We subtract the integrand at $k = 0$, which makes the integral
vanish more rapidly than $k^2$, and add $b k^2$ on the right, where $b$ is an
arbitrary finite constant.  The dominant terms are now the tree level term
$k^2$ and $bk^2$.  For the equation to be consistent, the subtraction term must
precisely cancel the tree-level term, so $b = -1$.  This gives  
\eqn\dsllandasa{\eqalign{
 C_L(k^2)^{1 + \a_L} 
= {  { 2 \hbar g^2N  } \over { C_T C_L \ (2\pi)^4  }  }
    \int d^4k_1 & \ 
		{ { k^2 k_1^2 - (k \cdot k_1)^2 } \over {(k_1^2)^{2 + \a_T} } }       
 \Big(   { {  \ 1 } \over { \ (k_1^2)^{\a_L} 
      \ k_1^2 } }
     - { {  \ 1 } \over {  \ (k_2^2)^{\a_L} 
      \ (k_2^2 + k^2) } } \Big) .   
}} 
This integral is also convergent in the infrared for $\a_T = - 2\a_L < 0$.
The right hand side now vanishes more rapidly than $k^2$.  This conclusion,
agrees with the ``horizon condition" \horizcon, and with the
confinement criterion of Kugo and Ojima in the BRST
framework~\kugoj,~\nakanoj.  Conversely we could have imposed the horizon
condition on the DS equation for $L(k^2)$, and derived the suppression of the
transverse propagator $1/T(k^2)$ at low momentum. 

  The subtracted expression on the right is most simply evaluated by
continuing in space-time dimension $d$.  In this case one can ignore the
subtraction term, and evaluate the unsubtracted integral
with dimensional regularization for $d < 4$, and continue the resulting
expression to $d = 4$,  
\eqn\dsllandassub{\eqalign{
 C_L(k^2)^{1 + \a_L} 
  = - \ {  { 2 \hbar g^2N  } \over { C_T C_L \ (2\pi)^d  }  }
   \int d^dk_1 \ { {  k^2 k_1^2 - (k \cdot k_1)^2 } \over
     { \ (k_1^2)^{2 + \a_T}  \ (k_2^2)^{\a_L} 
      \ (k_2^2 + k^2) } }.  
}}
The denominator $k_2^2 + k^2$ results from the non-local expression for the
vertex.  One obtains the corresponding equation for the ghost
propagator in Faddeev-Popov theory, eq.~(6.15) of \dznonpland,  from this
equation by the substitution 
${ {2} \over {k_2^2 + k^2} } \to { {1} \over {k_2^2} }$.

	By equating powers of $k$ for general space-time dimension $d$, 
one again gets \relcritexp, and we see that the DS equations for the
transverse and longitudinal parts are consistent.  The degree of divergence of
this integral is $2\a_L$, and after one subtraction its degree of divergence is
$2\a_L - 2$, so the subtracted integral is convergent provided that $\a_L < 1$,
or equivalently that $\a_T > - \ 2 - (4-d)/2$.  From this and our previous
bound, we conclude that for $d = 4$, this subtracted integral and
\dstas\ are both finite provided that $\a_L$ is in the range $1/2 < \a_L < 1$,
or equivalently that $\a_T$ is in the range $-2 < \a_T < -1$.  We write the
preceding equation as
\eqn\dsllandasil{\eqalign{
{ {C_T C_L^2} \over \hbar g^2 N}  = I_L,
}}
where $I_L$ is evaluated in Appendix~F.

	Upon comparison with \dstasit, we obtain
\eqn\evdsa{\eqalign{
I_T(\a_L) = I_L(\a_L)
}}
which determines the critical exponent $\a_L$.  From Appendix~F, this gives,
for
$d = 4$,
\eqn\evds{\eqalign{
     { { \Gamma^2(2 - \a_L) \ \Gamma(2\a_L - 1)} \over
      { \Gamma^2(1 + \a_L) \ \Gamma(4 - 2\a_L)} }  
 =   { { 3 \ (- \a_L^2 + 2\a_L + 2) \ \Gamma^2(1 - \a_L) \ \Gamma(2\a_L + 1)}
   \over { \a_L \ \Gamma(\a_L + 2) \Gamma(\a_L + 3) \Gamma(2 - 2\a_L) } }.
}}
Both expressions are finite and positive, as they should be, for $\a_L$ in
the interval $1/2 < \a_L < 1$.  Moreover at $\a_L = 1/2$, the left-hand side
diverges whereas the right is finite.  On the other hand at $\a_L = 1$,
the left-hand side is finite whereas the right diverges.  Consequently
there is at least one root in the interval $1/2 < \a_L < 1$.  After cancelling
$\Gamma$-functions, the last two equations give the quartic equation
\eqn\poly{\eqalign{
49\a_L^4 - 189\a_L^3 + 133\a_L^2 + 117\a_L - 74 = 0.
}}
From a numerical investigation it appears that there is only one root in the
interval $1/2 < \a_L < 1$, with the value
\eqn\critexp{\eqalign{
\a_L & \approx 0.5214602698  \cr
\a_T & = - 2\a_L \approx - 1.04292054.
}}

\newsec{Conclusion}

	We derived time-independent stochastic quantization from the principle
of gauge equivalence which states that probability distributions that give the
same expectation values for all gauge-invariant observables are physically
indistiguishable.  This quantization is expressed by an equation for the
Euclidean proabability distribution $P(A)$ that is of time-independent
Fokker-Planck form, with a corresponding equation for the Minkowski case.  
By making several changes of variable, we transformed this equation into an
equation of DS type, suitable for non-perturbative calculations. The
most novel of these changes of variable is accomplished when the equation for
the quantum effective action
$\Gamma$ is exchanged for an equation for the quantum effective drift force
${\cal Q}_x$.  We then adopted a truncation scheme and obtained a consistent
Landau gauge limit, $a \to 0$, and found, remarkably, that the longitudinal
propagator function $L(k^2)$ that appears in the longitudinal
part of the gluon propagator $D^L = a / L(k^2)$, does not decouple in the
$a \to 0$ limit, but plays a role similar to the ghost in Faddeev-Popov
theory.  

	We calculated the infrared critical exponents that characterize 
the asymptotic form at low momentum of the transverse and longitudinal
components of the gluon propagator in Landau gauge, 
$D^T \sim 1/(k^2)^{1 + \a_T}$, and 
$D^L \sim a/(k^2)^{1 + \a_L}$, 
and obtained the values
$\a_L \approx 0.5214602698$ and $\a_T = - 2\a_L \approx - 1.04292054$.
In the Landau-gauge limit $a \to 0$ only the transverse part survives.  As
a function of $k$, it vanishes at $k = 0$, albeit rather weakly,  
$D^T \sim (k^2)^{-1-\a_T} \sim (k^2)^{0.043}$.  On the
other hand, the longitudinal part of the propagator, is long range,
$D^L \sim a / (k^2)^{1.521}$.  Qualitatively similar
values have been obtained recently 
for the infrared critical exponents of the gluon and ghost propagators in
Landau gauge from the DS equation in Faddeev-Popov
theory, using a variety of approximations for the vertex,
\smekal, \atkinsona, \atkinsonb, \lerche\ and \dznonpland, in particular,
\lerche\ and \dznonpland,
$\a_T = - 2a_G = -1.1906$, and $a_G = 0.595353$ respectively.
As we have argued recently \dznonpland, these calculations in Faddeev-Popov
theory should be interpreted as including a cut-off at the Gribov horizon.
This makes them similar in spirit to the present calculation for which, as
shown in sec.~3, the probability also gets concentrated inside the
Gribov horizon in the Landau gauge limit $a \to 0$.  Reassuringly, the
solutions of the DS equation in Faddeev-Popov theory and in the present
time-independent stochastic method are in satisfactory agreement.

	We comment briefly on the physical significance of our results.  
(i)~We have avoided gauge fixing and instead derived the equation of
time-independent stochastic quantization from the principle of {\it gauge
equivalence}, thereby overcoming the Gribov critique.  
Since we do not gauge {\it fix}, we do not brutally eliminate ``unphysical"
variables and keep only ``physical" degrees of freedom, which would violate
Singer's theorem \singer.  Instead, we gently {\it tame} the gauge degrees of
freedom by exploiting the principle of gauge equivalence.   (ii) We
derived a set of equations of DS type  that was solved approximately but
non-perturbatively in Landau gauge asymptotically at low momentum.  (iii) The
values we obtained for the infra-red critical exponents of the gluon
propagator in Landau gauge are in satisfactory agreement with corresponding
values in Faddeev-Popov theory, and also with numerical simulations.  (iv)~A
striking result of this investigation is that the invariant longitudinal
propagator function
$L(k^2)$ does not decouple in stochastic quantization, even though the
longitudinal part of the gluon propagator 
$D^L = a/L(k^2)$ vanishes with the gauge
parameter $a$ in the Landau gauge limit $a \to 0$. Indeed, because some
vertices are of order $a^{-1}$, transverse gluons exchange longitudinal
gluons as virtual particles, with an amplitude that remains finite in the
limit $a \to 0$. Thus, while ghosts are absent in time-independent stochastic
quantization, they are replaced dynamically by the longitudinal part of the
gluon propagator in the Landau gauge limit.  In fact, the DS equations \dst\
and \dsllanda\ for
$T(k^2)$ and $L(k^2)$ bear a remarkable similarity to the DS equations for the
gluon and ghost propagators $D$ and $G$ in Landau gauge in Faddeev-Popov
theory, with the correspondences $D \leftrightarrow 1/T$ and $G
\leftrightarrow 1/L$.  In both cases, it is the ghost loop or
longitudinal-gluon loop that gives the dominant contribution to the
transverse-gluon inverse propagator in the infrared region, and causes
suppression of the would-be physical, transverse gluon propagator at $k = 0$,
a signal that the gluon has left the physical spectrum.  
	
	To conclude, we mention some challenging open problems.  (i)~The possibility
of comparison with numerical simulations is an essential and  promising aspect
of the present situation.  Any DS calculation involves a truncation which
remains an uncontrolled approximation, without further investigation.  It may
be controlled by varying the vertex function~\lerche, or by extending the
calculation to include the vertex self-consistently.  Fortunately, comparison
with numerical simulation provides an independent control.  In this regard we
note that the stochastic quantization used here may be and, in fact, has been
effected on the lattice in numerical simulations by Nakamura and collaborators
\nakamuraa,\nakamurab, \nakamurac, \nakamurad, \nakamurae.  A direct
comparison with this data would require a solution of the DS equations for
finite gauge parameter $a$, or extrapolation of lattice data to $a = 0$. 
Naturally, a comparison of numerical results with asymptotic infrared
calculations also requires control of finite-volume lattice artifacts. 
(ii)~Conversely the results of the DS calculations suggest new numerical
calculations.  In particular our prediction that, for small values of the
gauge parameter $a$, the longitudinal part of the gluon propagator is long
range,  should be tested numerically.  (iii)~The present scheme is not based on
a local action, but rather on the DS equations of time-independent stochastic
quantization.  Renormalizability follows from the indirect argument that
correlators of the 4-dimensional time-independent formulation used here
coincide with the equal-time correlators of a local, 5-dimensional theory
whose renormalizability has been established~\bgz.  The renormalization
constants of the 5-dimensional theory were calculated some time ago at the
one-loop level, and were found to yield the usual
$\beta$-function \munoz, \okano, \halperne.  However a direct proof of
renormalizability in the present  time-independent formulation remains a
challenge.  The Ward identity, derived in Appendix~C, is a first step.  
(iv)~One should extend the solution obtained here for the
asymptotic infrared region to finite momentum~$k$.  (v)~The Landau gauge is a
singular limit, $a \to 0$, of the DS equations for finite gauge parameter
$a$.  It would be valuable to also solve the DS equations for arbitrary,
finite~$a$.  (vi)~One should extend the solution of the DS equations to
include quarks.  (vii)~As we have explained, our results are intuitively
transparent and lend themselves to a simple confinement scenario
in which the would-be-physical transverse gluon leaves the physical spectrum. 
However it is clear that our discussion of confinement remains at the level of
a scenario because we have dealt here only with the gluon propagator which is
a gauge-dependent quantity.  This is only a first step in a program,
some of whose elements have just been indicated.  Clearly the goal is to
calculate gauge-invariant quantities.  Gauge-invariant states, the hadrons,
appear as intermediate states in gluon-gluon and quark-anti-quark scattering
amplitudes.  One must extend to this sector the solution of the DS equations
obtained here, and of the Bethe-Salpeter equations that follow from them.

\vskip .5cm
{\centerline{\bf Acknowledgments}}

It is a pleasure to thank Reinhard Alkofer, Laurent Baulieu, Richard Brandt,
Christian Fischer, Stanislaw Glazek, Martin Halpern, Alexander Rutenburg, Alan
Sokal, and Lorenz von Smekal for valuable discussions and correspondence. 
This research was partially supported by the National Science Foundation under
grant PHY-0099393.

\appendix A{Time-independent Fokker-Planck equation for quarks}

	We extend to quarks the derivation of time-independent stochastic quantization
from the principle of gauge equivalence.  Following the method of sec.~3, we
seek a weight function $P = P(A, \psi, \bar{\psi})$ that depends on the gluon
and quark and anti-quark fields.  We wish to establish a class of
gauge-equivalent normalized distributions that includes the formal
gauge-invariant weight 
$P = N \exp( - S)$ as a limiting case.
Here
\eqn\action{\eqalign{
S \equiv S_{\rm YM} + \int d^4x \ \bar{\psi} (m + \not\!\!{D})\psi,
}}  
is the Euclidean action of gluons and quarks, 
$\not\!\!\!{D} \equiv \g_\m D_\m = \g_\m(\p_\m + gA_\m^a t^a)$, 
where $\{ \g_\m, \g_\n \} = 2\d_{\m\n}$,
and the $t^a$
are the quark representation of the Lie algebra of the structure group, 
$[t^a, t^b] = f^{abc} t^c$.  We take $P$ to be the solution of 
$H_{\rm FP} P = 0$, where we now specify the extended Fokker-Planck
hamiltonian.

	As in  sec.~3, we take $H_{\rm FP}$ to be of the
form
\eqn\extfph{\eqalign{
H_{\rm FP} & = H_{\rm inv} - (v, G)^{\dag}   \cr
 & = H_{\rm inv} + 
   \int d^4x \ \Big( \ { {\d } \over {\d A_\m^a} } D_\m^{ac}  
  - g{ {\d } \over {\d \psi} } t^c\psi  
   +  g{ {\d } \over {\d \bar{\psi}} } (\bar{\psi}t^c) \ \Big) \ v^c, 
}}
where $H_{\rm inv}$ is a gauge-invariant operator, specified below, that has
$\exp(-S)$ as null vector, $H_{\rm inv}\exp(-S) = 0$, and the Grassmannian
deriveratives are left derivatives.  Here
\eqn\glgauge{\eqalign{
G^a(x) = - \ D_\m^{ac} { {\d } \over {\d A_\m^c(x)} }
   - \ g(t^a\psi){ {\d } \over {\d \psi(x)} }
  + \ g(\bar{\psi}t^a){ {\d } \over {\d \bar{\psi}(x)} }
}} 
is the generator of local gauge transformations that satisfies
\eqn\gaugetr{\eqalign{
[G^a(x), A_\m^b(y)] & = - D_\m^{ab} \ \d(x-y)   \cr
 [G^a(x), \psi(y)] & = - \ g \ t^a \psi(x) \ \d(x-y)   \cr
 [G^a(x), \bar{\psi}(y)] & = g \ \bar{\psi}(x) t^a \ \d(x-y) \cr
	[G^a(x),G^b(y)] & = \d(x-y) \ gf^{abc} G^c(x)  \cr
  [G^a(x), H_{\rm inv}] & = 0.
}}
With gauge-invariant observables defined by the condition
$G^a(x)W = 0$, the proof of sec.~3, that the expectation-value of gauge
invariant observables $\langle W \rangle = \int dA d\psi d\bar{\psi} \ W \ P$
is independent of~$v$, applies here as well. As explained in sec.~3, we take
$v^a(x) = a^{-1} \p_\l A_\l(x)$, where $a$ is a gauge parameter.  

	 There remains to specify $H_{\rm inv}$.    We
suppose that it is a sum of gluon and quark and anti-quark hamiltonians,
\eqn\hinvsum{\eqalign{
H_{\rm inv} = H_1 + H_2 + H_3
}}
where $H_1$ is the gauge-invariant gluon hamiltonian as in sec.~3, 
\eqn\hglue{\eqalign{
H_1 = \int d^4x{ {\d } \over {\d A} } 
\Big( - { {\d } \over {\d A} } - { {\d S } \over {\d A} }\Big).
}}
For the quark and anti-quark hamiltonians we take
\eqn\hquark{\eqalign{
H_2 & = \int d^4x 
  { {\d } \over {\d \psi} } \ N_2 \ 
   \Big({ {\d } \over {\d \bar{\psi}} } 
   + { {\d S } \over {\d \bar{\psi}} }\Big)  \cr 
H_3 & = \int d^4x
 { {\d } \over {\d \bar{\psi}} } \ N_3 \ 
   \Big({ {\d } \over {\d \psi} } 
    + { {\d S} \over {\d \psi} }\Big),
}}
where $N_2$ and $N_3$ are gauge-covariant kernels with engineering dimensions
of mass.  All terms in $H_{\rm FP}$ contain a derivative on the left,  which
assures that $H_{\rm FP}$ has a null eigenvalue, for we have
$\int dA d\psi d\bar{\psi} \ H_{\rm FP} F = 0$ for any $F$.  The corresponding
right eigenvector $P$ that satisfies $H_{\rm FP}P = 0$, is the physical
distribution that we seek, that depends on the gauge parameters.
Each of the operators $H_i$ satisfies $H_i \exp(-S) =0$, so also
$H_{\rm inv}\exp(-S) = 0$.  This assures the applicability of the proof of
sec.~3, namely that {\it the normalized solutions for different $v$ are gauge
equivalent, $P_v(A) \sim P_{v'}(A)$, and include 
$N\exp(-S)$ as a limiting distribution.}  

	We have obtained this result without any assumptions about the kernels
$N_2$ and $N_3$, apart from gauge covariance (and regularity).  
This would not be consistent unless {\it the normalized solutions for
different choices of the kernels give gauge-equivalent distributions 
$P_N(A) \sim P_{N'}(A)$ or, in other words, the parameters that specify
$N_2$ are and $N_3$ are gauge parameters.}  We prove this directly.

Let 
$\d N_2$ be an infinitesimal variation of $N_2$.  It induces a corresponding
change in $H_2$
\eqn\delhtwo{\eqalign{
\d H_2  = \int d^4x 
  { {\d } \over {\d \psi} } \ \d N_2 \ 
   \Big({ {\d } \over {\d \bar{\psi}} } 
   + { {\d S } \over {\d \bar{\psi}} }\Big).
}}
The corresponding change in $P$ satisfies 
$\d H_2 P + H_{\rm FP} \d P = 0$, so
$ \d P = - \ H_{\rm FP}^{-1} \d H_2 P$.  Let $W$ be a gauge-invariant
observable.  We have
\eqn\ntwo{\eqalign{
\d \langle W \rangle 
   & = \int dA d\psi d \bar{\psi} \ \d P \ W   \cr
    & = - \int dA d\psi d \bar{\psi} \ H_{\rm FP}^{-1} \d H_2 P \ W \cr
  &  = - \ \int dA d\psi d \bar{\psi} \ P 
    \ \d H_2^{\dag} \ (H_{\rm FP}^{\dag})^{-1} \ W   \cr
   & = - \ \int dA d\psi d \bar{\psi} \ P 
    \ \d H_2^{\dag} \ (H_{\rm inv}^{\dag})^{-1} \ W   \cr
  	& = - \ \int dA d\psi d \bar{\psi} \ \d H_2 P 
     \ (H_{\rm inv}^{\dag})^{-1} \ W,
}}
where we have used
$(H_{\rm FP}^{\dag})^{-1} \ W = (H_{\rm inv}^{\dag})^{-1} \ W$
which holds for a gauge-invariant observable, as was shown in sec.~3. 
Moreover 
$\d H_2^{\dag} \ (H_{\rm inv}^{\dag})^{-1} \ W$ is gauge invariant, so the last
expression is independent of the gauge parameter $a$, as was also shown in
sec.~3, and we may evaluate it for $a \to \infty$.  We have
\eqn\delhtwo{\eqalign{
\lim_{a \to \infty} \d H_2 P =  \int d^4x 
  { {\d } \over {\d \psi} } \ \d N_2 \ 
   \Big({ {\d } \over {\d \bar{\psi}} } 
   + { {\d S } \over {\d \bar{\psi}} }\Big)\ \lim_{a \to \infty}P = 0,
}}
because $\lim_{a \to \infty} P \sim N\exp(-S)$.  Thus 
$\d \langle W \rangle $ vanishes, as asserted.
    
	The quark action satisfies 
\eqn\quarkact{\eqalign{
S_{\rm qu} = \int d^4x \ \bar{\psi}(m + \not\!\!{D})\psi
  = - \  \int d^4x \ \psi C^{-1}(m + \not\!\!{D}) C\bar{\psi},
}}
where $C$ is a
numerical matrix that acts on spinor and group indices and satisfies
$C^{-1} \g_\m C = - \g_\m^{\rm tr}$
and $C^{-1} t^a C = - (t^a)^{\rm tr}$,
so we have
\eqn\hgluea{\eqalign{
{ {\d S} \over {\d \bar{\psi}(x)} } & = (m + \not\!\!{D})\psi(x) \cr
   { {\d S} \over {\d \psi(x)} } 
  & = - \ C^{-1}(m + \not\!\!{D})C\bar{\psi}(x),
}}
where the Grassmannian derivatives are left derivatives.  
The most general expressions for $N_2$ and $N_3$ that are local,
gauge-covariant, and have dimension of mass are 
\eqn\kernels{\eqalign{
N_2 & = m_2 - b_2 \not\!\!{D}  \cr
N_3 & = - \ C^{-1} (m_3 - b_3 \not\!\!{D}) C,
}}
where the $m_i$ and $b_i$ are gauge parameters.  We expect that the kernel
ultimately appears in the denominator in loop integrals so, to improve
convergence, we should take $b_2 \neq 0$ and $b_3 \neq 0$.  For the gauge
choice to respect charge-conjugation invariance, we take $b_2 = b_3$, and 
$m_1 = m_2 = c \ m$, which gives    
\eqn\hquark{\eqalign{
H_2 & =  \int d^4x 
  { {\d } \over {\d \psi} } \ (c \ m - b\not\!\!{D}) \ 
   \Big({ {\d } \over {\d \bar{\psi}} } + (m + \not\!\!{D})\psi\Big)  \cr
H_3 & =  \int d^4x
 { {\d } \over {\d \bar{\psi}} } \ (-1)C^{-1}(c \ m - b\not\!\!{D})C \ 
   \Big({ {\d } \over {\d \psi} } 
    + (-1)C^{-1}(m + \not\!\!{D})C\bar{\psi}\Big),
}}
where $b$ and $c$ are gauge parameters.  This gauge choice also respects chiral
symmetry in the limit $m \to 0$.  One may show that the eigenvalues
of $H_2$ and $H_3$ are the eigenvalues of fermi oscillators, with frequencies
$\l_n$ that are the eigenvalues of the operator 
$(c \ m - b\ \not\!\!{D})(\ m + \not\!\!{D})$, which for  
$b = c > 0$, simplifies to
$b( m^2 - \ \not\!\!{D}^2)$.  In this case $H_2$ and $H_3$ have the unique null
eigenvector $\exp(-S)$, and all other their eigenvalues are strictly
positive, as occurs for $H_1$.  Indeed $H_1$ satisfies
\eqn\hgluea{\eqalign{
\exp(S/2)H_1\exp(-S/2) =  \int d^4x \ 
\Big({ {\d } \over {\d A_x} } - (1/2){ {\d S } \over {\d A_x} }\Big)
\Big( - { {\d } \over {\d A_x} } - (1/2){ {\d S } \over {\d A_x} }\Big),
}}
where the operator on the right is manifestly positive, with the unique
null-vector $\exp(-S/2)$.  Thus $H_1$ has the unique null-vector
$\exp(-S)$, and all its other eigenvalues are strictly positive.  However we
expect that $b$ and $c$ must be kept as independent constants when needed as
renormalization counter-terms. 

	Altogether, the total Fokker-Planck hamiltonian, including quarks, is given
by  
\eqn\totham{\eqalign{
H_{\rm FP} = \int d^4x \Big[ &
  \ { {\d } \over {\d A_\m} } 
\Big( - { {\d } \over {\d A_\m} } - { {\d S } \over {\d A_\m} } \Big)    \cr
& +  { {\d } \over {\d \psi} } \ (c \ m - b \ \not\!\!{D})
   \Big({ {\d } \over {\d \bar{\psi}} } \psi
   + { {\d S } \over {\d \bar{\psi}} } \ \Big)  \cr
   & + { {\d } \over {\d \bar{\psi}} }  \ (-1)C^{-1}(c \ m - b \ \not\!\!{D})C
   \Big({ {\d } \over {\d \psi} } 
    + { {\d S} \over {\d \psi} }  \ \Big) \   \cr
   & + a^{-1} \ \Big( \ { {\d } \over {\d A_\m^a} } D_\m^{ac}  
  - \ g \ { {\d } \over {\d \psi} } (t^c\psi)  
   +  g \ { {\d } \over {\d \bar{\psi}} } (\bar{\psi}t^c) \ \Big)
   \ \p \cdot A^c  \ \Big],
}}
where $a > 0$, $b > 0$ and $c > 0$ are gauge parameters.

\appendix B{Time-independent stochastic quantization on the lattice}

	We briefly outline how to extend time-independent stochastic quantization to
lattice gauge theory. To each link $(x, \m)$ of the lattice is associated a
variable
$U_{x,\m} \in$ SU(N).  These variables are subject to the local gauge
transformation
$U_{x,\m} \to {^gU}_{x,\m} = g_x^{-1} U_{x,\m}g_{x + \hat{\m}}$,
where $g_x \in$ SU(N) is associated to the site $x$ of the lattice.
Observables $W(U)$ are invariant under this transformation,
$W({^gU}_{x,\m}) = W(U_{x,\m})$.  Expectation values are calculated by
$\langle W \rangle = \int dU \ W(U) \ P_W(U)$, where $dU$ is the product of
Haar measure over all link variables of the lattice, and 
$P_W= N \exp(-S_W)$ is the normalized probability distribution associated to
the gauge-invariant Wilson action $S_W$.

		We shall exhibit a Fokker-Planck hamiltonian $H_{\rm FP}$ for the lattice,
such that the positive normalized solutions $P$ to $H_{\rm FP}P = 0$ are gauge
equivalent to $P_W$, $P \sim P_W$.  Let $J_{x,\m}^a$ be the Lie differential
operator associated to the group variable on the link $(x,\m)$, that satisfies
the Lie algebra commutation relations
$[J_{x,\m}^a, J_{y,\n}^b] = \d_{xy} \ \d_{\m\n} f^{abc}J_{x,\m}^c$.  And let
$G_x$ be the generator of local gauge transformations that is defined by
$(1 + \sum_x \e_xG_x)F(U) = F({^gU})$, where $g_x = 1 + \e_x$ is an
infinitesimal local gauge transformation.  These generators
satisfie the Lie algebra commutation relations of the local gauge group of the
lattice
$[G_x^a, G_y^b] = \d_{xy} \ f^{abc}G_x^c$, and may be expressed as a linear
combination of the $J$'s.  A hamiltonian with the desired properties is
given by
\eqn\glgauge{\eqalign{
H_{\rm FP} & = H_{\rm inv} - (v, G)^{\dag}  \cr
H_{\rm inv} & = \sum_x J_{x,\m}( \ - J_{x,\m} - [J_{x,\m},S_W] \ )  \cr
(v, G) & = \sum_x v_x G_x,
}}
where $\dag$ is the adjoint with respect to the inner product $dU$, and
$v_x^a$ is a site variable with values in the Lie algebra.   Indeed,
the argument of sec.~3 holds here, with the substitution
$S_{\rm YM}(A) \to S_{\rm W}(U)$, and shows that the probability distributions 
$P_v$ for different $v$, defined by $H_{\rm FP}P = 0$,
are gauge equivalent to each other $P_v \sim P_{v'}$ and to
$P_W$.  As in sec.~3, we choose $v_x^a(U)$ so the infinitesimal gauge
transformation $g_x = 1 + \e \ t^a v_x^a$ is the direction of steepest descent
in gauge orbit directions of a minimizing functional $F(U)$.   A convenient
choice is 
$F(U) = \sum_{\langle xy \rangle}\tr( I - U_{\langle xy \rangle})$, where the
sum extends over all links $\langle xy \rangle$ of the lattice.

\appendix C{Ward Identity}

	In sec.~3 we showed that probability distributions $P_v(A)$ for different $v$
are gauge equivalent, $P_v(A) \sim P_{v'}(A)$.  Another way to make gauge
equivalent distributions is by making a local gauge transformation, 
because, for all gauge-invariant observables $W(A)$, this cannot change the expectation value 
$\int dA \ W(A) P_v(A) = \int dA \ W(A) P_v({^gA})$, and we have 
$P_v(A) \sim P_v({^gA})$.  If the class of gauge-equivalent distributions 
$P_v(A)$ that was introduced in sec.~3 is large enough, then the gauge
transformation corresponds to a change of $v$, 
\eqn\absorb{\eqalign{
P_v({^gA}) = P_{v'}(A)
}} 
for some $v'$.  This is in fact the case, and provides a Ward identity.

	To prove this, we apply the infinitesimal local gauge transformation
$1 + (\e, G)$, 
where $(\e, G) \equiv \int d^4x \ \e^a(x) G^a(x)$, 
to the time-independent Fokker-Planck equation \fpeq and \dech,
\eqn\absorb{\eqalign{
[ 1 + (\e, G)] \ [H_{\rm inv} + (G, v)] \ P_v = 0.
}}
From the commutation relations
\eqn\cre{\eqalign{
& [(\e, G), H_{\rm inv}] = 0   \cr
& [(\e, G), G^a(x)] = - f^{abc} \ \e^b(x) \ G^c(x),	 
}}
we obtain
\eqn\crevg{\eqalign{
[(\e, G), (G,v)] = (G, \d v),
}}
where 
\eqn\deltav{\eqalign{
\d v^a \equiv [(\e, G), v^a] + f^{abc} \e^b v^c,
}} 
and, to first order in $\e$,
\eqn\trfph{\eqalign{
[H_{\rm inv} + (G, v + \d v)] \ [1 + (\e, G)] \ P_v = 0.
}}
Note that while $v^a(x)$ and $\e^a(x)$ are both local gauge transformations, 
$v^a(x) = v^a(x, A)$ depends on $A$, but $\e^a(x)$, by assumption, does not.
By comparison with the defining equation for the probability distribution
$P_{v+\d v}$,
\eqn\trfpha{\eqalign{
[H_{\rm inv} + (G, v + \d v)] \  P_{v+\d v} = 0,
}}
we conclude that the gauge-transformed probability distribution
$[1 +  (\e, G)] \ P_v(A) = P_v(A + D\e)$ coincides with $P_{v + \d v}$, 
\eqn\relprob{\eqalign{
P_v(A + D\e) = P_{v + \d v}(A),
}}
where $\d v$ is given above.  This states how a gauge transformation is
absorbed by a change in $v$, and provides the Ward identity.   

This identity is inherited by the functionals we introduced, the quantum
effective action $\Gamma_v$ and the quantum effective drift force 
$Q_v$, and we have
\eqn\relgq{\eqalign{
\Gamma_v(A + D\e) & = \Gamma_{v + \d v}(A)     \cr
{\cal Q}_v^a(x,A + D\e) 
   & = {\cal Q}^a_{v + \d v}(x,A) - f^{abc}\e^b(x){\cal Q}^c_{v + \d v}(x,A). 
}}

	We now specialize to $v = a^{-1} \p \cdot A$, and obtain
\eqn\spdeltav{\eqalign{
\d v^a 
  & = a^{-1} \p \cdot D^{ac}(A) \ \e^c + a^{-1}f^{abc} \e^b \ \p \cdot A^c  
              \cr
  & = a^{-1} D^{ac}(A) \cdot \p \ \e^c 
   = a^{-1} [ \p^2 \e^a + f^{abc} A_\m^b \ \p_\m \e^c].
}}
Only the derivative of $\e$ appears here because, for 
$v = a^{-1} \p \cdot A$, the probability distribution $P_v(A)$ is invariant
under global (x-independent) gauge transformations.  We further specialize
to a linear dependence of $\e$ on $x$, $\e^a(x) = \eta^a_\m x_\m$, where the
$\eta^a_\m$ are infinitesimal constants.  In this case we have
\eqn\lineps{\eqalign{
\d v^a = a^{-1} f^{abc} A_\m^b \ \eta^c_\m.
}}
Although this breaks Lorentz invariance, it does not break translational
invariance, so the perturbed hamiltonian 
defined by
\eqn\deltahfp{\eqalign{
H_{\rm FP} + \d H_{\rm FP} & = H_{\rm inv} + (G, v) + (G, \d v)  \cr
   & = H_{\rm inv} + a^{-1}(G, \p \cdot A) 
   + a^{-1}(G, A_\m \times \eta_\m),
}}
where $(A_\m \times \eta_\m)^a \equiv f^{abc} A_\m^b \eta_\m^c$,
is translationally invariant,
even though 
$A_\m^a + D_\m^{ac} \e^c = A_\m^a + \eta^a_\m + f^{abc}A_\m^b \ \eta_\n^c
x_\n$, has an explicit $x$-dependence.  Moreover the inhomogeneous term 
$a^{-1}\p^2 \e$ in $\d v$ vanishes with this choice of $\e$, so $A = 0$ 
remains the classical vacuum.  Without further calculation we conclude that
the transformed quantum effective action 
$\Gamma_v(A + D\e) = \Gamma_{v + \d v}(A)$ 
is a translationally invariant functional of $A$ for $v = a^{-1}\p \cdot A$ and
$\e^a(x) = \eta^a_\m x_\m$, with $A = 0$ as classical vacuum.  

	More generally, we note that the gauge field $A_\m^a$ appears undifferentiated
in $\d v^a = a^{-1} f^{abc}A_\m^b \times \eta_\m^c$,  
whereas it is differentiated in $v^a = a^{-1} \p \cdot A^a$.  This
means that the perturbation $\d H_{\rm FP}$ 
is softer than the unperturbed hamiltonian $H_{\rm FP}$ or, in other words,
less divergent in the ultraviolet.  If we calculate with the original
hamiltonian, we get a certain number of divergent constants in the
correlators.  The result of a gauge transformation $\e^a = \eta^a_\m x_\m$ 
on these correlators must agree with a calculation using the soft
perturbation.  This constrains the divergent constants.

\appendix D{Solution for $\Gamma^{(4)}$ and $\Gamma^{(n)}$}

The solution for $\Gamma^{(4)}$ and higher coefficient functions is similar
to the solution for $\Gamma^{(3)}$ found in sec.~7.  We differentiate
\diffeqg\ with respect to $A_{x_i}$ four times and obtain, after 
setting~$A = 0$, 
\eqn\hjfour{\eqalign{
    & \Gamma^{(2)}_{x_1, x} 
\ ( \Gamma^{(4)}_{x,x_2,x_3, x_4} +  Q^{(3)}_{x;x_2,x_3, x_4} )    
 + \sum {\rm part}(4,1)   \cr
&  + \ \Gamma^{(3)}_{x_1, x_2, x} \ 
( \Gamma^{(3)}_{x,x_3,x_4}
+  Q^{(2)}_{x;x_3,x_4} ) 
 + \sum {\rm part}(4,2)  = 0,
 }}
where ${\Gamma^{(2)}}$ and ${\Gamma^{(3)}}$ are already known, and we have
again used ${\Gamma^{(2)}} = - {Q^{(1)}}$.  Here $\sum {\rm part}(n,n_1)$
is the sum over all partitions of the set of $n$ objects,
$x_1, x_2, ... x_n$, into subsets of $n_1$ and $n_2 = n - n_1$ objects.
In terms of the fourier transforms
\eqn\fourierc{\eqalign{
{Q^{(3)}}_{\m_1\m_2\m_3\m_4}^{a_1 a_2 a_3 a_4}(x_1;x_2,x_3, x_4)  
  = (2\pi)^{-12}
& \int d^4k_1 d^4k_3 d^4k_3 d^4k_4 
 \exp(i \sum_{i=1}^4 k_i \cdot x_i)  
\cr & \times \d(k_1 + k_2 + k_3 + k_4) \ 
 {\tilde{Q}^{(3)}}{_{\m_1 \m_2 \m_3 \m_4}^{a_1 a_2 a_3 a_4}}(k_1;k_2,k_3,k_4)  
}}
\eqn\fourierd{\eqalign{
{\Gamma^{(4)}}_{\m_1\m_2\m_3\m_4}^{a_1 a_2 a_3 a_4}(x_1,x_2,x_3, x_4)  
  = (2\pi)^{-12}
& \int d^4k_1 d^4k_3 d^4k_3 d^4k_4 
 \exp(i \sum_{i=1}^4 k_i \cdot x_i)  
\cr & \times \d(k_1 + k_2 + k_3 + k_4) \ 
 {\tilde{\Gamma}^{(4)}}{_{\m_1 \m_2 \m_3 \m_4}^{a_1 a_2 a_3
a_4}}(k_1,k_2,k_3,k_4),   }}
where 
${\tilde{Q}^{(3)}}{_{\m_1 \m_2 \m_3\m_4}^{a_1 a_2 a_3a_4}}(k_1;k_2,k_3,k_4)$
and
${\tilde{\Gamma}^{(4)}}{_{\m_1\m_2\m_3\m_4}^{a_1 a_2a_3a_4}}(k_1,k_2,k_3,k_4)$ 
are defined only for $k_1 + k_2 + k_3 + k_4 = 0$, the equation for
${\tilde{\Gamma}^{(4)}}$ reads
\eqn\hjthreeb{\eqalign{
   \tilde{\Gamma}^{(2)}_{\m_1\n_1}(k_1) \ 
{\tilde{\Gamma}^{(4)}}{_{\n_1\m_2\m_3\m_4}^{a_1a_2a_3a_4}}(k_1,k_2,k_3,k_4) 
   +  \sum{\rm part}(4,1)  
   = - \ {H^{(4)}}{_{\m_1\m_2\m_3\m_3}^{a_1 a_2 a_3a_4}}(k_1,k_2,k_3,k_4),
 }}
where
\eqn\efsfofq{\eqalign{
{H^{(4)}}{_{\m_1\m_2\m_3\m_4}^{a_1 a_2 a_3a_4}}(k_1,k_2,k_3,k_4)  \equiv
 & \ \tilde{\Gamma}^{(2)}_{\m_1\m}(k_1) \ 
{\tilde{Q}^{(3)}}{_{\m\m_2\m_3\m_4}^{a_1 a_2 a_3 a_4}}(k_1;k_2,k_3,k_4)
+  \sum{\rm part}(4,1)  \cr
& + {R^{(4)}}{_{\m_1\m_2\m_3\m_4}^{a_1 a_2 a_3a_4}}(k_1,k_2,k_3,k_4)  
}}
\eqn\efidel{\eqalign{
{R^{(4)}}{_{\m_1\m_2\m_3\m_4}^{a_1 a_2 a_3a_4}}(k_1,k_2,k_3,k_4) &  
 \equiv 
{\tilde{\Gamma}^{(3)}}{_{\m_1\m_2\m}^{a_1 a_2 a}}(k_1, k_2, -k_3-k_4) 
\cr 
& \times [ \ {\tilde{\Gamma}^{(3)}}
{_{\m\m_3\m_4}^{a a_3a_4}}(-k_1-k_2,k_3,k_4) 
+ {\tilde{Q}^{(2)}}{_{\m\m_3\m_4}^{a a_3 a_4}}(-k_1-k_2;k_3,k_4) \ ]  \cr 
& \ \ \ \ \ \ \ \ \ \ \ \ \ \ \ \ \ \ \ \ \ \ + \sum {\rm part}(4,2).
}}

	To solve \hjthreeb, we project on each argument with a transverse transverse
or longitudinal projector to obtain
\eqn\solvegfour{\eqalign{
{\tilde{\Gamma}^{(4){\rm TTTT}}}
{_{\m_1\m_2\m_3\m_4}^{a_1 a_2 a_3 a_4}}(k_1,k_2,k_3, k_4) = - \
 [T(k_1^2) & + T(k_2^2) + T(k_3^2) + T(k_4^2)]^{-1}  \cr
& \times
{H^{(4){\rm TTTT}}}
{_{\m_1\m_2\m_3\m_4}^{a_1 a_2 a_3 a_4}}(k_1,k_2,k_3, k_4),
}}
\eqn\solvegfoura{\eqalign{
{\tilde{\Gamma}^{(4){\rm LTTT}}}
{_{\m_1\m_2\m_3\m_4}^{a_1 a_2 a_3 a_4}}(k_1,k_2,k_3, k_4) = - \
 [a^{-1}L(k_1^2) & + T(k_2^2) + T(k_3^2) + T(k_4^2)]^{-1}  \cr
& \times
{H^{(4){\rm LTTT}}}
{_{\m_1\m_2\m_3\m_4}^{a_1 a_2 a_3 a_4}}(k_1,k_2,k_3, k_4),
}}
etc.

	The formula for $\Gamma^{(n)}$ for arbitrary $n$ is similar.  
Each $\Gamma^{(n)}$ is expressed explicity and uniquely in terms of 
$Q^{(n-1)}$ and of $\Gamma^{(2)}$ to $\Gamma^{(n-1)}$ which are already
known.  It is given by a symmetrized sum of products of two factors, as in
eqs.~\efsfofq\ and \efidel, to which is applied a transverse or longitudinal
projector onto each argument, and a division by 
$\sum_{i=1}^n \Gamma_i^{(2)}(k_i^2)$, where 
$\Gamma_i^{(2)}(k^2) = {\rm T}(k^2)$ or 
$\Gamma_i^{(2)}(k^2) = a^{-1}{\rm L}(k^2)$.  
This gives all the $\Gamma^{(n)}$ uniquely in terms of $Q^{(1)}$ to
$Q^{(n-1)}$. 

\appendix E{Evaluation of $\Gamma_{\rm gt}^{(3)}$} 

	We evaluate $\Gamma_{\rm gt}^{(3)}$, using the formulas of sec.~7, with the
substitutions~\subst.  From~\defsofq\ and \tktwo, we obtain
\eqn\defsymgt{\eqalign{
H^{(3)}_{\m_1\m_2\m_3}(k_1,k_2,k_3) 
   & = a^{-1}\tilde{\Gamma}^{(2)}_{\m_1\m}(k_1) \ 
  {\tilde{K}_{\rm gt}^{(2)}}{_{\m\m_2\m_3}}(k_1;k_2,k_3)  
   +  (\rm cyclic)     
	   \cr
    & = i a^{-1} g 
    \ \Big( \ (k_3)_{\m_3} \ [ \ \tilde{\Gamma}^{(2)}_{\m_1\m_2}(k_1)
      - (1 \leftrightarrow 2) \ ]
   +  (\rm cyclic) \ \Big) .
}}

	We apply transverse or longitudinal projectors to each Lorentz index, and use
$\tilde{\Gamma}^{(2)}_{\l\m}(k) = T(k^2) P_{\l\m}^{\rm T}(k)
+ a^{-1}L(k^2) P_{\l\m}^{\rm L}(k)$ to obtain
\eqn\hthrgt{\eqalign{
{H^{(3)}}
  {_{\m_1\m_2\m_3}^{\rm T \ T \ T}}(k_1,k_2,k_3) & = 0,  \cr
{H^{(3)}}
  {_{\m_1\m_2\m_3}^{\rm T \ T \ L}}(k_1,k_2,k_3) & = 
  i a^{-1}g \ (T_1 - T_2) \ (k_3)_{\m_3} 
  \ [P^{\rm T}(k_1)P^{\rm T}(k_2)]_{\m_1\m_2} ,         \cr
{H^{(3)}}
  {_{\m_1\m_2\m_3}^{\rm T \ L \ L}}(k_1,k_2,k_3) & = 
   i a^{-1}g \ \Big( \ (a^{-1}L_3 - T_1)
   \ (k_2)_{\m_2} \ [P^{\rm T}(k_1)P^{\rm L}(k_3)]_{\m_1\m_3}    
  -  (2 \leftrightarrow 3) \Big)    \cr
{H^{(3)}}
  {_{\m_1\m_2\m_3}^{\rm L \ L \ L}}(k_1,k_2,k_3) & = 
i a^{-2}g \ \Big( \  (L_2 - L_3) \ 
(k_1)_{\m_1} \ [P^{\rm L}(k_2)P^{\rm L}(k_3)]_{\m_2\m_3}  
  + {\rm cyclic} \Big) ,
}}
where we have used the notation $T_i \equiv T(k_i^2)$ and 
$L_i \equiv L(k_i^2)$.
From \solvegthr\ and \solvegthra, we obtain finally
\eqn\gthrgttt{\eqalign{
{\tilde{\Gamma}_{\rm gt}^{(3)}}
  {_{\m_1\m_2\m_3}^{\rm T \ T \ T}}(k_1,k_2,k_3) & = 0,    \cr
{\tilde{\Gamma}_{\rm gt}^{(3)}}
  {_{\m_1\m_2\m_3}^{\rm T \ T \ L}}(k_1,k_2,k_3) 
    & = - \ i g \ 
   { {T_1 - T_2} \over {aT_1 + aT_2 + L_3} }
   \ (k_3)_{\m_3}   \ [P^{\rm T}(k_1)P^{\rm T}(k_2)]_{\m_1\m_2},          \cr
{\tilde{\Gamma}_{\rm gt}^{(3)}}
    {_{\m_1\m_2\m_3}^{\rm T \ L \ L}}(k_1,k_2,k_3) 
   & = - \ ia^{-1}g  \ \Big( \ 
    { {L_3 - aT_1} \over {aT_1 + L_2 + L_3} }
   \ (k_2)_{\m_2} \ [P^{\rm T}(k_1)P^{\rm L}(k_3)]_{\m_1\m_3} 
       \cr
   & \ \ \ \ \ \ \ \ \ \   
    -  \ { {L_2 - aT_1} \over {aT_1 + L_2 + L_3} }
     \ (k_3)_{\m_3} 
    \ [P^{\rm T}(k_1)P^{\rm L}(k_2)]_{\m_1\m_2}   \ \Big),    \cr
{\tilde{\Gamma}_{\rm gt}^{(3)}}
     {_{\m_1\m_2\m_3}^{\rm L \ L \ L}}(k_1,k_2,k_3) & = 
    - \ i a^{-1}g \ \Big( \  
 { {L_2 - L_3} \over {L_1 + L_2 + L_3} }
    \ (k_1)_{\m_1} \ [P^{\rm L}(k_2)P^{\rm L}(k_3)]_{\m_2\m_3} 
  + ({\rm cyclic}) \Big) . }}

\appendix F{Evaluation of loop integrals}

	We evaluate the integral that appears in \dstasit\ namely
\eqn\evintt{\eqalign{
 I_T \equiv  { { 1 } \over {   (k^2)^{\a_T+2} \ (d-1) \ (2\pi)^d } }
  \int d^dk_1    
 \  { { \  k_1^2 \ k^2 - (k_1 \cdot k)^2 \  } \over 
     \ { \  (k_1^2)^{1 + \a_L} \ [(k - k_1)^2]^{1 + \a_L} \ } }.
}}
We write this as
\eqn\evintta{\eqalign{
 I_T =  { { 1 } \over 
     { (k^2)^{\a_T+2} \ (d-1) \ \Gamma^2(1 + \a_L) } }
  \ \int_0^\infty d\a d\b \ \a^{\a_L} \ \b^{\a_L} \ R_T,
}}
where
\eqn\evinttb{\eqalign{
 R_T \equiv  (2\pi)^{-d}\int d^dk_1    \ 
 [ \  k_1^2 \ k^2 - (k_1 \cdot k)^2 \  ] 
\ \exp[ \ - \ \a k_1^2 - \b(k_1 - k)^2 \ ]. 
}}
We complete the square in the exponent,
$$\a k_1^2 + \b(k_1 - k)^2 = (\a + \b)p^2 + (\a + \b)^{-1}\a\b k^2,$$
where $p = k_1 - (\a + \b)^{-1}\b k$, and obtain
\eqn\evinttc{\eqalign{
 R_T & = \exp[ \ - \ (\a + \b)^{-1}\a\b k^2 \ ] \ 
    (2\pi)^{-d}\int d^dp    \ 
 [ \  p^2 \ k^2 - (p \cdot k)^2 \  ] \ \exp[ \ - \ (\a + \b)p^2 \ ]    \cr
  & = { {(d-1) \ k^2} \over {2 \ (4\pi)^{d/2} \ (\a + \b)^{1 + d/2} } } \ 
      \exp[ \ - \ (\a + \b)^{-1}\a\b k^2 \ ].
}}
This gives
\eqn\evinttd{\eqalign{
I_T = { { 1} \over {2 \ (4\pi)^{d/2} (k^2)^{\a_T + 1} \ \Gamma^2(1 + \a_L) } } 
    \ S_T,
}}
where
\eqn\evintte{\eqalign{
S_T = \int_0^\infty d\a d\b \ 
   { {\a^{\a_L} \ \b^{\a_L} } \over { (\a + \b)^{1 + d/2} } }
    \ \exp[ \ - \ (\a + \b)^{-1}\a\b k^2 \ ].
}}
We insert the identity $1 = \int_0^\infty d \s \ \d(\a + \b - \s)$,
and change variables according to $\a = \s \a'$ and $\b = \s \b'$.  This
gives, after dropping primes,
\eqn\evinttf{\eqalign{
S_T & = \int_0^\infty d\a d\b d\s \ \d(\a + \b -1) \  \a^{\a_L} \ \b^{\a_L}
   \s^{2\a_L -d/2}
    \ \exp[ \ - \ \a\b \s k^2 \ ]   \cr
  & = (k^2)^{d/2 - 2\a_L -1} \ \Gamma(2\a_L + 1 - d/2)  
     \int_0^\infty d\a d\b \ \d(\a + \b -1) 
					 \ \a^{d/2 - \a_L -1} \ \b^{d/2 - \a_L -1}  \cr
 & = (k^2)^{d/2 - 2\a_L -1} \ { {\Gamma(2\a_L +1 - d/2) \ \Gamma^2(d/2 -\a_L)} 
      \over { \Gamma(d - 2\a_L) } }.
}}
We obtain finally,
\eqn\evinttg{\eqalign{
I_T & =  { { \Gamma(2\a_L + 1 - d/2) \ \Gamma^2(d/2 - \a_L) } 
     \over { 2 \ (4\pi)^{d/2} \ \Gamma^2(1 + \a_L) \ \Gamma(d - 2\a_L) } } \cr
  & =  { { \Gamma(2\a_L -1) \ \Gamma^2(2 - \a_L) } 
     \over { 2 \ (4\pi)^2 \ \Gamma^2(1 + \a_L) \ \Gamma(4 - 2\a_L) } }, 
}}
for $d = 4$.

	We also evaluate the integral that appears in \dsllandasil,
\eqn\evintl{\eqalign{
I_L \equiv { {- \ 2} \over { (k^2)^{1+\a_L} \ (2\pi)^d } } 
     \int d^dk_1 
   \ { { k^2 k_1^2 - (k \cdot k_1)^2 } \over
    { (k_1^2)^{2 + \a_T} \ [(k_1 - k)^2]^{\a_L} [ (k_1 - k)^2 + k^2 ]   } }.
}}
It contains the denominator $[ (k_1 - k)^2 + k^2 ]$ that comes from
the non-local vertex.  This integral is convergent in the ultraviolet for 
$d < 4 + 2\a_T + 2\a_L$.  We shall evaluate it for $d$ satisfying this
condition, and then continue in $d$.  We write it as
\eqn\evintla{\eqalign{
 I_L = 
  { {- \ 2} \over 
{ (k^2)^{1+\a_L} \Gamma(2 + \a_T) \ \Gamma(\a_L) } } 
  \ \int_0^\infty d\a d\b d\g \ \a^{1+\a_T} \ \b^{\a_L - 1} 
   \exp(- \g k^2) \ R_L,
}}
where
\eqn\evintlb{\eqalign{
 R_L \equiv  (2\pi)^{-d}\int d^dk_1    \ 
 [ \  k_1^2 \ k^2 - (k_1 \cdot k)^2 \  ] 
\ \exp[ \ - \ \a k_1^2 - (\b + \g)(k_1 - k)^2 \ ]. 
}}
We complete the square in the exponent,
$$\a k_1^2 + (\b + \g)(k_1 - k)^2 
   = (\a + \b + \g)p^2 + (\a + \b + \g)^{-1}\a(\b + \g) k^2,$$
where $p = k_1 - (\a + \b + \g)^{-1}(\b + \g) k$, and obtain
\eqn\evintlc{\eqalign{
 R_L & = \exp[ \ - \ (\a + \b + \g)^{-1}\a(\b + \g) \ k^2 \ ] 
                 \cr 
    & \ \ \ \ \ \ \ \ \ \ \ \ \ \ \ \times (2\pi)^{-d}\int d^dp    \ 
 [ \  p^2 \ k^2 - (p \cdot k)^2 \  ] \ \exp[ \ - \ (\a + \b + \g)p^2 \ ]    \cr
  & = { {(d-1) \ k^2} \over {2 \ (4\pi)^{d/2} \ (\a + \b + \g)^{1 + d/2} } } \ 
      \exp[ \ - \ (\a + \b + \g)^{-1}\a(\b+\g) k^2 \ ].
}}
This gives
\eqn\evintld{\eqalign{
I_L = - \ { {(d-1)} \over 
{ (k^2)^{\a_L} \ (4\pi)^{d/2} \ \Gamma(2 + \a_T) \Gamma(\a_L)} } 
\ S_L,
}}
where
\eqn\evintle{\eqalign{
S_L = \int_0^\infty d\a d\b d\g \ 
   { {\a^{1+\a_T} \ \b^{\a_L-1} } \over { (\a + \b + \g)^{1 + d/2} } }
    \ \exp[ \ - \ \g k^2 - \ (\a + \b + \g)^{-1}\a(\b + \g) k^2 \ ].
}}
We insert the identity $1 = \int_0^\infty d \s \ \d(\a + \b + \g - \s)$,
and change variables according to $\a = \s \a'$, $\b = \s \b'$
 and $\g = \s \g'$.  This
gives, after dropping primes,
\eqn\evintlf{\eqalign{
S_L & = \int_0^\infty d\a d\b d\g d\s \ \d(\a + \b + \g -1) 
\  \a^{1 + \a_T} \ \b^{\a_L - 1}
   \s^{1 + \a_T + \a_L -d/2}     \cr
   & \ \ \ \ \ \ \ \ \ \ \ \ \ \ \ \ \ \ \ \ \ \ \ \ \ \ \times
    \ \exp\{ \ - \s [\g + \a(\b + \g)] k^2 \}   \cr
  & = (k^2)^{-\a_T - \a_L - 2 + d/2} \ \Gamma(2 + \a_T +  \a_L -d/2) \ M_L ,
}} 
The argument of the $\Gamma$-function is positive in the region of
convergence of the integral, $d < 4 + 2\a_T + 2\a_L$. 
Here $M_L$ is the finite integral 
\eqn\evintlg{\eqalign{
  M_L & \equiv  \int_0^\infty d\a d\b d\g \ \d(\a + \b + \g -1) \  
					\a^{1 + \a_T} \ \b^{\a_L -1}[\a(\b + \g) + \g]^{\a_L}   \cr
    & = \int_0^1 d\b \int_0^{1-\b}d\a \ 
		\a^{1 + \a_T} \ \b^{\a_L -1}(1 - \a^2 - \b)^{\a_L},  
}}
where we have used $\a_T + 2 \a_L = - \ (4 - d)/2$.  
This gives
\eqn\evintlda{\eqalign{
I_L & = - \ { {(d-1)} \ \Gamma(- \a_L)\over 
   { \ (4\pi)^{d/2} \ \Gamma(2 + \a_T) \Gamma(\a_L)} } \ M_L   \cr
    & = \ { {(d-1)} \ \Gamma(1 - \a_L)\over 
    { \ (4\pi)^{d/2} \ \Gamma(2 + \a_T) \Gamma(1+\a_L)} } \ M_L.
}}
Note that $I_L$ is negative
in the region of convergence of the integral, but after the continuation in
$d$, it is positive.  

	To evaluate $M_L$ we change variable to 
$x = \a^2$, and obtain
\eqn\evintlh{\eqalign{
  M_L  = (1/2)\int_0^1 d\b \int_0^{(1-\b)^2} dx \ 
		x^{\a_T/2} \ \b^{\a_L -1}(1 - \b - x)^{\a_L},  
}}
and upon changing variables to $x = (1-\b)y$, we get
\eqn\evintli{\eqalign{
  M_L  = (1/2)\int_0^1 d\b \int_0^{1-\b} dy \ 
		y^{\a_T/2} \ \b^{\a_L -1}(1-\b)^{1 + \a_L + \a_T/2}(1 - y)^{\a_L}.  
}}
We again use $\a_T + 2 \a_L = - \ (4 - d)/2$ to write this as
\eqn\evintlj{\eqalign{
  M_L  = (1/2)\int_0^1 dy \ y^{\a_T/2} (1 - y)^{\a_L}
   \int_0^{1-y} d\b 
		 \ \b^{\a_L -1}(1-\b)^{d/4}.  
}}
This is integrable by quadrature for $d = 4$, and in this case it gives
\eqn\evintlk{\eqalign{
  M_L & = (1/2)\int_0^1 dy \ y^{- \a_L} 
   \Big( { {(1 - y)^{2\a_L} } \over {\a_L} } 
       - { {(1 - y)^{2\a_L+1} } \over {\a_L+1} } \Big)   \cr 
  & = (1/2) \Big( { {\Gamma(1-\a_L) \ \Gamma(1+ 2\a_L) }
    \over { \a_L \ \Gamma(2 + \a_L)} }
   - { {\Gamma(1-\a_L) \ \Gamma(2+ 2\a_L) }
    \over { (\a_L+1) \ \Gamma(3 + \a_L)} }   \Big)   \cr
  & = { { (-\a_L^2 + 2\a_L +2) \ \Gamma(1-\a_L) \ \Gamma(2\a_L+1)  }
    \over { 2\a_L(\a_L+1) \ \Gamma(\a_L+3)} },  
}}
where we used $\a_T = - 2 \a_L$.  This gives finally
\eqn\evintll{\eqalign{
I_L =
   { { 3 \ (-\a_L^2 + 2\a_L +2) \ \Gamma^2(1 - \a_L) \ \Gamma(2\a_L+1) } \over 
  { \ 2 \ (4\pi)^2 \ \a_L \ \Gamma(2-2\a_L) \Gamma(\a_L+2) \ \Gamma(\a_L+3)}}.
}}

\footatend\vfill\supereject\immediate\closeout\rfile\writestoppt
\baselineskip=14pt\centerline{{\bf References}}\bigskip{\frenchspacing%
\parindent=20pt\escapechar=` \input refs.tmp\vfill\eject}\nonfrenchspacing

%%%%%%%%%%%%%%%%%%%% FIGURES %%%%%%%%%%%%%%%%%%%%%%%%

\vfill\eject\immediate\closeout\ffile{\parindent40pt
\baselineskip14pt\centerline{{\bf Figure Captions}}\nobreak\medskip
\escapechar=` \input figs.tmp\vfill\eject}

%%%%%%%%%%%%%%%%%%%%%%%%%%%%%%%%%%%%%%%%%%%%%%%%%%%%%%%%%%%%%%%%%%%

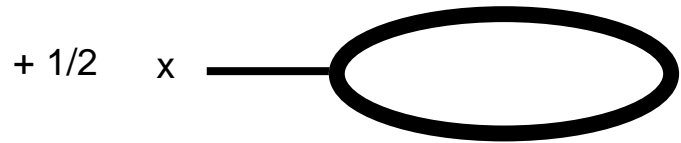
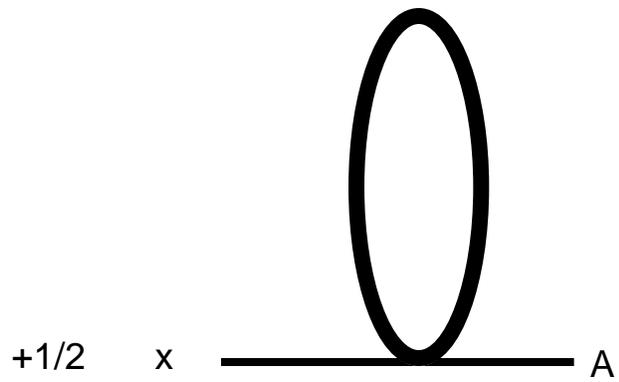
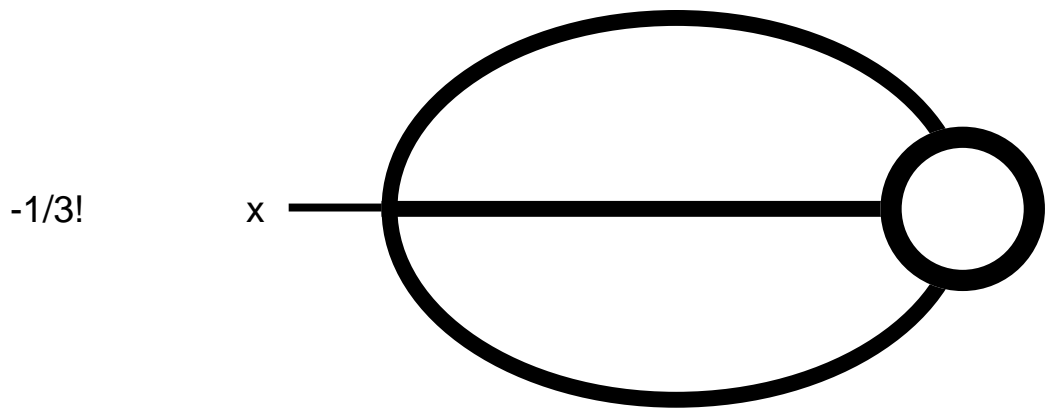
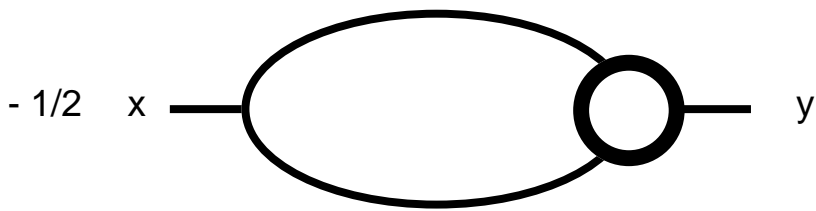
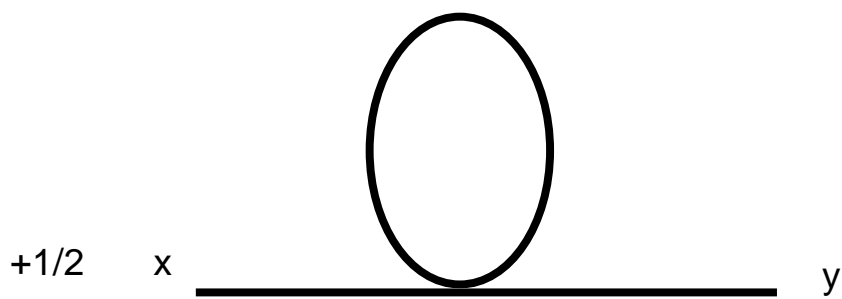
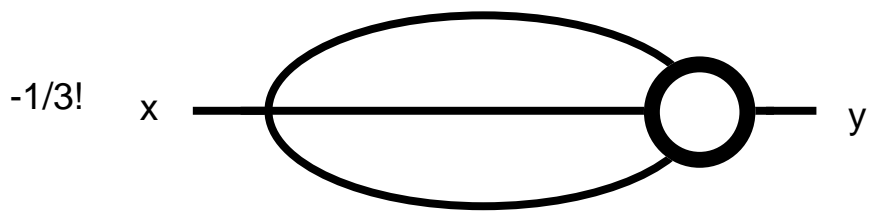
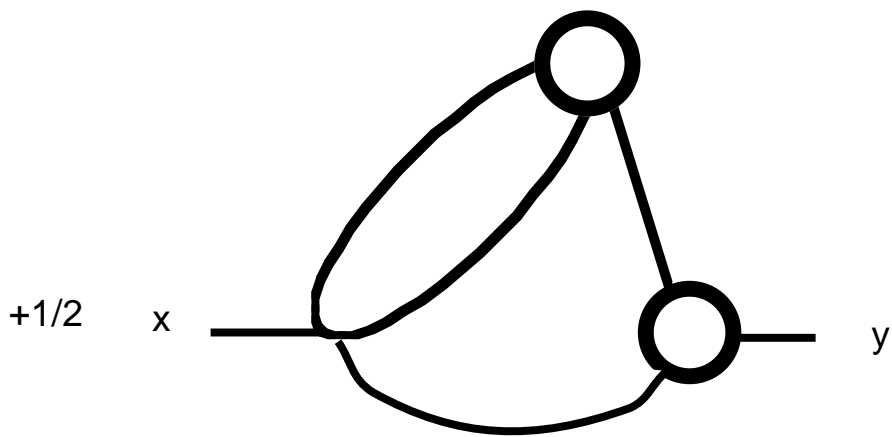
\bye